\documentclass[reprint,longbibliography,twocolumn,superscriptaddress,
showpacs,preprintnumbers,
notitlepage,amsmath,amssymb,
aps,
prx,
]{revtex4-1}

\usepackage{graphicx}
\usepackage{dcolumn}
\usepackage{bm}
\usepackage{diagbox}

\usepackage{graphicx}
\usepackage{enumerate}
\usepackage{subfigure}
\usepackage{epstopdf}
\usepackage[colorlinks=true,citecolor=blue]{hyperref}
\usepackage{amssymb}
\usepackage{framed}
\usepackage{tabularx}
\usepackage{booktabs}
\usepackage{float}
\usepackage[table]{xcolor}
\usepackage{array}
\usepackage{tikz}
\usetikzlibrary{shapes,snakes}
\hypersetup{colorlinks=true}

\begin{document}
\title{Tensor network representations of fermionic crystalline topological phases on two-dimensional lattices}
\author{Jian-Hao Zhang}
\email{jianhaozhang11@cuhk.edu.hk}
\affiliation{Department of Physics, The Chinese University of Hong Kong, Shatin, New Territories, Hong Kong, China}
\author{Shuo Yang}
\email{shuoyang@tsinghua.edu.cn}
\affiliation{State Key Laboratory of Low Dimensional Quantum Physics and Department of Physics, Tsinghua University, Beijing 100084, China}
\affiliation{Frontier Science Center for Quantum Information, Beijing 100084, China}

\begin{abstract}
We investigate the tensor network representations of fermionic crystalline symmetry-protected topological (SPT) phases on two-dimensional lattices. As a mapping from virtual indices to physical indices, projected entangled-pair state (PEPS) serves as a concrete way to construct the wavefunctions of 2D crystalline fermionic SPT (fSPT) phases protected by 17 wallpaper group symmetries, for both spinless and spin-1/2 fermions. Based on PEPS, the full classification of 2D crystalline fSPT phases with wallpaper groups can be obtained. Tensor network states provide a natural framework for studying 2D crystalline fSPT phases. 
\end{abstract}

\newcommand{\lra}{\longrightarrow}
\newcommand{\xra}{\xrightarrow}
\newcommand{\ra}{\rightarrow}
\newcommand{\bs}{\boldsymbol}
\newcommand{\ul}{\underline}
\newcommand{\1}{\text{\uppercase\expandafter{\romannumeral1}}}
\newcommand{\2}{\text{\uppercase\expandafter{\romannumeral2}}}
\newcommand{\3}{\text{\uppercase\expandafter{\romannumeral3}}}
\newcommand{\4}{\text{\uppercase\expandafter{\romannumeral4}}}
\newcommand{\5}{\text{\uppercase\expandafter{\romannumeral5}}}
\newcommand{\6}{\text{\uppercase\expandafter{\romannumeral6}}}

\maketitle
\tableofcontents

\section{Introduction}
Since the discovery of Haldane phase in one-dimensional spin chain \cite{Haldane,AKLT}, the interplay between symmetry and topology has been one of the most crucial issues in condensed matter physics during the past few decades. Representative examples including topological insulators (TI) and superconductors (TSC) have been intensively studied theoretically and experimentally \cite{KaneRMP,ZhangRMP}. In recent years, a large class of nontrivial topological phases require symmetry protections: they can be smoothly deformed to a trivial phase in the absence of global symmetry. Such symmetry-protected topological (SPT) phases are systematically constructed and classified \cite{pollmann10,chen11a,chen11b,XieChenScience,cohomology,invertible1,invertible2,invertible3,Kapustin2014,wen15,special,general1,general2,Kapustin2015,Kapustin2017,fidkowski10, fidkowski11,wangc-science,invertible2,ChongWang2014,Witten,LevinGu,Gu-Levin,gauging1,threeloop,ran14,wangj15,wangcj15,lin15,gauging3,dimensionalreduction,gauging2,2DFSPT,braiding,Ashvin2013,ChongWang2013,XieChen2015,ChenjieWang2016,XLQi2013,Senthil2013,Lukasz2013,XieChen2014,ChongWang2014} in both interacting bosonic and fermionic systems. 

On the other hand, tensor networks are extremely suitable for describing SPT phases because nonlocal topological characters of a system are captured by the symmetries of local tensors \cite{Pollman2012,CZX,schuch11,chen11a,Cirac2013,Schuch2014,Read2015,Frank2016,Eisert2017,Frank2017,Cirac2018,Kapustin2018,Frank2021}.  In one-dimensional (1D) systems, all SPT states can be constructed and classified by matrix product state (MPS). In 2D systems, bosonic SPT (bSPT) phases are constructed by projected entangled-pair states (PEPS) and classified by matrix product operators (MPO) acting on arbitrary boundary of PEPS; fermionic SPT (fSPT) phases are constructed by PEPS with graded structure who characterizes the fermion parity. 

In recent years, crystalline symmetry plays a more important role in studying SPT phases because each monocrystal in condensed matter systems corresponds to a specific space group in arbitrary dimension. Crystalline SPT phases are not only of conceptual importance \cite{TCI,Fu2012,ITCI,reduction,building,correspondence,SET,230,BCSPT,Jiang2017,Kane2017,Shiozaki2018,ZDSong2018,defect,realspace,KenX,rotation,LuX,YMLu2018,Cheng2018,Hermele2018,Huang2020PRR,Huang2021PRR,Ning2021} but also have a great possibility to be experimentally realized. As the simplest example of crystalline SPT phases, the crystalline TI is first proposed in free fermion systems and then be realized in many different materials \cite{TCIrealization1,TCIrealization2,TCIrealization3,TCIrealization4}. For free fermion systems, the crystalline SPT phases are systematically constructed and classified by \textit{symmetry indicators}, which are roughly the symmetry representations of band structures at high-symmetry momenta. In interacting bosonic and fermionic systems, a systematic \textit{real-space construction} is established by decorating different phases on lower-dimensional blocks \cite{reduction, realspace,dihedral,wallpaper}. Furthermore, it was pointed out that the classification of crystalline SPT phases are closely related to the SPT phases with internal symmetries. In Ref.~\cite{correspondence}, 
a ``\textit{crystalline equivalence principle}'' is proposed with rigorous mathematical proof: crystalline topological phases with space group symmetry $G$ are in one-to-one correspondence with topological phases protected by the same internal symmetry $G$ but acting in a twisted way. If an element of $G$ is a mirror reflection (orientation-reversing symmetry), it should be regarded as time-reversal symmetry (anti-unitary symmetry). This principle has been confirmed in bosonic \cite{realspace} and 2D interacting fermionic systems \cite{dihedral,wallpaper}. 

In this paper, we apply the tensor network method as a more straightforward and comprehensible way to characterize the crystalline fSPT phases in 2D interacting fermionic systems for both spinless and spin-1/2 fermions. Firstly we define PEPS tensors with graded structures on 2D lattices with wallpaper group symmetry as maps from virtual indices to physical indices, where physical/virtual indices are aligned on vertices/links of 2D lattices. Distinct from the systems with on-site symmetry, the local symmetry of physical and virtual indices might be different in crystalline SPT states. Subsequently, by investigating the virtual indices and physical indices of fermionic MPOs (fMPOs), we obtain all classification data as possible PEPS tensors on the lattice. Then by utilizing the injectivity condition, some PEPS tensors may not be well-defined and some others may correspond to trivial SPT phases, we call them \textit{obstruction} and \textit{trivialization}. An obstruction and trivialization-free PEPS corresponds to a nontrivial 2D crystalline fSPT state, and all of them form the classification group. We use the lattice with \#9 wallpaper group ($cmm$) symmetry as a representative example to highlight this paradigm. Finally, there might be some superposition rules of PEPS tensors, which leads to a nontrivial group structure of the classification of 2D crystalline fSPT phases. According to this paradigm, we obtain the full classifications of 2D crystalline fSPT phases with wallpaper group symmetries (see Table \ref{classification}), confirmed to the results in Refs. \cite{wallpaper} and \cite{resolution} which are yielded by alternative methods (real-space construction and Atiyah-Hirzebruch spectral sequence \cite{AHSS}). 

Tensor network is a powerful tool not only for directly constructing the PEPSs as the explicit ground states of crystalline fSPT phases, but also for studying quantum phase transitions between different topological phases. Therefore, constructing tensor network representations of crystalline fSPT phases lays the important foundation for investigating the topological quantum phase transitions of crystalline fSPT phases. 

The rest of the paper is organized as follows: In Sec. \ref{review}, we review the tensor networks with graded structures for describing interacting fermionic states and their application to 2D lattices with wallpaper group symmetries. In Sec. \ref{general}, we introduce the general paradigm of constructing the PEPS tensor network representations of 2D crystalline fSPT states. In Sec. \ref{example}, we explicitly construct the PEPS tensor network states of 2D crystalline fSPT phases with \#9 wallpaper group symmetry $cmm$ as a concrete example, for both spinless and spin-1/2 fermions. All results are summarized in Table \ref{classification}. Finally, the conclusions and discussions about further applications of tensor network representations of 2D crystalline fSPT phases are presented in Sec. \ref{conclusion}. In Appendix \ref{SES}, we review the mathematical characterization and physical meaning of spins of fermions. In Appendix \ref{point group center}, we summarize all possible physical indices of PEPS tensors we might use. All physical indices should be aligned at the center of a specific point group. In Appendix \ref{AppMPO}, we review the superposition and pentagon equation fMPOs with graded structure.

\section{Tensor networks with the graded structure on the 2D lattice \label{review}}
For fermionic systems, the fermion parity is always conserved and is described by the fermion parity symmetry $\mathbb{Z}_2^f$. To describe this symmetry, we should consider the tensor networks with the graded structure \cite{Frank2017}: A super vector space $V$ has a natural direct sum structure:
\begin{align}
V=V^0\oplus V^1
\end{align}
where vectors in $V^0$ and $V^1$ are called homogeneous vectors, the vector $v\in V^0/V^1$ is said to have even/odd fermion parity. 

\begin{figure}
\begin{tikzpicture}
\tikzstyle{sergio}=[rectangle,draw=none]
	 \draw[thick] (-2,0.5) -- (-1,0.5);
        \draw[thick] (-2.5,-0.5) -- (-1.5,-0.5);
        \draw[thick] (-2,0.5) -- (-2.5,-0.5);
		\draw[thick] (-1,0.5) -- (-1.5,-0.5);
        \draw[thick] (-2.25,0) -- (-2.75,0);
        \draw[thick] (-1.25,0) -- (-0.75,0);
        \path (-3,0) node [style=sergio] {$\tau_1$};
        \path (-0.5,0) node [style=sergio] {$\tau_3$};
        \draw[thick] (-1.5,0.5) -- (-1.25,1);
        \path (-1.25,1.25) node [style=sergio] {$\tau_4$};
		\draw[thick] (-2,-0.5) -- (-2.25,-1);
        \path (-2.25,-1.25) node [style=sergio] {$\tau_2$};
		\draw[densely dashed, thick] (-1.75,0) -- (-1.75,1);
        \path (-2,1) node [style=sergio] {$\mu$};
        \path (-1,-0.5) node [style=sergio] {$A_\mu$};
\path (-3,1) node [style=sergio] {$(a)$};
	 \draw[thick] (1.5,0.5) -- (2.25,0.5);
        \draw[thick] (1.25,0) -- (2,0);
        \draw[thick] (1.5,0.5) -- (1.25,0);
		\draw[thick] (2.25,0.5) -- (2,0);
        \draw[thick] (1.375,0.25) -- (0.75,0.25);
        \draw[thick] (2.125,0.25) -- (2.875,0.25);
        \draw[thick] (1.875,0.5) -- (2.125,1);
		\draw[thick] (1.625,0) -- (1.375,-0.5);
		\draw[densely dashed, thick] (1.75,0.25) -- (1.75,0.75);
\draw[ultra thick,lightgray] (1.25,0.75) -- (4.25,0.75);
\draw[ultra thick,lightgray] (1.25,0.75) -- (0.25,-1.25);
\draw[ultra thick,lightgray] (3.25,-1.25) -- (0.25,-1.25);
\draw[ultra thick,lightgray] (3.25,-1.25) -- (4.25,0.75);
\path (0.25,1) node [style=sergio] {$(b)$};
\path (2.25,-0.25) node [style=sergio] {$R$};
\path (1.75,-1.5) node [style=sergio] {$\partial R$};
		\draw[thick] (1.75,-0.5) -- (1,-0.5);
		\draw[thick] (1.5,-1) -- (0.75,-1);
\draw[thick] (1,-0.5) -- (0.75,-1);
\draw[thick] (1.75,-0.5) -- (1.5,-1);
\draw[thick] (1.125,-1) -- (0.875,-1.5);
\draw[thick] (3,0.5) -- (2.75,0);
\draw[densely dashed, thick] (1.25,-0.75) -- (1.25,-0.25);
\draw[thick] (0.875,-0.75) -- (0.25,-0.75);
\draw[thick] (1.625,-0.75) -- (2.375,-0.75);
\draw[thick] (2.5,-0.5) -- (2.25,-1);
\draw[thick] (2.25,-1) -- (3,-1);
\draw[thick] (2.5,-0.5) -- (3.25,-0.5);
\draw[thick] (2.75,0) -- (3.5,0);
\draw[thick] (3,0.5) -- (3.75,0.5);
\draw[thick] (3.25,-0.5) -- (3,-1);
\draw[thick] (3.75,0.5) -- (3.5,0);
\draw[thick] (3.75,-0.75) -- (3.125,-0.75);
\draw[thick] (4.25,0.25) -- (3.625,0.25);
\draw[thick] (2.625,-1) -- (2.375,-1.5);
\draw[thick] (3.125,0) -- (2.875,-0.5);
\draw[thick] (3.375,0.5) -- (3.625,1);
\draw[densely dashed, thick] (3.25,0.25) -- (3.25,0.75);
\draw[densely dashed, thick] (2.75,-0.75) -- (2.75,-0.25);
\end{tikzpicture}
\caption{(a). A PEPS tensor on a quadrate vertex; (b). The PEPS map $A_R$ from virtual indices on edges in $\partial R$ (the boundary of the region $R$) to the physical indices on vertices in $R$.}
\label{PEPS}
\end{figure}
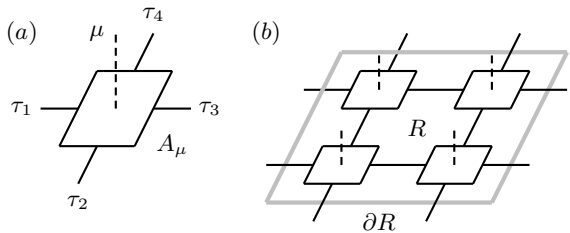

A 2D PEPS can be defined on any lattice $\Gamma$:
\begin{align}
A_\mu=\sum\limits_{\mu=1}^d\sum\limits_{\{\tau_j\}=1}^{\{D_j\}}\left(A_\mu\right)_{\{\tau_j\}}^\mu|\mu\rangle\bigotimes\limits_{\tau_j\in E_\mu}(\tau_j|
\label{PEPS tensor}
\end{align}
for $\forall\mu\in\Gamma$. $E_\mu$ is the set of edges with $\mu$ as an endpoint. Here $\mu$ is the physical index running over the basis for the Hilbert space of a site $\mathbb{C}^d$, and $\tau_j$ is the virtual index of dimension $D_j$ along with the bond $\tau_j$, see Fig. \ref{PEPS}(a). 

Then consider a region $R\in\Gamma$ whose boundary $\partial R$ forms a contractible closed loop. Then we can define a PEPS map in this region as:
\begin{align}
A_R=\bigotimes\limits_{\tau_j\in\partial R}\mathbb{C}^{D_j}\rightarrow\bigotimes\limits_{\mu_k\in R}\mathbb{C}^{d_k}
\end{align}
Here $\tau_j$ is the virtual index across $\partial R$ of dimension $D_j$, and $\mu_k$ is the physical index inside the region $R$ of dimension $d_k$, see Fig. \ref{PEPS}(b). We depict that the region $R$ includes an integer number of unit cells.

We note that the PEPS on the lattice with wallpaper group symmetry is slightly different from the PEPS with on-site symmetry. For PEPS with on-site symmetry, the symmetry groups on both physical and virtual indices are identical; for PEPS on the lattice with wallpaper group symmetry, the symmetry groups on physical indices and virtual indices might be different. This is because the ``on-site'' symmetry groups of physical and virtual indices are given by the wallpaper group symmetry acting internally. We label the effective ``on-site'' symmetry of physical/virtual indices by $G_p$ and $G_v$, respectively. Usually $G_v\subset G_p$.

In order to describe the crystalline SPT phases, we should consider the MPO-injective PEPS with single-blocked condition \cite{Frank2016,Frank2021}. Then for PEPS $A_{\mu_k}$ on the lattice site $\mu_k$ with virtual indices $\tau_j$ [$j=1,2,3,4$, see Fig. \ref{PEPS}(a)], there are some subtleties for investigating the invariances under wallpaper group that all come from the difference between the effective ``on-site'' symmetry on physical and virtual indices ($G_p\ne G_v$). For example, we suppose the PEPS indicated in Fig. \ref{PEPS}(a) is assigned on the lattice site as the center of 2-order dihedral group symmetry $D_2$ which is generated by two reflection symmetry generators $\bs{M}_1$ and $\bs{M}_2$. Some of the virtual indices might be related by operations $g\in D_2$:
\begin{align}
\begin{aligned}
&\bs{M}_1:~\left(\tau_1,\tau_2,\tau_3,\tau_4\right)\mapsto\left(\tau_1,\tau_4,\tau_3,\tau_2\right)\\
&\bs{M}_2:~\left(\tau_1,\tau_2,\tau_3,\tau_4\right)\mapsto\left(\tau_3,\tau_2,\tau_1,\tau_4\right)
\end{aligned}
\end{align}
Hence the effective ``on-site'' symmetries of physical/virtual indices are $G_p=\mathbb{Z}_2\rtimes\mathbb{Z}_2$, $G_v=\mathbb{Z}_2$ (for virtual indices $\tau_1$ and $\tau_3$, $\bs{M}_1$ acts internally; for virtual indices $\tau_2$ and $\tau_4$, $\bs{M}_2$ acts internally). In fact, we have the following short exact sequence:
\begin{align}
0\rightarrow G_v\rightarrow G_p \rightarrow G_p/G_v\rightarrow0
\label{extension}
\end{align}
For $\forall g\in G_p$, we can express it in terms of a product of two group elements: $g=g_0h$, where $g_0\in G_v$ acts on the virtual indices internally, and $h\in G_p/G_v$ transforms a virtual index to another. 

The graded structure of the physical indices is characterized by a map $n_1: G_p\rightarrow\mathbb{Z}_2=\{0,1\}$ and classified by 1-cohomology $\mathcal{H}^1(G_p,\mathbb{Z}_2)$; The graded structure of the virtual indices is characterized by another map $n_2: G_v\times G_v\rightarrow\mathbb{Z}_2=\{0,1\}$ and classified by 2-cohomology $\mathcal{H}^2(G_v,\mathbb{Z}_2)$. This is because a virtual bond always connects two virtual degrees of freedom, but a physical bond is only relevant to a single degree of freedom at the lattice site. Furthermore, the fermion parity of each PEPS tensor should be even (because the fermion parity symmetry cannot be broken under any morphisms, such as $A_\mu$), hence the above maps to $\mathbb{Z}_2$, $n_1$ and $n_2$ are 1-cocycle/2-cocycle \cite{Frank2017}. 

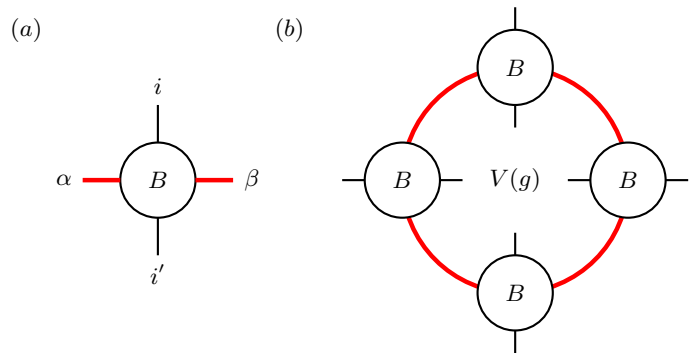
\begin{figure}
\begin{tikzpicture}
\tikzstyle{sergio}=[rectangle,draw=none]
\filldraw[fill=none, draw=black, thick]  (-1.75,0) ellipse (0.5 and 0.5);
        \draw[line width=0.6mm,color=red] (-2.25,0) -- (-2.75,0);
        \draw[line width=0.6mm,color=red] (-1.25,0) -- (-0.75,0);
        \path (-3,0) node [style=sergio] {$\alpha$};
        \path (-0.5,0) node [style=sergio] {$\beta$};
        \draw[thick] (-1.75,0.5) -- (-1.75,1);
        \path (-1.75,1.25) node [style=sergio] {$i$};
		\draw[thick] (-1.75,-0.5) -- (-1.75,-1);
        \path (-1.75,-1.25) node [style=sergio] {$i'$};
                \path (-1.75,0) node [style=sergio] {$B$};
\path (-3.5,2) node [style=sergio] {$(a)$};
\path (0,2) node [style=sergio] {$(b)$};
\filldraw[fill=none, draw=red, line width=0.6mm]  (3,0) ellipse (1.5 and 1.5);
\draw[thick] (3.7,0) -- (4.2,0);
\draw[thick] (5.3,0) -- (4.8,0);
\draw[thick] (3,-1.2) -- (3,-0.7);
\draw[thick] (3,-2.3) -- (3,-1.8);
\draw[thick] (3,1.8) -- (3,2.3);
\draw[thick] (3,1.2) -- (3,0.7);
\draw[thick] (0.7,0) -- (1.2,0);
\draw[thick] (1.8,0) -- (2.3,0);
\filldraw[fill=white, draw=black, thick]  (3,-1.5) ellipse (0.5 and 0.5);
\filldraw[fill=white, draw=black, thick]  (4.5,0) ellipse (0.5 and 0.5);
\filldraw[fill=white, draw=black, thick]  (3,1.5) ellipse (0.5 and 0.5);
\filldraw[fill=white, draw=black, thick]  (1.5,0) ellipse (0.5 and 0.5);
\path (3,0) node [style=sergio] {$V(g)$};
\path (1.5,0) node [style=sergio] {$B$};
\path (3,-1.5) node [style=sergio] {$B$};
\path (3,1.5) node [style=sergio] {$B$};
\path (4.5,0) node [style=sergio] {$B$};
\end{tikzpicture}
\caption{Graphical representation of (a). An MPO tensor; (b). The MPO symmetry operator $V(g)$ on the virtual indices connected to the lattice site $\mu$. Solid black/red links label the virtual indices of PEPSs/MPOs.}
\label{MPO}
\end{figure}

With the aforementioned arguments, we are ready to investigate the wallpaper group symmetry properties on PEPS. Firstly we consider the PEPS $A_\mu$ on a single lattice site $\mu$, as indicated in Fig. \ref{PEPS}(a). Physical symmetry acting on the physical index should form a linear representation of $G_p$ with the notation $U(g),~g\in G_p$. On the virtual indices, the symmetry action $V(g)$ is a fermionic matrix product operator (fMPO) acting on the virtual indices. Repeatedly because $G_v\ne G_p$, The fMPO for wallpaper group symmetry is slightly different from the cases with on-site symmetry. The MPO $V(g)$ associated with the PEPS we discussed in this paragraph is:
\begin{align}
V(g)=\sum\limits_{\{i_n\}=1}^{\{D_n\}}\sum\limits_{\{i_n'\}=1}^{\{D_n\}}&\mathrm{Tr}\left[B_{\tau_1}^{i_1,i_1'}B_{\tau_2}^{i_2,i_2'}B_{\tau_3}^{i_3,i_3'}B_{\tau_4}^{i_4,i_4'}\right]\nonumber\\
&\times|i_1,i_2,i_3,i_4\rangle\langle i_1',i_2',i_3',i_4'|
\end{align}
where $D_n$ is the dimension of the virtual index $\tau_n$ (might be different for various $n$), and $\left(B^{i,i'}_{\tau}\right)_{a,b}$ is a $\chi\times\chi$ matrix, as shown in Fig. \ref{MPO}. Hence the symmetry of the PEPS $A_\mu$ is of the form:
\begin{align}
U(g)A_\mu=A_\mu V(g)
\label{symmetry}
\end{align}

In general, for the region $R\in\Gamma$, The symmetry of the PEPS $A_R$ is of the form:
\begin{align}
\bigotimes\limits_{\mu_k\in R}U_k(g_k)A_R=A_RV^{\partial R}
\label{symmetry R}
\end{align}
where $g_k\in G_p^{\mu_k}$ is a group element of the symmetry acting on the lattice site $\mu_k$ internally, $U_k$ is the corresponding linear representation; the fMPO $V^{\partial R}$ is of the form:
\begin{align}
V^{\partial R}=\sum\limits_{\{i_n\}=1}^{\{D_n\}}\sum\limits_{\{i_n'\}=1}^{\{D_n\}}&\mathrm{Tr}\left[B_{\tau_1}^{i_1,i_1'}\cdot\cdot\cdot B_{\tau_N}^{i_N,i_N'}\right]\nonumber\\
&\times|i_1,\cdot\cdot\cdot,i_N\rangle\langle i_1',\cdot\cdot\cdot,i_N'|
\end{align}
where the virtual indices crossing $\partial R$ are ordered from $1$ to $N=|\partial R|_e$ (the number of virtual indices crossing $\partial R$). 

A unique property of the 2D lattice with wallpaper group symmetry is that there is only one possible symmetry operation acting on the bond of lattice, i.e., the reflection symmetry operation. Equivalently, the effective ``on-site'' symmetry of the virtual indices $G_v$ is either no or $\mathbb{Z}_2$. We will discuss them separately.

\subsection{Virtual indices without symmetry}
For the virtual index crossing $\partial R$ without effective ``on-site'' symmetry, the connecting fermionic MPO (fMPO) can be illustrated as Fig. \ref{MPO}(a). Superposing two fMPOs gives a new fMPO, with the following graphical representation:
\begin{align}
\begin{tikzpicture}
\tikzstyle{sergio}=[rectangle,draw=none]
\filldraw[fill=none, draw=black, thick]  (-1.5,0) ellipse (0.28 and 0.28);
        \draw[color=red, line width=0.6mm] (-1.78,0) -- (-2.25,0);
        \draw[color=red, line width=0.6mm] (-1.22,0) -- (-0.75,0);
        \path (-2.5,0) node [style=sergio] {$\alpha_1$};
        \path (-0.5,0) node [style=sergio] {$\beta_1$};
        \draw[thick] (-1.5,0.28) -- (-1.5,0.75);
        \path (-1.25,0.5) node [style=sergio] {$i$};
		\draw[thick] (-1.5,-0.28) -- (-1.5,-0.72);
        \path (-1.25,-0.5) node [style=sergio] {$i''$};
                \path (-1.5,0) node [style=sergio] {$B_1$};
\filldraw[fill=none, draw=black, thick]  (-1.5,-1) ellipse (0.28 and 0.28);
\path (-1.5,-1) node [style=sergio] {$B_2$};
        \draw[color=red, line width=0.6mm] (-1.78,-1) -- (-2.25,-1);
        \draw[color=red, line width=0.6mm] (-1.22,-1) -- (-0.75,-1);
        \path (-2.5,-1) node [style=sergio] {$\alpha_2$};
        \path (-0.5,-1) node [style=sergio] {$\beta_2$};
                \draw[thick] (-1.5,-1.75) -- (-1.5,-1.28);
        \path (-1.25,-1.5) node [style=sergio] {$i'$};
\path (0,-0.5) node [style=sergio] {$=$};
\filldraw[fill=none, draw=black, thick]  (1.75,-0.5) ellipse (0.3 and 0.3);
        \draw[color=red, line width=0.6mm] (1.45,-0.5) -- (1,-0.5);
        \draw[color=red, line width=0.6mm] (2.05,-0.5) -- (2.5,-0.5);
        \path (0.7,-0.5) node [style=sergio] {$\alpha_{12}$};
        \path (2.8,-0.5) node [style=sergio] {$\beta_{12}$};
        \path (1.75,-0.5) node [style=sergio] {$\scriptsize B_{12}$};
                \draw[thick] (1.75,-1.25) -- (1.75,-0.8);
        \path (2,-1) node [style=sergio] {$i'$};
                \draw[thick] (1.75,-0.2) -- (1.75,0.25);
        \path (2,0) node [style=sergio] {$i$};
\end{tikzpicture}
\label{superpose}
\end{align}
This can be realized by defining a projection operator with the graded structure (see Appendix \ref{AppMPO} for more details):
\begin{align}
\begin{tikzpicture}
\tikzstyle{sergio}=[rectangle,draw=none]
	 \draw[thick] (-2,0.5) -- (-2,-0.5);
\draw[thick] (-2,0.5) -- (-1.5,0);
\draw[thick] (-1.5,0) -- (-2,-0.5);
\path (-1.82,0) node [style=sergio] {$X$};
\draw[color=red, line width=0.6mm] (-2.5,0.25) -- (-2,0.25);
\draw[color=red, line width=0.6mm] (-2.5,-0.25) -- (-2,-0.25);
\draw[color=red, line width=0.6mm] (-1.5,0) -- (-1,0);
\path (-2.75,0.25) node [style=sergio] {$\alpha_{1}$};
\path (-2.75,-0.25) node [style=sergio] {$\alpha_{2}$};
\path (-0.5,0) node [style=sergio] {$\alpha_{12}$};
\end{tikzpicture}
\label{projection}
\end{align}
In Appendix \ref{AppMPO}, we note that for the translation symmetry breaking system with on-site symmetry $G$, the different fSPT phases are labeled by 3-order group super-cohomology with the following indices:
\begin{align}
\left\{
\begin{aligned}
&n_1\in\mathcal{H}^1(G,\mathbb{Z}_2)\\
&n_2\in\mathcal{H}^2(G,\mathbb{Z}_2)\\
&\nu_3\in\mathcal{H}^3[G,U(1)]
\end{aligned}
\right.
\label{supercohomology}
\end{align}
which are given by the $F$-moves and super pentagon equations of the fMPOs with graded structure. Nevertheless, for arbitrary indices connected a fMPO as indicated in Fig. \ref{MPO}(a), there is no effective ``on-site'' symmetry. As the consequence, all parameters $(n_1,n_2,\nu_3)$ in Eq. (\ref{supercohomology}) are trivial.

\subsection{Virtual indices with reflection symmetry}
For the virtual index crossing $\partial R$ at which the reflection symmetry acting internally, the connecting fMPO can still be illustrated as Fig. \ref{MPO}(a). We demonstrate that there is a subtle difference between the true $\mathbb{Z}_2$ on-site symmetry and the reflection symmetry acting internally. Here the virtual indices $i$ and $i'$ are aligned on the reflection axis, and the reflection symmetry operation exchanges the indices $\alpha$ and $\beta$. Equivalently, there is no effective ``on-site'' symmetry on the indices $\alpha$ and $\beta$ which is distinct from the case with true $\mathbb{Z}_2$ on-site symmetry. 

The superposition of fMPOs for this case can still be described by Eq. (\ref{superpose}) and realized by the projection operator defined in Eq. (\ref{projection}). Nevertheless, there is no effective ``on-site'' symmetry on the indices $\alpha$ and $\beta$, hence all parameters in Eq. (\ref{supercohomology}) are also trivial. 

We conclude that the fMPOs connected to $\partial R$ do not give rise to any nontrivial fSPT phases for both cases (with and without reflection symmetry).

\subsection{The role of translation symmetry}
In Ref. \onlinecite{chen11a} the authors demonstrated that for a bosonic system, each virtual bond in Fig. \ref{PEPS} represents an entanglement pair between a projective representation of the symmetry group $G$ and its inverse that is classified by 2-cohomology $\mathcal{H}^2[G,U(1)]$. For a fermionic system, each virtual bond has a graded structure representing the fermion parity which is classified by 1-cohomology $\mathcal{H}^1(G,\mathbb{Z}_2)$; furthermore, the virtual bond in a fermionic system can also represent the Majorana entanglement pair (MEP) that is formed by two Majorana fermions $\gamma_1$ and $\gamma_2$: $i\gamma_1\gamma_2$ \cite{Majorana,1Dfermion,fidkowski10,fidkowski11} which contribute another $\mathbb{Z}_2$. Therefore, the virtual indices of the PEPS tensor in the fermionic systems are characterized by the following three parameters:
\begin{align}
\left\{
\begin{aligned}
&n_0\in\mathbb{Z}_2=\mathcal{H}^0(G,\mathbb{Z}_2)\\
&n_1\in\mathcal{H}^1(G,\mathbb{Z}_2)\\
&\nu_2\in\mathcal{H}^2[G,U(1)]
\end{aligned}
\right.
\label{2-supercohomology}
\end{align}
with twisted cocycle conditions \cite{special,general1,general2} ($g_1,g_2,g_3\in G$):
\begin{align}
\begin{aligned}
&n_1(g_1)+n_1(g_2)-n_1(g_1g_2)=\omega_2\smile n_0\\
&\frac{\nu_2(g_1,g_2)\nu_2(g_1g_2,g_3)}{\nu_2(g_1,g_2g_3)\nu_2(g_2,g_3)}=(-1)^{\omega_2\smile{n_1}(g_1,g_2,g_3)}
\end{aligned}
\label{2-twisted}
\end{align}
Therefore, if we truncate all virtual bonds crossing $\partial R$, there is an unpaired dangling mode characterized by parameters $(n_0,n_1,\nu_2)$ [cf. Eq. (\ref{2-supercohomology})] on each virtual bond crossing $\partial R$, with the effective ``on-site'' symmetry group $G_v=0$ or $\mathbb{Z}_2$ \cite{footnote}. 

With the absence of translation symmetry, boundary modes can be combined by renormalization and change the parameters $(n_0,n_1,\nu_2)$ from one to another and, in particular, to the trivial class. On the other hand, if we restore the translation symmetry, each boundary mode is well-defined. The parameters $(n_0,n_1,\nu_2)$ do label different phases because the renormalization procedure breaks the translation symmetry.

\section{General paradigm of classifying crystalline SPT phases\label{general}}
In this section, we highlight the general paradigm of the classification of the crystalline fSPT phases on 2D lattice with wallpaper group symmetry. There are several major steps:
\begin{enumerate}[1.]
\item The physical indices with graded structure: consider a 2D PEPS tensor network state on a torus who has no open virtual index (because of the absence of the boundary) with translation symmetry, a gapped state $|\psi\rangle$ that do not break the two symmetries must transform as ($\forall g\in G_p$):
\begin{align}
U(g)\otimes\cdot\cdot\cdot\otimes U(g)|\psi\rangle=[\alpha(g)]^N
\end{align}
Here $U(g)$ is the linear representation of $G_p$ acting on each site, $N$ is the number of PEPS tensor, and $\alpha(g)$ is an 1D linear representation of $G_p$ which is classified by 1-cohomology $\mathcal{H}^1[G_p,U(1)]$. Together with the $\mathbb{Z}_2$ from graded structure, the physical indices of the fermionic PEPS tensor are characterized by the following two parameters:
\begin{align}
n_0\in\mathbb{Z}_2,~~\nu_1\in\mathcal{H}^1[G_p,U(1)]
\label{1-supercohomology}
\end{align}
with the twisted cocycle conditions \cite{special,general1,general2} ($g_1,g_2\in G_p$):
\begin{align}
\frac{\nu_1(g_1)\nu_1(g_2)}{\nu_1(g_1g_2)}=(-1)^{\omega_2(g_1,g_2)}
\label{1-twisted}
\end{align}
Alternatively, these two parameters can be unified to a single 1-cocycle in 1-cohomology of the \textit{total} symmetry group $G_p^f=G_p\times_{\omega_2}\mathbb{Z}_2^f$: $\mathcal{H}^1\left[G_p^f,U(1)\right]$. All possible cases for 2D systems are summarized in Appendix \ref{point group center}.
\item The virtual indices with graded structure: characterized by Eq. (\ref{2-supercohomology}) with twisted cocycle conditions (\ref{2-twisted}). For 2D systems, there are only two possibilities: $G_v=\mathbb{Z}_1$ (virtual bond away from reflection axis) or $\mathbb{Z}_2$ (virtual bond on the reflection axis). For spinless fermions, the virtual bond with $G_v=\mathbb{Z}_1$ has only one nontrivial index: MEP; the virtual bond with $G_v=\mathbb{Z}_2$ has two nontrivial indices: MEP and double MEPs. For spin-1/2 fermions, the virtual bond with $G_v=\mathbb{Z}_1$ has only one nontrivial index: MEP; the virtual bond with $G_v=\mathbb{Z}_2$ does not have any nontrivial index.
\item Obstruction: Each virtual bond connects a dangling mode characterized by Eq. (\ref{2-supercohomology}) and its inverse. By definition, each PEPS tensor on a lattice site [see Fig. \ref{PEPS}(a)] is a map from the connecting virtual indices to the physical index, hence for a valid SPT phase, the dangling modes at the endpoints of virtual bonds should form a linear representation of $G_p^f$. If not, we say the corresponding PEPS tensor is \textit{obstructed}.
\item Trivialization: Nontrivial PEPSs do not warrant nontrivial crystalline SPT phases, we should further consider possible trivializations that deform nontrivial PEPSs to trivial states. Some tensor-network states can be mutually deformed by some symmetric finite-depth quantum circuits, hence they are topologically equivalent. Equivalently, trivializations of tensor-network states are characterized by the superposition of PEPS tensors.
\item Group structure of the classification: The accurate group structure of
classification will be obtained after investigating the composite rule of different PEPS tensors. More precisely, stacking several copies of a PEPS tensor may leads to another nontrivial PEPS tensor.
\end{enumerate}

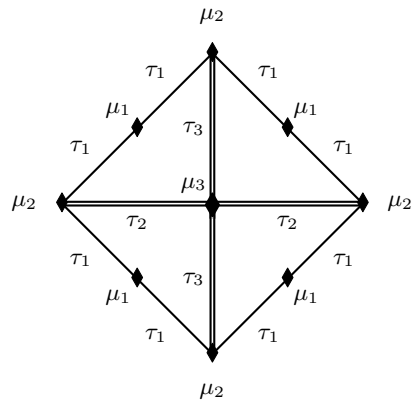
\begin{figure}
\usetikzlibrary{arrows}
\begin{tikzpicture}
\tikzstyle{sergio}=[rectangle,draw=none]
\draw[double, thick] (-1*2,0.5*2)--(0,0.5*2);
\draw[double, thick] (1*2,0.5*2)--(0,0.5*2);
\draw[double, thick] (0,-0.5*2)--(0,0.5*2);
\draw[double, thick] (0,3)--(0,0.8);
\draw[diamond-, thick](0,1.15);
\draw[thick] (-0.5*2,1*2)--(0,1.5*2);
\draw[thick] (-1*2,0.5*2) -- (-0.5*2,1*2);
\draw[thick] (0.5*2,1*2)--(0,1.5*2);
\draw[thick] (1*2,0.5*2) -- (0.5*2,1*2);
\draw[thick] (0.5*2,0)--(0,-0.5*2);
\draw[thick] (0.5*2,0)--(1*2,0.5*2);
\draw[thick] (-0.5*2,0*2)--(0,-0.5*2);
\draw[thick] (-0.5*2,0)--(-1*2,0.5*2);
\draw[diamond-] (-2,1.15);
\draw[diamond-] (-1,2.15);
\draw[diamond-] (0,3.15);
\draw[diamond-] (1,2.15);
\draw[diamond-] (2,1.15);
\draw[diamond-] (1,0.15);
\draw[diamond-] (0,-0.85);
\draw[diamond-] (-1,0.15);
\path (-1.25,2.25) node [style=sergio] {$\mu_1$};
\path (1.25,2.25) node [style=sergio] {$\mu_1$};
\path (1.25,-0.25) node [style=sergio] {$\mu_1$};
\path (-1.25,-0.25) node [style=sergio] {$\mu_1$};
\path (-2.5,1) node [style=sergio] {$\mu_2$};
\path (2.5,1) node [style=sergio] {$\mu_2$};
\path (0,3.5) node [style=sergio] {$\mu_2$};
\path (0,-1.5) node [style=sergio] {$\mu_2$};
\path (-0.25,1.25) node [style=sergio] {$\mu_3$};
\path (-1,0.75) node [style=sergio] {$\tau_2$};
\path (1,0.75) node [style=sergio] {$\tau_2$};
\path (-0.25,2) node [style=sergio] {$\tau_3$};
\path (-0.25,0) node [style=sergio] {$\tau_3$};
\path (-1.75,0.25) node [style=sergio] {$\tau_1$};
\path (1.75,0.25) node [style=sergio] {$\tau_1$};
\path (1.75,1.75) node [style=sergio] {$\tau_1$};
\path (-1.75,1.75) node [style=sergio] {$\tau_1$};
\path (-0.75,-0.75) node [style=sergio] {$\tau_1$};
\path (0.75,-0.75) node [style=sergio] {$\tau_1$};
\path (0.75,2.75) node [style=sergio] {$\tau_1$};
\path (-0.75,2.75) node [style=sergio] {$\tau_1$};
\end{tikzpicture}
\caption{The unit cell of \#9 wallpaper group $cmm$. Each double solid line represents a reflection axis and each solid diamond represents a center of 2-fold rotation symmetry.}
\label{cmm}
\end{figure}

\section{An explicit example of crystalline fSPT phases\label{example}}
With the general paradigm of constructing and classifying the symmetric tensor-network states of crystalline fSPT phases on 2D lattice with wallpaper group symmetry, we demonstrate a concrete and representative example in this section: the crystalline fSPT phases on 2D lattice with $\#9$ wallpaper group symmetry labeled by $cmm$. 

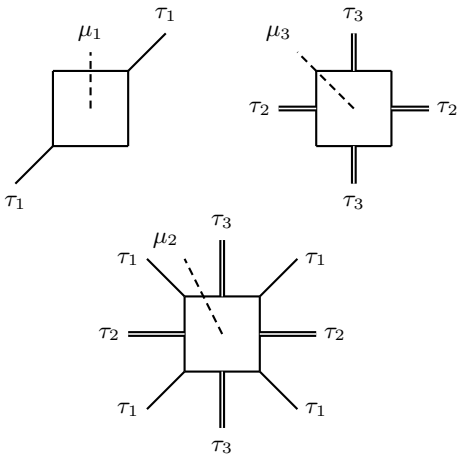
\begin{figure}
\begin{tikzpicture}
\tikzstyle{sergio}=[rectangle,draw=none]
	 \draw[thick] (-2,1.5) -- (-1,1.5);
        \draw[thick] (-2,0.5) -- (-1,0.5);
        \draw[thick] (-2,1.5) -- (-2,0.5);
		\draw[thick] (-1,1.5) -- (-1,0.5);
        \draw[thick] (-1,1.5) -- (-0.5,2);
        \path (-0.5,2.25) node [style=sergio] {$\tau_1$};
		\draw[thick] (-2,0.5) -- (-2.5,0);
        \path (-2.5,-0.25) node [style=sergio] {$\tau_1$};
		\draw[densely dashed, thick] (-1.5,1) -- (-1.5,1.75);
        \path (-1.5,2) node [style=sergio] {$\mu_1$};
        \tikzstyle{sergio}=[rectangle,draw=none]
	 \draw[thick] (1.5,1.5) -- (2.5,1.5);
        \draw[thick] (1.5,0.5) -- (2.5,0.5);
        \draw[thick] (1.5,1.5) -- (1.5,0.5);
		\draw[thick] (2.5,1.5) -- (2.5,0.5);
        \draw[double,thick] (1.5,1) -- (1,1);
        \draw[double,thick] (2.5,1) -- (3,1);
        \path (0.75,1) node [style=sergio] {$\tau_2$};
        \path (3.25,1) node [style=sergio] {$\tau_2$};
        \draw[double,thick] (2,1.5) -- (2,2);
        \path (2,2.25) node [style=sergio] {$\tau_3$};
		\draw[double,thick] (2,0.5) -- (2,0);
        \path (2,-0.25) node [style=sergio] {$\tau_3$};
		\draw[densely dashed, thick] (2,1) -- (1.25,1.75);
        \path (1,2) node [style=sergio] {$\mu_3$};
	 \draw[thick] (-0.25,-1.5) -- (0.75,-1.5);
        \draw[thick] (-0.25,-2.5) -- (0.75,-2.5);
        \draw[thick] (-0.25,-1.5) -- (-0.25,-2.5);
		\draw[thick] (0.75,-1.5) -- (0.75,-2.5);
        \draw[double,thick] (0.25,-1.5) -- (0.25,-0.75);
        \path (0.25,-0.5) node [style=sergio] {$\tau_3$};
		\draw[double,thick] (0.25,-2.5) -- (0.25,-3.25);
        \path (0.25,-3.5) node [style=sergio] {$\tau_3$};
		\draw[densely dashed, thick] (0.25,-2) -- (-0.25,-1);
        \path (-0.5,-0.75) node [style=sergio] {$\mu_2$};
        \draw[thick] (-0.25,-2.5) -- (-0.75,-3);
                \draw[thick] (0.75,-1.5) -- (1.25,-1);
                        \draw[thick] (0.75,-2.5) -- (1.25,-3);
        \draw[double,thick] (-0.25,-2) -- (-1,-2);
                \draw[double,thick] (0.75,-2) -- (1.5,-2);
                        \draw[thick] (-0.25,-1.5) -- (-0.75,-1);
        \path (1.75,-2) node [style=sergio] {$\tau_2$};
         \path (-1,-1) node [style=sergio] {$\tau_1$};
          \path (-1.25,-2) node [style=sergio] {$\tau_2$};
\path (1.5,-1) node [style=sergio] {$\tau_1$};
         \path (-1,-3) node [style=sergio] {$\tau_1$};
                  \path (1.5,-3) node [style=sergio] {$\tau_1$};
\end{tikzpicture}
\caption{Independent PEPS tensors $A_{\mu_j}$ of $cmm$ symmetric crystalline fSPT phases. $\mu_j$/$\tau_j$ represent physical/virtual indices ($j=1,2,3$).}
\label{cmm PEPS}
\end{figure}

From translation symmetry, the PEPS tensors in different unit cells are identical. Equivalently, it is enough to investigate the PEPS tensors in a specific unit cell. Fig. \ref{cmm} illustrates the unit cell of $cmm$ symmetry group. As a consequence, there are only three independent PEPS tensors as illustrated in Fig. \ref{cmm PEPS}.

We discuss the spinless fermions and spin-1/2 fermions separately.

\subsection{Spinless fermions}
Firstly we investigate the physical indices of $A_{\mu_j}$ ($j=1,2,3$) that are characterized by Eqs. (\ref{1-supercohomology}) and (\ref{1-twisted}). Dimension of the physical index of $A_{\mu_1}$ is $d_1=4$ with two generators: complex fermion ($c$) and eigenvalue $-1$ of 2-fold rotation ($r$); dimension of the physical index of $A_{\mu_2}/A_{\mu_3}$ is $d_{2,3}=8$ with three generators: complex fermion ($c$) and eigenstates of two reflection generators of $D_2$ group with eigenvalues $-1$ ($m_1$ and $m_2$). 

Subsequently we investigate the virtual indices of $A_{\mu_j}$ that are labeled by $\tau_k$ ($k=1,2,3$) and characterized by Eqs. (\ref{2-supercohomology}) and (\ref{2-twisted}). Dimension of $\tau_1$ is $D_1=2$ with one generator: MEP [$n_0$ in Eq. (\ref{2-supercohomology})]; dimension of $\tau_{2,3}$ is $D_{2,3}=4$ with two generators: MEP and double MEPs [$n_0$ and $n_1$ in Eq. (\ref{2-supercohomology})].

Some of the virtual indices may be obstructed: If there is an MEP on each virtual bond $\tau_1$, there will be two Majorana fermions at each physical index $\mu_1$ with the following 2-fold rotation property:
\begin{align}
\bs{R}\in C_2:~\gamma_1\leftrightarrow\gamma_2
\end{align}
Nevertheless, $\gamma_1$ and $\gamma_2$ form a projective representation of the symmetry group $\mathbb{Z}_2^f\times\mathbb{Z}_2$ on $\mu_1$: define a complex fermion from $\gamma_1$ and $\gamma_2$: $c^\dag=(\gamma_1+i\gamma_2)/2$ that span a 2D Hilbert space formed by $|0\rangle$ and $c^\dag|0\rangle$. In this Hilbert space, $\gamma_1=\sigma^x$, $\gamma_2=\sigma^y$, fermion parity operator $P_f=\sigma^z$ and $\bs{R}=(\sigma^x+\sigma^y)/\sqrt{2}$, where $\sigma^x$, $\sigma^y$ and $\sigma^z$ are Pauli matrices. This representation satisfies the spinless condition of fermions: $\bs{R}^2=1$. It is easy to verify that the fermion parity $P_f$ and 2-fold rotation $\bs{R}$ are anticommute: $P_f\bs{R}=-\bs{R}P_f$ that is the sufficient condition manifesting that the Hilbert space is a projective representation of $\mathbb{Z}_2^f\times\mathbb{Z}_2$. Accordingly, the MEP on each virtual bond $\tau_1$ is obstructed. Similarly, the MEP on each virtual bond $\tau_2/\tau_3$ is obstructed because it leads a projective representation of the local symmetry group $G_{\mu_2}^f/G_{\mu_3}^f$ of the physical index $\mu_2/\mu_3$. 

If there are double MEPs on each virtual bond $\tau_2$, there will be four Majorana fermions at each physical index $\mu_2$ with the following properties under $D_2$ symmetry:
\begin{align}
\begin{aligned}
&\bs{M}_1:~\left(\gamma_1,\gamma_1',\gamma_2,\gamma_2'\right)\mapsto\left(\gamma_1',\gamma_1,\gamma_2',\gamma_2\right)\\
&\bs{M}_2:~\left(\gamma_1,\gamma_1',\gamma_2,\gamma_2'\right)\mapsto\left(\gamma_2,\gamma_2',\gamma_1,\gamma_1'\right)
\end{aligned}
\end{align}
Here $\bs{M}_1$ and $\bs{M}_2$ are two reflection generators of $D_2$ with horizontal and vertical axes, respectively. Define two complex fermions from these 4 Majorana fermions, with $D_2$ symmetry properties:
\begin{align}
\left\{
\begin{aligned}
&c_1^\dag=\frac{1}{2}(\gamma_1+i\gamma_1')\\
&c_2^\dag=\frac{1}{2}(\gamma_2+i\gamma_2')
\end{aligned}
\right.,~\left\{
\begin{aligned}
&\bs{M}_1:\left(c_1^\dag,c_2^\dag\right)\mapsto\left(ic_1,ic_2\right)\\
&\bs{M}_2:\left(c_1^\dag,c_2^\dag\right)\mapsto\left(c_2^\dag,c_1^\dag\right)
\end{aligned}
\right.
\end{align}
We denote the fermion number operators $n_1=c_1^\dag c_1$ and $n_2=c_2^\dag c_2$. Firstly we consider a Hamiltonian of Hubbard interaction ($U>0$):
\begin{align}
H_U=U\left(n_1-\frac{1}{2}\right)\left(n_2-\frac{1}{2}\right)
\end{align}
It is easy to verify that $H_U$ respect the $D_2$ symmetry. There is a 2-fold ground-state degeneracy from $(n_1,n_2)=(1,0)~\mathrm{or}~(0,1)$ that can be viewed as a spin-1/2 degree of freedom:
\begin{align}
\tau_{12}^\mu=\left(c_1^\dag,c_2^\dag\right)\sigma^\mu\left(
\begin{array}{ccc}
c_1\\
c_2
\end{array}
\right)
\end{align}
In order to investigate that whether the degenerate ground states can be gapped out, we focus on the projective Hilbert space spanned by two states $c_1^\dag|0\rangle$ and $c_2^\dag|0\rangle$. In this projective Hilbert space, $\bs{M}_1=\sigma^y$ and $\bs{M}_2=\sigma^x$ and they are anticommuting:
\begin{align}
\bs{M}_1\bs{M}_2=-\bs{M}_2\bs{M}_1
\end{align}
Thus the projective Hilbert space is a projective representation of the $D_2$ group. Accordingly, the double MEPs on each virtual bond $\tau_2$ is obstructed. Similar to the virtual bond $\tau_3$, and they are obstructed. 

There is one exception: if there are double MEPs on both $\tau_2$ and $\tau_3$, there are two copies of the aforementioned projective representations of $D_2$ on each physical bond $\mu_2$ or $\mu_3$. They can form a linear representation of $D_2$ because there is only one nontrivial projective representation of $D_2$ guaranteed by the 2-cohomology $\mathcal{H}^2[D_2,U(1)]=\mathbb{Z}_2$. As the consequence, this exception is obstruction-free. We claim that it can only be obstruction-free with interactions, otherwise it is still obstructed.

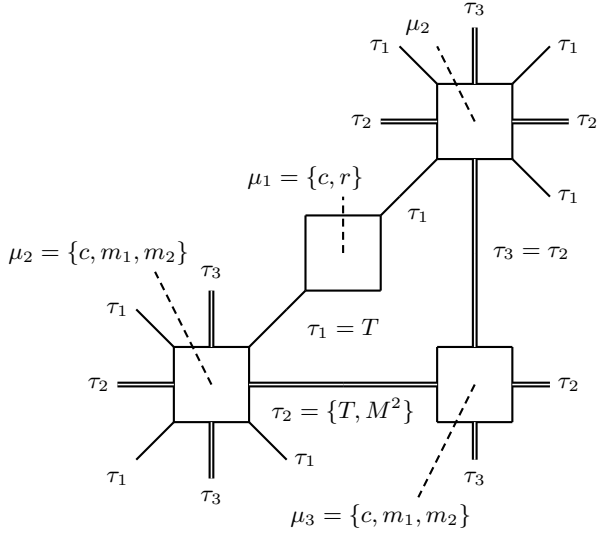
\begin{figure}
\begin{tikzpicture}
\tikzstyle{sergio}=[rectangle,draw=none]
	 \draw[thick] (-2,1.5) -- (-1,1.5);
        \draw[thick] (-2,0.5) -- (-1,0.5);
        \draw[thick] (-2,1.5) -- (-2,0.5);
		\draw[thick] (-1,1.5) -- (-1,0.5);
        \draw[thick] (-1,1.5) -- (-0.75,1.75);
        \path (-0.5,1.5) node [style=sergio] {$\tau_1$};
		\draw[thick] (-2,0.5) -- (-2.25,0.25);
        \path (-1.5,0) node [style=sergio] {$\tau_1=T$};
		\draw[densely dashed, thick] (-1.5,1) -- (-1.5,1.75);
        \path (-2,2) node [style=sergio] {$\mu_1=\{c,r\}$};
        \tikzstyle{sergio}=[rectangle,draw=none]
	 \draw[thick] (-0.25,-0.25) -- (0.75,-0.25);
        \draw[thick] (-0.25,-1.25) -- (0.75,-1.25);
        \draw[thick] (-0.25,-0.25) -- (-0.25,-1.25);
		\draw[thick] (0.75,-0.25) -- (0.75,-1.25);
        \draw[double,thick] (-0.25,-0.75) -- (-1.5,-0.75);
        \draw[double,thick] (0.75,-0.75) -- (1.25,-0.75);
        \path (-1.5,-1.1) node [style=sergio] {$\tau_2=\{T,M^2\}$};
        \path (1.5,-0.75) node [style=sergio] {$\tau_2$};
        \draw[double,thick] (0.25,-0.25) -- (0.25,1);
        \path (1,1) node [style=sergio] {$\tau_3=\tau_2$};
        \path (0.25,-2) node [style=sergio] {$\tau_3$};
		\draw[double,thick] (0.25,-1.25) -- (0.25,-1.75);
		\draw[densely dashed, thick] (0.25,-0.75) -- (-0.5,-2.25);
        \path (-1,-2.5) node [style=sergio] {$\mu_3=\{c,m_1,m_2\}$};
	 \draw[thick] (-3.75,-0.25) -- (-2.75,-0.25);
        \draw[thick] (-3.75,-1.25) -- (-2.75,-1.25);
        \draw[thick] (-3.75,-0.25) -- (-3.75,-1.25);
		\draw[thick] (-2.75,-0.25) -- (-2.75,-1.25);
        \draw[double,thick] (-3.25,-0.25) -- (-3.25,0.5);
        \path (-3.25,0.75) node [style=sergio] {$\tau_3$};
		\draw[double,thick] (-3.25,-1.25) -- (-3.25,-2);
        \path (-3.25,-2.25) node [style=sergio] {$\tau_3$};
		\draw[densely dashed, thick] (-3.25,-0.75) -- (-4,0.75);
        \path (-4.75,1) node [style=sergio] {$\mu_2=\{c,m_1,m_2\}$};
        \draw[thick] (-3.75,-1.25) -- (-4.25,-1.75);
                \draw[thick] (-2.75,-0.25) -- (-2.25,0.25);
                        \draw[thick] (-2.75,-1.25) -- (-2.25,-1.75);
        \draw[double,thick] (-3.75,-0.75) -- (-4.5,-0.75);
                \draw[double,thick] (-2.75,-0.75) -- (-1.5,-0.75);
                        \draw[thick] (-3.75,-0.25) -- (-4.25,0.25);
         \path (-4.5,0.25) node [style=sergio] {$\tau_1$};
          \path (-4.75,-0.75) node [style=sergio] {$\tau_2$};
         \path (-4.5,-2) node [style=sergio] {$\tau_1$};
                  \path (-2,-1.75) node [style=sergio] {$\tau_1$};
\draw[thick] (-0.25,3.25) -- (0.75,3.25);
        \draw[thick] (-0.25,2.25) -- (0.75,2.25);
        \draw[thick] (-0.25,3.25) -- (-0.25,2.25);
		\draw[thick] (0.75,3.25) -- (0.75,2.25);
        \draw[double,thick] (0.25,3.25) -- (0.25,4);
        \path (0.25,4.25) node [style=sergio] {$\tau_3$};
		\draw[double,thick] (0.25,2.25) -- (0.25,1);
		\draw[densely dashed, thick] (0.25,2.75) -- (-0.25,3.75);
        \path (-0.5,4) node [style=sergio] {$\mu_2$};
        \draw[thick] (-0.25,2.25) -- (-0.75,1.75);
                \draw[thick] (0.75,3.25) -- (1.25,3.75);
                        \draw[thick] (0.75,2.25) -- (1.25,1.75);
        \draw[double,thick] (-0.25,2.75) -- (-1,2.75);
                \draw[double,thick] (0.75,2.75) -- (1.5,2.75);
                        \draw[thick] (-0.25,3.25) -- (-0.75,3.75);
        \path (1.75,2.75) node [style=sergio] {$\tau_2$};
         \path (-1,3.75) node [style=sergio] {$\tau_1$};
          \path (-1.25,2.75) node [style=sergio] {$\tau_2$};
\path (1.5,3.75) node [style=sergio] {$\tau_1$};
                  \path (1.5,1.75) node [style=sergio] {$\tau_1$};
\end{tikzpicture}
\caption{All obstruction-free PEPS tensors in a unit cell of $cmm$-symmetric lattice for spinless fermions. Here $t/T$ represents the trivial physical/virtual index, $M^2$ represents the double MEPs, and the virtual indices $\tau_2$ and $\tau_3$ should be identical.}
\label{spinless_fig}
\end{figure}

Summarize above discussions, all obstruction-free PEPS tensors form a $\mathbb{Z}_2^9$ group, $\mathbb{Z}_2^8$ is from the physical indices $\mu_1$, $\mu_2$, and $\mu_3$, and the remaining $\mathbb{Z}_2$ is from the double MEPs on virtual bonds $\tau_2$ and $\tau_3$. All obstruction-free PEPS tensors in a unit cell are illustrated in Fig. \ref{spinless_fig}.

Next, we investigate the possible trivializations of the PEPS tensors. There are several possibilities:
\begin{enumerate}[1.]
\item We have demonstrated that there is no nontrivial and obstruction-free virtual index on $\tau_1$. Nevertheless, there is a subtlety that both vacuum and entanglement pair of two complex fermions on $\tau_1$ are trivial virtual indices [$n_1=0$ in Eq. (\ref{2-supercohomology}) with $G_{\tau_1}=0$], but they give different physical indices $\mu_1$: the former case leaves nothing on all physical indices, but the later case leaves two complex fermions $c_1^\dag$, $c_2^\dag$ on the physical index $\mu_1$ forming an atomic insulator $|\psi\rangle_{\mu_1}=c_1^\dag c_2^\dag|0\rangle$ and four complex fermions $c_j'^\dag$ ($j=1,2,3,4$) on the physical index $\mu_2$ forming another atomic insulator $|\psi\rangle_{\mu_2}=c_1'^\dag c_2'^\dag c_3'^\dag c_4'^\dag|0\rangle$, with the following symmetry properties:
\begin{align}
\begin{aligned}
&\bs{R}|\psi\rangle_{\mu_1}=c_2^\dag c_1^\dag|0\rangle=-|\psi\rangle_{\mu_1}\\
&\bs{M}_1|\psi\rangle_{\mu_2}=c_2'^\dag c_1'^\dag c_4'^\dag c_3'^\dag|0\rangle=|\psi\rangle_{\mu_2}\\
&\bs{M}_2|\psi\rangle_{\mu_2}=c_4'^\dag c_3'^\dag c_2'^\dag c_1'^\dag|0\rangle=|\psi\rangle_{\mu_2}
\end{aligned}
\end{align}
i.e., $|\psi\rangle_{\mu_1}$ represents the state of eigenvalue $-1$ under 2-fold rotation symmetry, labeled by $r$. On the other hand, by definition, the PEPS tensors within one unit cell $\oplus_{j=1}^3A_{\mu_j}$ should be an injective map from connecting virtual indices to a physical indices (injective condition), hence the atomic insulator $|\psi\rangle_{\mu_1}$ corresponds to trivial physical index at $\mu_1$. Equivalently, the physical index $r$ at $\mu_1$ is \textit{trivialized}.
\item Similar to above case, both vacuum and entanglement pair of two complex fermions on $\tau_2$ are trivial virtual indices [$n_1=0$ in Eq. (\ref{2-supercohomology}) with $G_{\tau_2}=\mathbb{Z}_2$], but give different physical indices $\mu_2$ and $\mu_3$: the former case leaves nothing on all physical indices, but the later case leaves two complex fermions on the physical index $\mu_2$ forming an atomic insulator $|\phi\rangle_{\mu_2}=c_1^\dag c_2^\dag|0\rangle$ and two complex fermions on the physical index $\mu_3$ forming another atomic insulator $|\phi\rangle_{\mu_3}=c_1'^\dag c_2'^\dag|0\rangle$. Under vertical reflection $\bs{M}_2$:
\begin{align}
\begin{aligned}
&\bs{M}_2|\phi\rangle_{\mu_2}=c_2^\dag c_1^\dag|0\rangle=-|\phi\rangle_{\mu_2}\\
&\bs{M}_2|\phi\rangle_{\mu_3}=c_2'^\dag c_1'^\dag|0\rangle=-|\phi\rangle_{\mu_3}
\end{aligned}
\end{align}
i.e., $|\phi\rangle_{\mu_2,\mu_3}$ represents the state with eigenvalue $-1$ of $\bs{M}_2$, labeled by $m_2$. According to the injective condition of the PEPS tensors within one unit cell, the atomic insulators $|\phi\rangle_{\mu_2}$ and $|\phi\rangle_{\mu_3}$ together correspond to trivial physical indices at $\mu_2$ and $\mu_3$. Equivalently, the physical indices $m_2$ at both $\mu_2$ and $\mu_3$ are trivialized.
\item Repeatedly, both vacuum and entanglement pair of two complex fermions on $\tau_3$ are trivial virtual indices, but give different physical indices $\mu_2$ and $\mu_3$: the former case leaves nothing on all physical indices, but the later case leaves two complex fermions on the physical index $\mu_2$ forming an atomic insulator $|\eta\rangle_{\mu_2}=c_1^\dag c_2^\dag|0\rangle$ and two complex fermions on the physical index $\mu_3$ forming another atomic insulator $|\eta\rangle_{\mu_3}=c_1'^\dag c_2'^\dag|0\rangle$. Under horizontal reflection $\bs{M}_1$:
\begin{align}
\begin{aligned}
&\bs{M}_1|\eta\rangle_{\mu_2}=c_2^\dag c_1^\dag|0\rangle=-|\eta\rangle_{\mu_2}\\
&\bs{M}_1|\eta\rangle_{\mu_3}=c_2'^\dag c_1'^\dag|0\rangle=-|\eta\rangle_{\mu_3}
\end{aligned}
\end{align}
i.e., $|\eta\rangle_{\mu_2,\mu_3}$ represents the state of eigenvalue $-1$ of $\bs{M}_1$, labeled by $m_1$. According to the injective condition of the PEPS tensors within one unit cell, the atomic insulators $|\eta\rangle_{\mu_2}$ and $|\eta\rangle_{\mu_3}$ together correspond to trivial physical indices at $\mu_2$ and $\mu_3$. Equivalently, the physical indices $m_2$ at both $\mu_2$ and $\mu_3$ are trivialized. 
\item Again, both vacuum and double MEPs on $\tau_1$ are trivial virtual indices, but give different physical indices $\mu_1$: the former case leaves nothing on all physical indices, but for the latter case, double MEPs can be illustrated as:
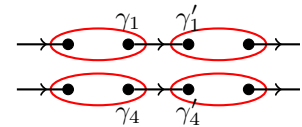
\begin{align}
\tikzstyle{sergio}=[rectangle,draw=none]
\begin{tikzpicture}
	\filldraw[fill=none, draw=red, thick] (-2.4,0.3) ellipse (18pt and 6pt);
	\filldraw[fill=none, draw=red, thick] (-2.4,-0.3) ellipse (18pt and 6pt);
	\filldraw[fill=none, draw=red, thick] (-0.8,0.3) ellipse (18pt and 6pt);
	\filldraw[fill=none, draw=red, thick] (-0.8,-0.3) ellipse (18pt and 6pt);
        \filldraw[fill=black, draw=black] (-2,0.3)circle (2pt);
        \filldraw[fill=black, draw=black] (0.4-1.6,0.3) circle (2pt);
        \draw[->,thick] (-0.322-1.6,0.3) -- (0.08-1.6,0.3);
        \draw[thick] (0.06-1.6,0.3) -- (0.322-1.6,0.3);
        \filldraw[fill=black, draw=black] (-0.4-1.6,-0.3)circle (2pt);
        \filldraw[fill=black, draw=black] (0.4-1.6,-0.3) circle (2pt);
        \draw[->,thick] (-0.322-1.6,-0.3) -- (0.08-1.6,-0.3);
        \draw[thick] (0.06-1.6,-0.3) -- (0.322-1.6,-0.3);
        \path (-0.4-1.6,0.6) node [style=sergio] {$\gamma_1$};
        \path (0.4-1.6,0.65) node [style=sergio] {$\gamma_1'$};
        \path (-0.4-1.6,-0.65) node [style=sergio] {$\gamma_4$};
        \path (0.4-1.6,-0.6) node [style=sergio] {$\gamma_4'$};
        	\filldraw[fill=black, draw=black] (-0.4-1.6-0.8,0.3)circle (2pt);
        	\filldraw[fill=black, draw=black] (-0.4-1.6-0.8,-0.3)circle (2pt);
        	\filldraw[fill=black, draw=black] (0.4-1.6+0.8,0.3)circle (2pt);
        	\filldraw[fill=black, draw=black] (0.4-1.6+0.8,-0.3)circle (2pt);
		\draw[->,thick] (-0.322-1.6-1.56,0.3) -- (0.08-1.6-1.56,0.3);
        \draw[thick] (0.06-1.6-1.56,0.3) -- (0.322-1.6-1.56,0.3);
		\draw[->,thick] (-0.322-1.6-1.56,-0.3) -- (0.08-1.6-1.56,-0.3);
        \draw[thick] (0.06-1.6-1.56,-0.3) -- (0.322-1.6-1.56,-0.3);
		\draw[->,thick] (-0.322-1.6+1.56,0.3) -- (0.08-1.6+1.56,0.3);
        \draw[thick] (0.06-1.6+1.56,0.3) -- (0.322-1.6+1.56,0.3);
		\draw[->,thick] (-0.322-1.6+1.56,-0.3) -- (0.08-1.6+1.56,-0.3);
        \draw[thick] (0.06-1.6+1.56,-0.3) -- (0.322-1.6+1.56,-0.3);
\end{tikzpicture}
\nonumber
\end{align}
Where the red ellipses represent the physical bonds and the arrows represent the Majorana entanglement pairs. 
Consider the Hamiltonian with a parameter $\theta$:
\begin{align}
H(\theta)=&\cos\theta(-i\gamma_1\gamma_1'-i\gamma_4\gamma_4')\nonumber\\
&+\sin\theta(i\gamma_1\gamma_4-i\gamma_1'\gamma_4')
\label{disentangle}
\end{align}
The eigenvalues of $H(\theta)$ are independent with $\theta$, hence the topological properties of $H(\theta)$ keep invariant for different parameter $\theta$. In particular, $\theta=0$ corresponds to two Majorana entanglement pairs and $\theta=\pi/2$ corresponds to the disentangled state with the form:
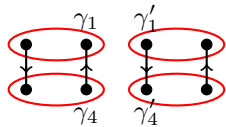
\begin{align}
\tikzstyle{sergio}=[rectangle,draw=none]
\begin{tikzpicture}
	\filldraw[fill=none, draw=red, thick] (-2.4,0.3) ellipse (18pt and 6pt);
	\filldraw[fill=none, draw=red, thick] (-2.4,-0.3) ellipse (18pt and 6pt);
	\filldraw[fill=none, draw=red, thick] (-0.8,0.3) ellipse (18pt and 6pt);
	\filldraw[fill=none, draw=red, thick] (-0.8,-0.3) ellipse (18pt and 6pt);
        \filldraw[fill=black, draw=black] (-0.4-1.6,0.3)circle (2pt);
        \filldraw[fill=black, draw=black] (0.4-1.6,0.3) circle (2pt);
        \filldraw[fill=black, draw=black] (-0.4-1.6,-0.3)circle (2pt);
        \filldraw[fill=black, draw=black] (0.4-1.6,-0.3) circle (2pt);
        \path (-0.4-1.6,0.6) node [style=sergio] {$\gamma_1$};
        \path (0.4-1.6,0.65) node [style=sergio] {$\gamma_1'$};
        \path (-0.4-1.6,-0.65) node [style=sergio] {$\gamma_4$};
        \path (0.4-1.6,-0.6) node [style=sergio] {$\gamma_4'$};
        	\filldraw[fill=black, draw=black] (-0.4-1.6-0.8,0.3)circle (2pt);
        	\filldraw[fill=black, draw=black] (-0.4-1.6-0.8,-0.3)circle (2pt);
        	\filldraw[fill=black, draw=black] (0.4-1.6+0.8,0.3)circle (2pt);
        	\filldraw[fill=black, draw=black] (0.4-1.6+0.8,-0.3)circle (2pt);
        	\draw[->,thick] (-0.4-1.6-0.8,0.3) -- (-0.4-1.6-0.8,-0.08);
        	\draw[thick] (-0.4-1.6-0.8,0) -- (-0.4-1.6-0.8,-0.3);
			\draw[->,thick] (0.4-1.6+0.8,-0.3) -- (0.4-1.6+0.8,0.04);
        	\draw[thick] (0.4-1.6+0.8,0) -- (0.4-1.6+0.8,0.3);
			\draw[->,thick] (-0.4-1.6,-0.3) -- (-0.4-1.6,0.04);
        	\draw[thick] (-0.4-1.6,0) -- (-0.4-1.6,0.3);
			\draw[->,thick] (0.4-1.6,0.3) -- (0.4-1.6,-0.08);
        	\draw[thick] (0.4-1.6,0) -- (0.4-1.6,-0.3);
\end{tikzpicture}
\nonumber
\end{align}
An important issue is that in this disentangled state, there is a Majorana chain with the periodic boundary condition (PBC) surrounding each lattice site $\mu$. It is well-known that the fermion parity of the Majorana chain with PBC is odd \cite{general1}, hence the double MEPs on each $\tau_1$ leaves a fermion parity odd state on each physical index $\mu_1$. 

\begin{figure}
\begin{tikzpicture}
\tikzstyle{sergio}=[rectangle,draw=none]
\path (-3.5,0) node [style=sergio] {$(a)$};
\filldraw[fill=black, draw=black] (-2,0)circle (1.5pt);
\filldraw[fill=black, draw=black] (-1.5,0)circle (1.5pt);
\filldraw[fill=black, draw=black] (-2,-1)circle (1.5pt);
\filldraw[fill=black, draw=black] (-1.5,-1)circle (1.5pt);
\draw[thick] (-2,0) -- (-2,-1);
\draw[thick] (-1.5,0) -- (-1.5,-1);
\filldraw[fill=black, draw=black] (-2.5,-1.5)circle (1.5pt);
\filldraw[fill=black, draw=black] (-2.5,-2)circle (1.5pt);
\filldraw[fill=black, draw=black] (-3.5,-1.5)circle (1.5pt);
\filldraw[fill=black, draw=black] (-3.5,-2)circle (1.5pt);
\draw[thick] (-3.5,-1.5) -- (-2.5,-1.5);
\draw[thick] (-3.5,-2) -- (-2.5,-2);
\filldraw[fill=black, draw=black] (0,-1.5)circle (1.5pt);
\filldraw[fill=black, draw=black] (0,-2)circle (1.5pt);
\filldraw[fill=black, draw=black] (-1,-1.5)circle (1.5pt);
\filldraw[fill=black, draw=black] (-1,-2)circle (1.5pt);
\draw[thick] (-1,-1.5) -- (0,-1.5);
\draw[thick] (-1,-2) -- (0,-2);
\filldraw[fill=black, draw=black] (-2,-2.5)circle (1.5pt);
\filldraw[fill=black, draw=black] (-1.5,-2.5)circle (1.5pt);
\filldraw[fill=black, draw=black] (-2,-3.5)circle (1.5pt);
\filldraw[fill=black, draw=black] (-1.5,-3.5)circle (1.5pt);
\draw[thick] (-2,-2.5) -- (-2,-3.5);
\draw[thick] (-1.5,-2.5) -- (-1.5,-3.5);
\path (1,0) node [style=sergio] {$(b)$};
\filldraw[fill=black, draw=black] (2.5,0)circle (1.5pt);
\filldraw[fill=black, draw=black] (3,0)circle (1.5pt);
\filldraw[fill=black, draw=black] (2.5,-1)circle (1.5pt);
\filldraw[fill=black, draw=black] (3,-1)circle (1.5pt);
\draw[thick] (2,-1.5) -- (2.5,-1);
\draw[thick] (1,-1.5) -- (2.5,0);
\filldraw[fill=black, draw=black] (2,-1.5)circle (1.5pt);
\filldraw[fill=black, draw=black] (2,-2)circle (1.5pt);
\filldraw[fill=black, draw=black] (1,-1.5)circle (1.5pt);
\filldraw[fill=black, draw=black] (1,-2)circle (1.5pt);
\filldraw[fill=black, draw=black] (2.5,-2.5)circle (1.5pt);
\filldraw[fill=black, draw=black] (3,-2.5)circle (1.5pt);
\filldraw[fill=black, draw=black] (2.5,-3.5)circle (1.5pt);
\filldraw[fill=black, draw=black] (3,-3.5)circle (1.5pt);
\filldraw[fill=black, draw=black] (4.5,-1.5)circle (1.5pt);
\filldraw[fill=black, draw=black] (4.5,-2)circle (1.5pt);
\filldraw[fill=black, draw=black] (3.5,-1.5)circle (1.5pt);
\filldraw[fill=black, draw=black] (3.5,-2)circle (1.5pt);
\draw[thick] (3.5,-1.5) -- (3,-1);
\draw[thick] (3.5,-2) -- (3,-2.5);
\draw[thick] (2.5,-2.5) -- (2,-2);
\draw[thick] (2.5,-3.5) -- (1,-2);
\draw[thick] (3,-3.5) -- (4.5,-2);
\draw[thick] (4.5,-1.5) -- (3,0);
\end{tikzpicture}
\caption{Deformation of double MEPs on each virtual bond $\mu_2/\mu_3$. (a)/(b) is the state before/after deformation, respectively. Here each black dot represents a Majorana fermion and each solid line represents an MEP.}
\label{double}
\end{figure}
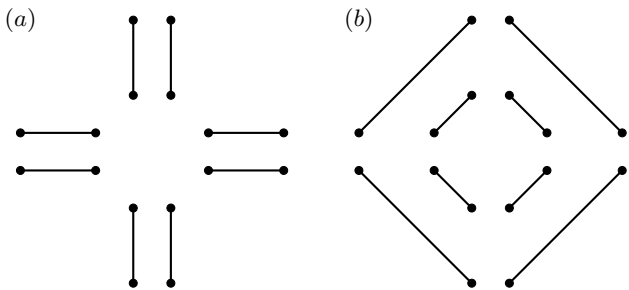

On the other hand, consider double MEPs on each virtual bond $\tau_2/\tau_3$ illustrated as Fig. \ref{double}(a). We have demonstrated that this block-state is obstruction-free. Repeatedly apply the disentangling Hamiltonian (\ref{disentangle}), it can be deformed to Fig. \ref{double}(b) with double MEPs on each virtual bond $\tau_1$, as the above case that has been deformed to Majorana chains with PBCs surrounding 0D blocks labeled by $\mu_1$. Furthermore, we should identify that 8 Majorana fermions on each physical index $\mu_2/\mu_3$ are bound by interactions, hence they do not form a Majorana chain with PBC that can change the fermion parity of the corresponding physical index. As the consequence, the PEPS tensors with physical index $c$ at each $\mu_1$ and double MEPs on each $\tau_2/\tau_3$ are trivialized. 
\end{enumerate}
To summarize these trivializations, we conclude that the trivialized non-vacuum PEPS tensors form a $\mathbb{Z}_2^4$ group. As the consequence, the ultimate classification for spinless fermions on 2D lattice with cmm symmetry is (the superscript represents the spin of fermions):
\begin{align}
\mathcal{G}_{cmm}^{0}=\mathbb{Z}_2^5
\end{align}

\begin{figure}
\begin{tikzpicture}
\tikzstyle{sergio}=[rectangle,draw=none]
	 \draw[thick] (-2,1.5) -- (-1,1.5);
        \draw[thick] (-2,0.5) -- (-1,0.5);
        \draw[thick] (-2,1.5) -- (-2,0.5);
		\draw[thick] (-1,1.5) -- (-1,0.5);
        \draw[thick] (-1,1.5) -- (-0.75,1.75);
        \path (-0.5,1.5) node [style=sergio] {$\tau_1$};
		\draw[thick] (-2,0.5) -- (-2.25,0.25);
        \path (-1.3,0) node [style=sergio] {$\tau_1=\{T,M\}$};
		\draw[densely dashed, thick] (-1.5,1) -- (-1.5,1.75);
        \path (-2,2) node [style=sergio] {$\mu_1=\{f^n\}$};
        \tikzstyle{sergio}=[rectangle,draw=none]
	 \draw[thick] (-0.25,-0.25) -- (0.75,-0.25);
        \draw[thick] (-0.25,-1.25) -- (0.75,-1.25);
        \draw[thick] (-0.25,-0.25) -- (-0.25,-1.25);
		\draw[thick] (0.75,-0.25) -- (0.75,-1.25);
        \draw[double,thick] (-0.25,-0.75) -- (-1.5,-0.75);
        \draw[double,thick] (0.75,-0.75) -- (1.25,-0.75);
        \path (-1.5,-1.1) node [style=sergio] {$\tau_2=T$};
        \path (1.5,-0.75) node [style=sergio] {$\tau_2$};
        \draw[double,thick] (0.25,-0.25) -- (0.25,1);
        \path (1,1) node [style=sergio] {$\tau_3=T$};
        \path (0.25,-2) node [style=sergio] {$\tau_3$};
		\draw[double,thick] (0.25,-1.25) -- (0.25,-1.75);
		\draw[densely dashed, thick] (0.25,-0.75) -- (-0.5,-2.25);
        \path (-1,-2.5) node [style=sergio] {$\mu_3=\{m_1,m_2\}$};
	 \draw[thick] (-3.75,-0.25) -- (-2.75,-0.25);
        \draw[thick] (-3.75,-1.25) -- (-2.75,-1.25);
        \draw[thick] (-3.75,-0.25) -- (-3.75,-1.25);
		\draw[thick] (-2.75,-0.25) -- (-2.75,-1.25);
        \draw[double,thick] (-3.25,-0.25) -- (-3.25,0.5);
        \path (-3.25,0.75) node [style=sergio] {$\tau_3$};
		\draw[double,thick] (-3.25,-1.25) -- (-3.25,-2);
        \path (-3.25,-2.25) node [style=sergio] {$\tau_3$};
		\draw[densely dashed, thick] (-3.25,-0.75) -- (-4,0.75);
        \path (-4.75,1) node [style=sergio] {$\mu_2=\{m_1,m_2\}$};
        \draw[thick] (-3.75,-1.25) -- (-4.25,-1.75);
                \draw[thick] (-2.75,-0.25) -- (-2.25,0.25);
                        \draw[thick] (-2.75,-1.25) -- (-2.25,-1.75);
        \draw[double,thick] (-3.75,-0.75) -- (-4.5,-0.75);
                \draw[double,thick] (-2.75,-0.75) -- (-1.5,-0.75);
                        \draw[thick] (-3.75,-0.25) -- (-4.25,0.25);
         \path (-4.5,0.25) node [style=sergio] {$\tau_1$};
          \path (-4.75,-0.75) node [style=sergio] {$\tau_2$};
         \path (-4.5,-2) node [style=sergio] {$\tau_1$};
                  \path (-2,-1.75) node [style=sergio] {$\tau_1$};
\draw[thick] (-0.25,3.25) -- (0.75,3.25);
        \draw[thick] (-0.25,2.25) -- (0.75,2.25);
        \draw[thick] (-0.25,3.25) -- (-0.25,2.25);
		\draw[thick] (0.75,3.25) -- (0.75,2.25);
        \draw[double,thick] (0.25,3.25) -- (0.25,4);
        \path (0.25,4.25) node [style=sergio] {$\tau_3$};
		\draw[double,thick] (0.25,2.25) -- (0.25,1);
		\draw[densely dashed, thick] (0.25,2.75) -- (-0.25,3.75);
        \path (-0.5,4) node [style=sergio] {$\mu_2$};
        \draw[thick] (-0.25,2.25) -- (-0.75,1.75);
                \draw[thick] (0.75,3.25) -- (1.25,3.75);
                        \draw[thick] (0.75,2.25) -- (1.25,1.75);
        \draw[double,thick] (-0.25,2.75) -- (-1,2.75);
                \draw[double,thick] (0.75,2.75) -- (1.5,2.75);
                        \draw[thick] (-0.25,3.25) -- (-0.75,3.75);
        \path (1.75,2.75) node [style=sergio] {$\tau_2$};
         \path (-1,3.75) node [style=sergio] {$\tau_1$};
          \path (-1.25,2.75) node [style=sergio] {$\tau_2$};
\path (1.5,3.75) node [style=sergio] {$\tau_1$};
                  \path (1.5,1.75) node [style=sergio] {$\tau_1$};
\end{tikzpicture}
\caption{All obstruction free PEPS tensors in a unit cell of $cmm$-symmetric lattice for spin-1/2 fermions. Here $M$ represents the MEP, $t/T$ represents the trivial physical/virtual index and $n=0,1,2,3$.}
\label{spin-1/2_fig}
\end{figure}
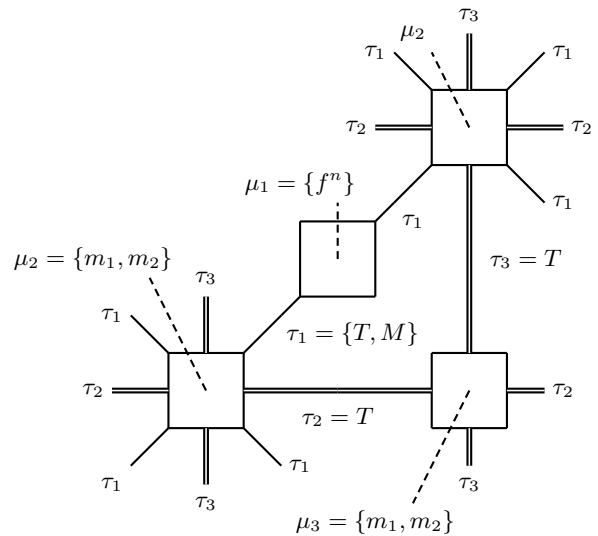

\subsection{Spin-1/2 fermions}
Firstly we investigate the physical indices of $A_{\mu_j}$ ($j=1,2,3$) that are characterized by Eqs. (\ref{1-supercohomology}) and (\ref{1-twisted}). Dimnsion of the physical index of $A_{\mu_1}$ is $d_1=4$ with one generator $f$ characterizing the eigenvalues of $\mathbb{Z}_4^f$, satisfying $f^4=1$; dimension of the physical index of $A_{\mu_2}/A_{\mu_3}$ is $d_{2,3}=4$ with two generators: eigenvalue $-1$ of $\bs{M}_1$ and $\bs{M}_2$ ($m_1$ and $m_2$). 

Subsequently we investigate the virtual indices $\tau_k$ ($k=1,2,3$) that are characterized by Eqs. (\ref{2-supercohomology}) and (\ref{2-twisted}). Dimension of $\tau_1$ is $D_1=2$ with one generator: MEP; dimension of $\tau_{2,3}$ is $D_{2,3}=1$. 

\begin{figure}
\begin{tikzpicture}
\tikzstyle{sergio}=[rectangle,draw=none]
	 \draw[thick] (7.5,1.5) -- (8.5,1.5);
        \draw[thick] (7.5,0.5) -- (8.5,0.5);
        \draw[thick] (7.5,1.5) -- (7.5,0.5);
		\draw[thick] (8.5,1.5) -- (8.5,0.5);
        \draw[thick] (8.5,1.5) -- (9,2);
        \path (8.75,2) node [style=sergio] {$\tau_1$};
		\draw[thick] (7.5,0.5) -- (7,0);
        \path (7.75,0) node [style=sergio] {$\tau_1=M$};
		\draw[densely dashed, thick] (8,1) -- (8,1.75);
        \path (7.5,2) node [style=sergio] {$\mu_1=t$};
\path (6.5,1) node [style=sergio] {$\oplus$};
	 \draw[thick] (4.5,1.5) -- (5.5,1.5);
        \draw[thick] (4.5,0.5) -- (5.5,0.5);
        \draw[thick] (4.5,1.5) -- (4.5,0.5);
		\draw[thick] (5.5,1.5) -- (5.5,0.5);
        \draw[thick] (5.5,1.5) -- (6,2);
        \path (5.75,2) node [style=sergio] {$\tau_1$};
		\draw[thick] (4.5,0.5) -- (4,0);
        \path (4.75,0) node [style=sergio] {$\tau_1=M$};
		\draw[densely dashed, thick] (5,1) -- (5,1.75);
        \path (4.5,2) node [style=sergio] {$\mu_1=t$};
\path (6.5,1) node [style=sergio] {$\oplus$};
\draw[->,thick] (9.25,1) -- (10.25,1);
 \draw[thick] (11,1.5) -- (12,1.5);
        \draw[thick] (11,0.5) -- (12,0.5);
        \draw[thick] (11,1.5) -- (11,0.5);
		\draw[thick] (12,1.5) -- (12,0.5);
        \draw[thick] (12,1.5) -- (12.5,2);
        \path (12.25,2) node [style=sergio] {$\tau_1$};
		\draw[thick] (11,0.5) -- (10.5,0);
        \path (11.25,0) node [style=sergio] {$\tau_1=T$};
		\draw[densely dashed, thick] (11.5,1) -- (11.5,1.75);
        \path (11,2) node [style=sergio] {$\mu_1=f$};
\end{tikzpicture}
\caption{Composite rule of the PEPS tensors $A_{\mu_1}$. ``$T$'' represents the trivial virtual index, ``$t$'' represents the trivial physical index, ``$M$'' represents the MEP as the virtual index, and ``$f$'' represents the physical index with odd fermion parity.}
\label{composite}
\end{figure}
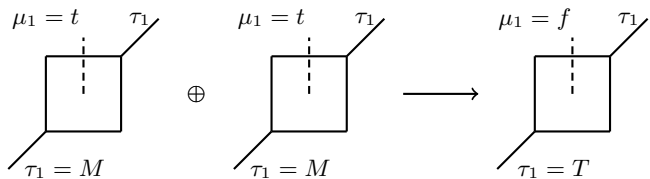

It is straightforward to verify that there is no obstruction and trivialization. Hence all obstruction and trivialization free PEPS tensors are illustrated in Fig. \ref{spin-1/2_fig}. We should further consider the group structure
of the classification: As aforementioned, double MEPs on each $\tau_1$ is a trivial virtual index but changes the fermion parity of the physical index $\tau_1$. It provides a composite rule of PEPS tensors that two copies of MEP on each $\tau_1$ leads to a complex fermion on each $\mu_1$ (see Fig. \ref{composite}). 

This composite rule indicates that there is a nontrivial stacking between physical indices and virtual indices that leads to a nontrivial group extension of the classification, characterized by the following short exact sequence (see Appendix \ref{SES} for more details):
\begin{align}
0\rightarrow\mathbb{Z}_2\rightarrow\mathbb{Z}_8\times\mathbb{Z}_2^4\rightarrow\mathbb{Z}_4\times\mathbb{Z}_2^4\rightarrow0
\end{align}
Finally, the ultimate classification of the crystalline SPT phases with spin-1/2 fermions on the lattice with $cmm$ symmetry is:
\begin{align}
\mathcal{G}_{cmm}^{1/2}=\mathbb{Z}_8\times\mathbb{Z}_2^4
\end{align}

\section{Conclusion and discussion\label{conclusion}}
In this work, we establish the tensor-network representation of 2D crystalline fSPT phases and derive the classification by investigating the various degrees of freedom of PEPS tensors per unit cell. For a 2D lattice with a specific wallpaper group symmetry, we first put a PEPS tensor defined as a map from the connecting virtual indices to the physical index [cf. Eq. (\ref{PEPS tensor})] on each lattice site, whose physical and virtual indices have graded structure representing the fermion parity of the corresponding degrees of freedom. To describe the 2D crystalline fSPT phases, the PEPS tensors within a unit cell should be injective. Distinct from the SPT phases with on-site symmetry, the effective ``on-site'' symmetry (by crystalline symmetry acting internally) of the physical and virtual indices of PEPS tensors for describing the crystalline fSPT phases, $G_p$ and $G_v$, might be different. 

Then the symmetry properties of the PEPS tensors are shown in Eqs. (\ref{symmetry}) and (\ref{symmetry R}), where $U$ is the linear representation of $G_p$ and $V$ is the fMPO defined on the virtual indices crossing the edge of the regime under investigation. A partial classification might be from the $F$-moves and super pentagon equations of the fMPOs with graded structure. Nevertheless, we have elucidated that fMPOs do not contribute nontrivial 2D crystalline fSPT phases with wallpaper group symmetry via Eq. (\ref{supercohomology}). In addition, due to the presence of the translation symmetry, the contributions from virtual and physical indices should also be taken into account, because the tensor renormalization procedure breaks the translation symmetry. The contributions from virtual indices are characterized by Eqs. (\ref{2-supercohomology}) and (\ref{2-twisted}), and the contributions from physical indices are characterized by Eqs. (\ref{1-supercohomology}) and (\ref{1-twisted}), and they form the classification data together.

Subsequently, the possible obstructions and trivializations should be taken into account. By definition, the PEPS tensors defined in a unit cell should be an injective map from virtual indices to physical indices. If the connecting virtual indices of a PEPS tensor cannot be mapped to a valid physical index (i.e., a linear representation of $G_p$), we call the corresponding PEPS tensor is \textit{obstructed}; with the condition of the injectivity, the physical indices shaped differently which are mapped from the same virtual indices should be equivalent. 

\begin{figure*}
\begin{tikzpicture}
\usetikzlibrary{shapes.geometric}
\tikzstyle{sergio}=[rectangle,draw=none]
\draw[thin] (-5.75,-7.3) -- (-6.25,-7.3);
\draw[thin] (-6.25,-7.8) -- (-6.25,-7.3);
\draw[thin] (-5.75,-7.3) -- (-5.75,-7.8);
\draw[thin] (-6.25,-7.8) -- (-5.75,-7.8);
\draw[double] (-5.75,-7.3) -- (-5.25,-7.15);
\draw[double] (-5.75,-7.8) -- (-5.25,-7.95);
\draw[double] (-6.25,-7.3) -- (-6.75,-7.15);
\draw[double] (-6.25,-7.8) -- (-6.75,-7.95);
\draw[double] (-6,-7.3) -- (-6,-6.8);
\draw[double] (-6,-8.65) -- (-6,-7.8);
\draw[thin] (-7.25,-7.95) -- (-6.75,-7.95);
\draw[thin] (-7.25,-7.95) -- (-7.25,-8.45);
\draw[thin] (-6.75,-8.45) -- (-6.75,-7.95);
\draw[thin] (-6.75,-8.45) -- (-7.25,-8.45);
\draw[double] (-7.75,-7.8) -- (-7.25,-7.95);
\draw[double] (-7,-9.3) -- (-7,-8.45);
\draw[thin] (-5.75,-8.65) -- (-6.25,-8.65);
\draw[thin] (-6.25,-9.15) -- (-6.25,-8.65);
\draw[thin] (-5.75,-8.65) -- (-5.75,-9.15);
\draw[thin] (-6.25,-9.15) -- (-5.75,-9.15);
\draw[double] (-6.25,-9.15) -- (-6.75,-9.3);
\draw[double] (-5.75,-9.15) -- (-5.25,-9.3);
\draw[thin] (-6.75,-9.3) -- (-6.75,-9.8);
\draw[thin] (-7.25,-9.8) -- (-6.75,-9.8);
\draw[thin] (-6.75,-9.3) -- (-7.25,-9.3);
\draw[thin] (-7.25,-9.8) -- (-7.25,-9.3);
\draw[double] (-6.25,-9.95) -- (-6.75,-9.8);
\draw[double] (-7.25,-9.3) -- (-7.75,-9.15);
\draw[double] (-7.75,-9.95) -- (-7.25,-9.8);
\draw[double] (-7,-10.3) -- (-7,-9.8);
\draw[double] (-6.75,-8.45) -- (-6.25,-8.65);
\draw[double] (-7.25,-8.45) -- (-7.75,-8.65);
\draw[double] (-7,-7.45) -- (-7,-7.95);
\draw[double] (-5.25,-8.45) -- (-5.75,-8.65);
\draw[double] (-6,-9.65) -- (-6,-9.15);
\path (-6.5,-11.1) node [style=sergio] {$14.~p3m1$};
\path (-6,-7.55) node [style=sergio] {\scriptsize$\mu_1$};
\path (-6,-8.9) node [style=sergio] {\scriptsize$\mu_2$};
\path (-7,-9.55) node [style=sergio] {\scriptsize$\mu_1$};
\path (-7,-8.2) node [style=sergio] {\scriptsize$\mu_3$};
\path (-5.75,-8.3) node [style=sergio] {\scriptsize$\tau_3$};
\path (-6.75,-7) node [style=sergio] {\scriptsize$\tau_3$};
\path (-5.25,-7) node [style=sergio] {\scriptsize$\tau_3$};
\path (-5.25,-7.8) node [style=sergio] {\scriptsize$\tau_2$};
\path (-5.8132,-6.8018) node [style=sergio] {\scriptsize$\tau_2$};
\path (-6.3857,-8.0206) node [style=sergio] {\scriptsize$\tau_2$};
\path (-7.7535,-8.0146) node [style=sergio] {\scriptsize$\tau_2$};
\path (-7.1625,-8.6794) node [style=sergio] {\scriptsize$\tau_2$};
\path (-7.75,-10.1) node [style=sergio] {\scriptsize$\tau_2$};
\path (-6.25,-10.1) node [style=sergio] {\scriptsize$\tau_2$};
\path (-7.75,-9.3) node [style=sergio] {\scriptsize$\tau_3$};
\path (-6.75,-10.15) node [style=sergio] {\scriptsize$\tau_3$};
\path (-6.516,-9.0687) node [style=sergio] {\scriptsize$\tau_3$};
\path (-5.25,-9.5) node [style=sergio] {\scriptsize$\tau_3$};
\path (-7.1987,-7.4817) node [style=sergio] {\scriptsize$\tau_1$};
\path (-7.75,-8.8) node [style=sergio] {\scriptsize$\tau_1$};
\path (-6.5708,-8.7098) node [style=sergio] {\scriptsize$\tau_1$};
\path (-5.25,-8.3) node [style=sergio] {\scriptsize$\tau_1$};
\path (-5.8163,-9.4879) node [style=sergio] {\scriptsize$\tau_1$};
\draw[thin] (-11.25,-7.95) -- (-11.75,-7.95);
\draw[thin] (-11.75,-8.45) -- (-11.75,-7.95);
\draw[thin] (-11.25,-7.95) -- (-11.25,-8.45);
\draw[thin] (-11.75,-8.45) -- (-11.25,-8.45);
\draw[thin] (-11.25,-8.1) -- (-10.75,-7.95);
\draw[thin] (-11.25,-8.3) -- (-10.75,-8.45);
\draw[thin] (-11.75,-8.1) -- (-12.25,-7.95);
\draw[thin] (-11.75,-8.3) -- (-12.25,-8.45);
\draw[thin] (-11.5,-7.95) -- (-11.5,-7.45);
\draw[thin] (-11.5,-8.95) -- (-11.5,-8.45);
\draw[thin] (-12.75,-8.45) -- (-12.25,-8.45);
\draw[thin] (-12.75,-8.45) -- (-12.75,-8.95);
\draw[thin] (-12.25,-8.95) -- (-12.25,-8.45);
\draw[thin] (-12.25,-8.95) -- (-12.75,-8.95);
\draw[thin] (-13.25,-8.3) -- (-12.75,-8.45);
\draw[thin] (-12.5,-9.45) -- (-12.5,-8.95);
\draw[thin] (-11.25,-8.95) -- (-11.75,-8.95);
\draw[thin] (-11.75,-9.45) -- (-11.75,-8.95);
\draw[thin] (-11.25,-8.95) -- (-11.25,-9.45);
\draw[thin] (-11.75,-9.45) -- (-11.25,-9.45);
\draw[thin] (-11.75,-9.45) -- (-12.25,-9.6);
\draw[thin] (-11.25,-9.45) -- (-10.75,-9.6);
\draw[thin] (-12.25,-9.45) -- (-12.25,-9.95);
\draw[thin] (-12.75,-9.95) -- (-12.25,-9.95);
\draw[thin] (-12.25,-9.45) -- (-12.75,-9.45);
\draw[thin] (-12.75,-9.95) -- (-12.75,-9.45);
\draw[thin] (-11.75,-9.95) -- (-12.25,-9.8);
\draw[thin] (-12.75,-9.6) -- (-13.25,-9.45);
\draw[thin] (-13.25,-9.95) -- (-12.75,-9.8);
\draw[thin] (-12.5,-10.45) -- (-12.5,-9.95);
\path (-12,-11.1) node [style=sergio] {$13.~p3$};
\path (-11.5,-8.2) node [style=sergio] {\scriptsize$\mu_1$};
\path (-11.5,-9.2) node [style=sergio] {\scriptsize$\mu_2$};
\path (-12.5,-9.7) node [style=sergio] {\scriptsize$\mu_1$};
\path (-12.5,-8.7) node [style=sergio] {\scriptsize$\mu_3$};
\path (-11.3239,-8.6892) node [style=sergio] {\scriptsize$\tau_2$};
\path (-12.25,-7.8) node [style=sergio] {\scriptsize$\tau_2$};
\path (-10.75,-7.8) node [style=sergio] {\scriptsize$\tau_2$};
\path (-10.7511,-8.6161) node [style=sergio] {\scriptsize$\tau_1$};
\path (-11.3132,-7.4518) node [style=sergio] {\scriptsize$\tau_1$};
\path (-11.9781,-8.5137) node [style=sergio] {\scriptsize$\tau_1$};
\path (-13.2535,-8.5146) node [style=sergio] {\scriptsize$\tau_1$};
\path (-12.6625,-9.1794) node [style=sergio] {\scriptsize$\tau_1$};
\path (-13.2524,-10.102) node [style=sergio] {\scriptsize$\tau_1$};
\path (-11.7483,-10.1121) node [style=sergio] {\scriptsize$\tau_1$};
\path (-13.25,-9.3) node [style=sergio] {\scriptsize$\tau_2$};
\path (-12.25,-10.3) node [style=sergio] {\scriptsize$\tau_2$};
\path (-11.9816,-9.3627) node [style=sergio] {\scriptsize$\tau_2$};
\path (-10.75,-9.8) node [style=sergio] {\scriptsize$\tau_2$};
\draw[double] (-0.463,-10.328) -- (-0.25,-9.9);
\draw[double] (0.4545,-10.328) -- (0.25,-9.9);
\draw[double] (0.85,-9.5) -- (0.25,-9.5);
\draw[double] (0.4545,-8.71) -- (0.25,-9.1);
\draw  (-0.25,-9.1) rectangle (0.25,-9.9);
\draw[thin] (0,-8.5) -- (0,-9.1);
\draw[thin] (0,-10.6) -- (0,-9.9);
\draw[thin] (-0.25,-9.7) -- (-0.75,-9.9);
\draw[thin] (-0.25,-9.3) -- (-0.75,-9.1);
\draw[thin] (0.75,-9.9) -- (0.25,-9.7);
\draw[thin] (0.75,-9.1) -- (0.25,-9.3);
\draw  (-1.25,-8.6) rectangle (-0.75,-9.1);
\draw[double] (-2.85,-9.5) -- (-2.25,-9.5);
\draw[double] (-2.463,-10.328) -- (-2.25,-9.9);
\draw[double] (-1.5455,-10.328) -- (-1.75,-9.9);
\draw[double] (-0.25,-9.5) -- (-1.75,-9.5);
\draw[double] (-2.4655,-8.71) -- (-2.25,-9.1);
\draw  (-2.25,-9.1) rectangle (-1.75,-9.9);
\draw[thin] (-2,-8.5) -- (-2,-9.1);
\draw[thin] (-2,-10.6) -- (-2,-9.9);
\draw[thin] (-2.25,-9.7) -- (-2.75,-9.9);
\draw[thin] (-2.25,-9.3) -- (-2.75,-9.1);
\draw[thin] (-1.25,-9.9) -- (-1.75,-9.7);
\draw[thin] (-1.25,-9.1) -- (-1.75,-9.3);
\draw[double] (-1.85,-7.5) -- (-1.25,-7.5);
\draw[double] (-1.75,-9.1) -- (-1.25,-7.9);
\draw[double] (-0.25,-9.1) -- (-0.75,-7.9);
\draw[double] (-0.15,-7.5) -- (-0.75,-7.5);
\draw[double] (-0.5455,-6.71) -- (-0.75,-7.1);
\draw[double] (-1.4655,-6.71) -- (-1.25,-7.1);
\draw  (-1.25,-7.1) rectangle (-0.75,-7.9);
\draw[thin] (-1,-6.5) -- (-1,-7.1);
\draw[thin] (-1,-8.6) -- (-1,-7.9);
\draw[thin] (-1.25,-7.7) -- (-1.75,-7.9);
\draw[thin] (-1.25,-7.3) -- (-1.75,-7.1);
\draw[thin] (-0.25,-7.9) -- (-0.75,-7.7);
\draw[thin] (-0.25,-7.1) -- (-0.75,-7.3);
\path (-1,-11.1) node [style=sergio] {$15.~p31m$};
\path (-2,-9.5) node [style=sergio] {\scriptsize$\mu_1$};
\path (0,-9.5) node [style=sergio] {\scriptsize$\mu_1$};
\path (-1,-7.5) node [style=sergio] {\scriptsize$\mu_1$};
\path (-1,-8.85) node [style=sergio] {\scriptsize$\mu_2$};
\path (-0.8551,-8.2607) node [style=sergio] {\scriptsize$\tau_2$};
\path (-1.4853,-9.0023) node [style=sergio] {\scriptsize$\tau_2$};
\path (-0.522,-9.003) node [style=sergio] {\scriptsize$\tau_2$};
\path (-0.0019,-8.3254) node [style=sergio] {\scriptsize$\tau_2$};
\path (0.1991,-10.525) node [style=sergio] {\scriptsize$\tau_2$};
\path (-0.7525,-10.0873) node [style=sergio] {\scriptsize$\tau_2$};
\path (0.7532,-8.9363) node [style=sergio] {\scriptsize$\tau_2$};
\path (0.7511,-10.0778) node [style=sergio] {\scriptsize$\tau_2$};
\path (-1.2549,-10.0878) node [style=sergio] {\scriptsize$\tau_2$};
\path (-2.1799,-10.4395) node [style=sergio] {\scriptsize$\tau_2$};
\path (-2.7414,-10.0969) node [style=sergio] {\scriptsize$\tau_2$};
\path (-2.7499,-8.9052) node [style=sergio] {\scriptsize$\tau_2$};
\path (-2.2,-8.55) node [style=sergio] {\scriptsize$\tau_2$};
\path (-1.6345,-8.0183) node [style=sergio] {\scriptsize$\tau_2$};
\path (-0.188,-7.7328) node [style=sergio] {\scriptsize$\tau_2$};
\path (-0.25,-6.9) node [style=sergio] {\scriptsize$\tau_2$};
\path (-1.1483,-6.6801) node [style=sergio] {\scriptsize$\tau_2$};
\path (-1.5882,-7.005) node [style=sergio] {\scriptsize$\tau_2$};
\path (-1.5706,-6.5561) node [style=sergio] {\scriptsize$\tau_1$};
\path (-0.4558,-6.5672) node [style=sergio] {\scriptsize$\tau_1$};
\path (-2.0274,-7.4983) node [style=sergio] {\scriptsize$\tau_1$};
\path (0.0296,-7.4998) node [style=sergio] {\scriptsize$\tau_1$};
\path (-1.6753,-8.4214) node [style=sergio] {\scriptsize$\tau_1$};
\path (-0.3335,-8.4119) node [style=sergio] {\scriptsize$\tau_1$};
\path (0.515,-8.5466) node [style=sergio] {\scriptsize$\tau_1$};
\path (-1.0186,-9.6586) node [style=sergio] {\scriptsize$\tau_1$};
\path (0.5516,-10.477) node [style=sergio] {\scriptsize$\tau_1$};
\path (1.0479,-9.4983) node [style=sergio] {\scriptsize$\tau_1$};
\path (-3.0076,-9.4872) node [style=sergio] {\scriptsize$\tau_1$};
\path (0.5516,-10.477) node [style=sergio] {\scriptsize$\tau_1$};
\draw  (-1.5,-16) rectangle (-1,-16.8);
\draw[double] (-0.7,-15.5) -- (-1,-16);
\draw[double] (-2.25,-14.75) -- (-1.5,-16);
\draw[double] (-0.4,-16.4) -- (-1,-16.4);
\draw[double] (-3.2392,-16.3954) -- (-1.5,-16.4);
\draw[double] (-0.7,-17.3) -- (-1,-16.8);
\draw[double] (-1.8,-17.3) -- (-1.5,-16.8);
\draw[double] (-2,-16.9) -- (-1.5,-16.6);
\draw[double] (-0.5,-16.9) -- (-1,-16.6);
\draw[double] (-1.5,-16.2) -- (-3.2506,-15.396);
\draw[double] (-1,-16.2) -- (-0.5,-15.9);
\draw[double] (-1.25,-15.4) -- (-1.25,-16);
\draw[double] (-1.25,-17.4) -- (-1.25,-16.8);
\draw  (-6,-16) rectangle (-5.5,-16.8);
\draw[double] (-4.75,-14.75) -- (-5.5,-16);
\draw[double] (-6.3,-15.5) -- (-6,-16);
\draw[double] (-3.7392,-16.3954) -- (-5.5,-16.4);
\draw[double] (-6.6,-16.4) -- (-6,-16.4);
\draw[double] (-5.2,-17.3) -- (-5.5,-16.8);
\draw[double] (-6.3,-17.3) -- (-6,-16.8);
\draw[double] (-6.5,-16.9) -- (-6,-16.6);
\draw[double] (-5,-16.9) -- (-5.5,-16.6);
\draw[double] (-6,-16.2) -- (-6.5,-15.9);
\draw[double] (-5.5,-16.2) -- (-3.7624,-15.3949);
\draw[double] (-5.75,-15.4) -- (-5.75,-16);
\draw[double] (-5.75,-17.4) -- (-5.75,-16.8);
\draw  (-3.75,-12.6) rectangle (-3.25,-13.4);
\draw[double] (-2.95,-12.1) -- (-3.25,-12.6);
\draw[double] (-4.05,-12.1) -- (-3.75,-12.6);
\draw[double] (-2.65,-13) -- (-3.25,-13);
\draw[double] (-4.35,-13) -- (-3.75,-13);
\draw[double] (-2.75,-14.25) -- (-3.25,-13.4);
\draw[double] (-4.25,-14.25) -- (-3.75,-13.4);
\draw[double] (-4.25,-13.5) -- (-3.75,-13.2);
\draw[double] (-2.75,-13.5) -- (-3.25,-13.2);
\draw[double] (-3.75,-12.8) -- (-4.25,-12.5);
\draw[double] (-3.25,-12.8) -- (-2.75,-12.5);
\draw[double] (-3.5,-12) -- (-3.5,-12.6);
\draw[double] (-3.5069,-14.9039) -- (-3.5,-13.4);
\draw[double] (-3.7676,-14.9093) -- (-4.2491,-14.7584);
\draw[double] (-3.2688,-14.9093) -- (-2.7525,-14.7462);
\draw[double] (-3.5,-15.4) -- (-3.5,-16.15);
\draw[double] (-5.186,-13.9995) -- (-4.748,-14.2351);
\draw[double] (-2.2658,-14.255) -- (-1.7791,-14.0039);
\draw  (-4.75,-14.25) rectangle (-4.25,-14.75);
\draw  (-2.75,-14.25) rectangle (-2.25,-14.75);
\draw  (-3.7392,-16.1454) rectangle (-3.2392,-16.6454);
\draw  (-3.7569,-14.9039) rectangle (-3.2569,-15.4039);
\path (-3.5,-18) node [style=sergio] {$17.~p6m$};
\path (-3.5,-13) node [style=sergio] {\scriptsize$\mu_1$};
\path (-5.75,-16.4) node [style=sergio] {\scriptsize$\mu_1$};
\path (-1.25,-16.4) node [style=sergio] {\scriptsize$\mu_1$};
\path (-3.4892,-16.3954) node [style=sergio] {\scriptsize$\mu_2$};
\path (-3.5069,-15.1539) node [style=sergio] {\scriptsize$\mu_3$};
\path (-4.5,-14.5) node [style=sergio] {\scriptsize$\mu_2$};
\path (-2.5,-14.5) node [style=sergio] {\scriptsize$\mu_2$};
\path (-3.736,-14.1669) node [style=sergio] {\scriptsize$\tau_2$};
\path (-4.45,-13.4) node [style=sergio] {\scriptsize$\tau_2$};
\path (-4.45,-12.4) node [style=sergio] {\scriptsize$\tau_2$};
\path (-3.6555,-12.109) node [style=sergio] {\scriptsize$\tau_2$};
\path (-2.5848,-12.3896) node [style=sergio] {\scriptsize$\tau_2$};
\path (-2.5579,-13.5579) node [style=sergio] {\scriptsize$\tau_2$};
\path (-4.5,-15.5) node [style=sergio] {\scriptsize$\tau_2$};
\path (-2.5,-15.5) node [style=sergio] {\scriptsize$\tau_2$};
\path (-2.2,-17) node [style=sergio] {\scriptsize$\tau_2$};
\path (-0.3067,-17.0369) node [style=sergio] {\scriptsize$\tau_2$};
\path (-0.3443,-15.8179) node [style=sergio] {\scriptsize$\tau_2$};
\path (-1.2659,-15.2324) node [style=sergio] {\scriptsize$\tau_2$};
\path (-1.4285,-17.2055) node [style=sergio] {\scriptsize$\tau_2$};
\path (-4.7663,-16.9827) node [style=sergio] {\scriptsize$\tau_2$};
\path (-5.5847,-17.2164) node [style=sergio] {\scriptsize$\tau_2$};
\path (-6.7,-17) node [style=sergio] {\scriptsize$\tau_2$};
\path (-6.6865,-15.8031) node [style=sergio] {\scriptsize$\tau_2$};
\path (-5.9086,-15.5881) node [style=sergio] {\scriptsize$\tau_2$};
\path (-6.3938,-15.3477) node [style=sergio] {\scriptsize$\tau_1$};
\path (-6.4507,-16.2061) node [style=sergio] {\scriptsize$\tau_1$};
\path (-6.0252,-17.2124) node [style=sergio] {\scriptsize$\tau_1$};
\path (-5.0738,-17.1601) node [style=sergio] {\scriptsize$\tau_1$};
\path (-4.5122,-16.1836) node [style=sergio] {\scriptsize$\tau_1$};
\path (-2.5512,-16.181) node [style=sergio] {\scriptsize$\tau_1$};
\path (-0.5486,-16.5716) node [style=sergio] {\scriptsize$\tau_1$};
\path (-0.7312,-15.2966) node [style=sergio] {\scriptsize$\tau_1$};
\path (-1.5335,-15.611) node [style=sergio] {\scriptsize$\tau_1$};
\path (-1.9888,-17.3347) node [style=sergio] {\scriptsize$\tau_1$};
\path (-0.9389,-17.2655) node [style=sergio] {\scriptsize$\tau_1$};
\path (-5.4939,-15.6838) node [style=sergio] {\scriptsize$\tau_1$};
\path (-4.0095,-13.6317) node [style=sergio] {\scriptsize$\tau_1$};
\path (-2.9904,-13.6642) node [style=sergio] {\scriptsize$\tau_1$};
\path (-2.45,-12.9) node [style=sergio] {\scriptsize$\tau_1$};
\path (-4.5688,-12.9846) node [style=sergio] {\scriptsize$\tau_1$};
\path (-4.2171,-12.1194) node [style=sergio] {\scriptsize$\tau_1$};
\path (-3.192,-12.1065) node [style=sergio] {\scriptsize$\tau_1$};
\path (-5.5,-14) node [style=sergio] {\scriptsize$\tau_3$};
\path (-1.5,-14) node [style=sergio] {\scriptsize$\tau_3$};
\path (-3.9647,-14.6451) node [style=sergio] {\scriptsize$\tau_3$};
\path (-3.0665,-14.6449) node [style=sergio] {\scriptsize$\tau_3$};
\path (-3.6871,-15.7886) node [style=sergio] {\scriptsize$\tau_3$};
\draw  (-9.75,-15.45) rectangle (-9.25,-16.25);
\draw[thin] (-8.95,-14.95) -- (-9.25,-15.45);
\draw[thin] (-10.05,-14.95) -- (-9.75,-15.45);
\draw[thin] (-8.65,-15.85) -- (-9.25,-15.85);
\draw[thin] (-10.75,-15.85) -- (-9.75,-15.85);
\draw[thin] (-8.95,-16.75) -- (-9.25,-16.25);
\draw[thin] (-10.05,-16.75) -- (-9.75,-16.25);
\draw[thin] (-10.25,-16.35) -- (-9.75,-16.05);
\draw[thin] (-8.75,-16.35) -- (-9.25,-16.05);
\draw[thin] (-9.75,-15.65) -- (-10.7437,-15.2421);
\draw[thin] (-9.25,-15.65) -- (-8.75,-15.35);
\draw[thin] (-9.5,-14.85) -- (-9.5,-15.45);
\draw[thin] (-9.5,-16.85) -- (-9.5,-16.25);
\draw  (-12.75,-15.45) rectangle (-12.25,-16.25);
\draw[thin] (-11.95,-14.95) -- (-12.25,-15.45);
\draw[thin] (-13.05,-14.95) -- (-12.75,-15.45);
\draw[thin] (-11.25,-15.85) -- (-12.25,-15.85);
\draw[thin] (-13.35,-15.85) -- (-12.75,-15.85);
\draw[thin] (-11.95,-16.75) -- (-12.25,-16.25);
\draw[thin] (-13.05,-16.75) -- (-12.75,-16.25);
\draw[thin] (-13.25,-16.35) -- (-12.75,-16.05);
\draw[thin] (-11.75,-16.35) -- (-12.25,-16.05);
\draw[thin] (-12.75,-15.65) -- (-13.25,-15.35);
\draw[thin] (-12.25,-15.65) -- (-11.2555,-15.241);
\draw[thin] (-12.5,-14.85) -- (-12.5,-15.45);
\draw[thin] (-12.5,-16.85) -- (-12.5,-16.25);
\draw  (-11.25,-13.15) rectangle (-10.75,-13.95);
\draw[thin] (-10.45,-12.65) -- (-10.75,-13.15);
\draw[thin] (-11.55,-12.65) -- (-11.25,-13.15);
\draw[thin] (-10.15,-13.55) -- (-10.75,-13.55);
\draw[thin] (-11.85,-13.55) -- (-11.25,-13.55);
\draw[thin] (-10.55,-14.45) -- (-10.75,-13.95);
\draw[thin] (-11.45,-14.45) -- (-11.25,-13.95);
\draw[thin] (-11.75,-14.05) -- (-11.25,-13.75);
\draw[thin] (-10.25,-14.05) -- (-10.75,-13.75);
\draw[thin] (-11.25,-13.35) -- (-11.75,-13.05);
\draw[thin] (-10.75,-13.35) -- (-10.25,-13.05);
\draw[thin] (-11,-12.55) -- (-11,-13.15);
\draw[thin] (-11,-14.75) -- (-11,-13.95);
\draw  (-11.95,-14.45) rectangle (-11.45,-14.95);
\draw  (-10.55,-14.45) rectangle (-10.05,-14.95);
\draw  (-11.25,-15.6) rectangle (-10.75,-16.1);
\draw  (-11.25,-14.75) rectangle (-10.75,-15.25);
\path (-11,-18) node [style=sergio] {$16.~p6$};
\path (-11,-13.55) node [style=sergio] {\scriptsize$\mu_1$};
\path (-12.5,-15.85) node [style=sergio] {\scriptsize$\mu_1$};
\path (-9.5,-15.85) node [style=sergio] {\scriptsize$\mu_1$};
\path (-11,-15.85) node [style=sergio] {\scriptsize$\mu_2$};
\path (-11,-15) node [style=sergio] {\scriptsize$\mu_3$};
\path (-11.7,-14.7) node [style=sergio] {\scriptsize$\mu_2$};
\path (-10.3,-14.7) node [style=sergio] {\scriptsize$\mu_2$};
\path (-11.1536,-14.3444) node [style=sergio] {\scriptsize$\tau_2$};
\path (-11.95,-13.95) node [style=sergio] {\scriptsize$\tau_2$};
\path (-11.95,-12.95) node [style=sergio] {\scriptsize$\tau_2$};
\path (-11.1555,-12.659) node [style=sergio] {\scriptsize$\tau_2$};
\path (-10.0848,-12.9396) node [style=sergio] {\scriptsize$\tau_2$};
\path (-10.0579,-14.1079) node [style=sergio] {\scriptsize$\tau_2$};
\path (-11.5974,-15.2137) node [style=sergio] {\scriptsize$\tau_2$};
\path (-10.3672,-15.2293) node [style=sergio] {\scriptsize$\tau_2$};
\path (-10.45,-16.45) node [style=sergio] {\scriptsize$\tau_2$};
\path (-8.5567,-16.4869) node [style=sergio] {\scriptsize$\tau_2$};
\path (-8.5943,-15.2679) node [style=sergio] {\scriptsize$\tau_2$};
\path (-9.5159,-14.6824) node [style=sergio] {\scriptsize$\tau_2$};
\path (-9.6785,-16.6555) node [style=sergio] {\scriptsize$\tau_2$};
\path (-11.5163,-16.4327) node [style=sergio] {\scriptsize$\tau_2$};
\path (-12.3347,-16.6664) node [style=sergio] {\scriptsize$\tau_2$};
\path (-13.45,-16.45) node [style=sergio] {\scriptsize$\tau_2$};
\path (-13.4365,-15.2531) node [style=sergio] {\scriptsize$\tau_2$};
\path (-12.6586,-15.0381) node [style=sergio] {\scriptsize$\tau_2$};
\path (-13.1438,-14.7977) node [style=sergio] {\scriptsize$\tau_1$};
\path (-13.2007,-15.6561) node [style=sergio] {\scriptsize$\tau_1$};
\path (-12.7752,-16.6624) node [style=sergio] {\scriptsize$\tau_1$};
\path (-11.8238,-16.6101) node [style=sergio] {\scriptsize$\tau_1$};
\path (-11.7588,-15.9813) node [style=sergio] {\scriptsize$\tau_1$};
\path (-10.26,-15.9902) node [style=sergio] {\scriptsize$\tau_1$};
\path (-8.7986,-16.0216) node [style=sergio] {\scriptsize$\tau_1$};
\path (-8.9812,-14.7466) node [style=sergio] {\scriptsize$\tau_1$};
\path (-9.7835,-15.061) node [style=sergio] {\scriptsize$\tau_1$};
\path (-10.2388,-16.7847) node [style=sergio] {\scriptsize$\tau_1$};
\path (-9.1889,-16.7155) node [style=sergio] {\scriptsize$\tau_1$};
\path (-12.2439,-15.1338) node [style=sergio] {\scriptsize$\tau_1$};
\path (-11.5095,-14.1817) node [style=sergio] {\scriptsize$\tau_1$};
\path (-10.4904,-14.2142) node [style=sergio] {\scriptsize$\tau_1$};
\path (-9.95,-13.45) node [style=sergio] {\scriptsize$\tau_1$};
\path (-12.0688,-13.5346) node [style=sergio] {\scriptsize$\tau_1$};
\path (-11.7171,-12.6694) node [style=sergio] {\scriptsize$\tau_1$};
\path (-10.692,-12.6565) node [style=sergio] {\scriptsize$\tau_1$};
\end{tikzpicture}
\caption{The corresponding PEPS tensors for \#13 to \#17 wallpaper groups. Here $\mu$'s label different physical indices, and $\tau$'s label different virtual indices.}
\label{summarize2}
\end{figure*}
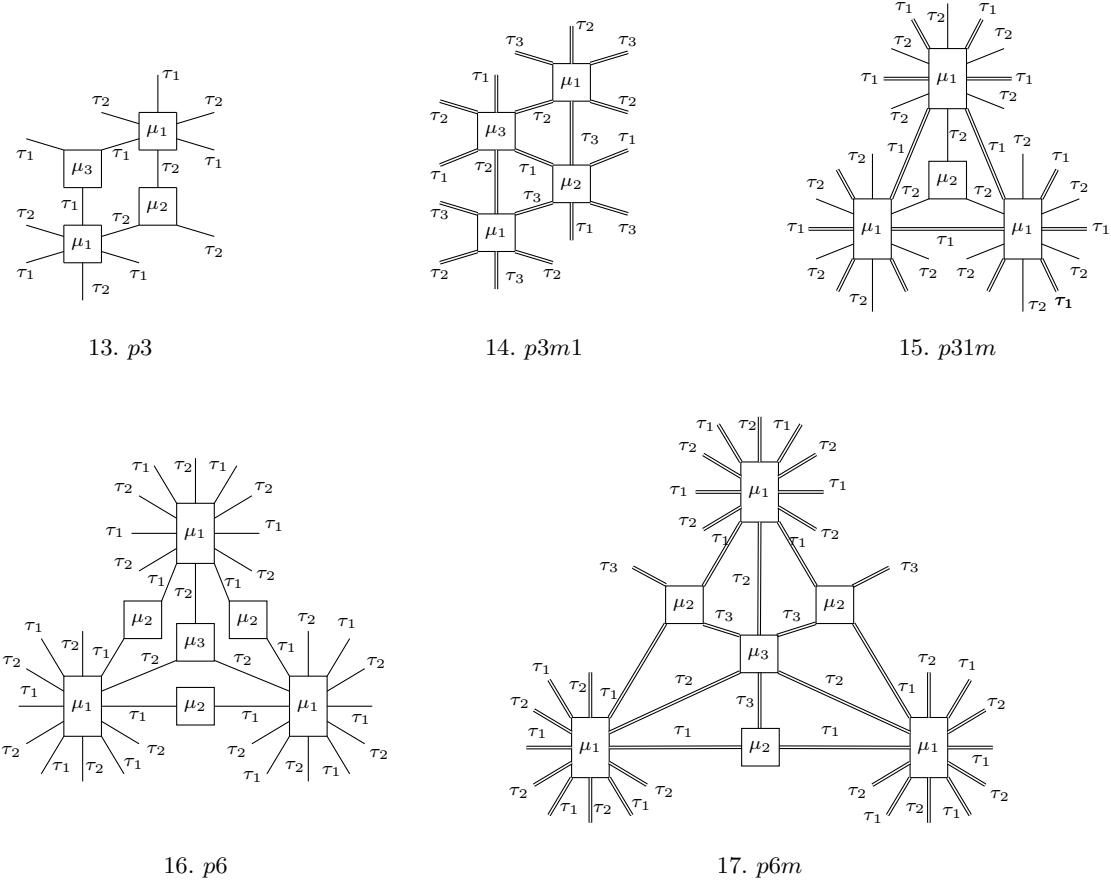

Finally, by investigating the composite rule of different PEPS tensors, we can calculate the accurate group structure of the classification of 2D crystalline fSPT phases with wallpaper group symmetry. All results of classifications for both spinless and spin-1/2 fermions are summarized in Table \ref{classification}, and the corresponding PEPS tensors for all wallpaper groups within one unit cell are summarized in Figs. \ref{summarize1} and \ref{summarize2}. Here we use single/double solid lines to label the virtual indices lying on/away from the reflection axis, equivalently, the effective on-site symmetry groups of virtual bonds labeled by single/double solid lines are $\mathbb{Z}_1$/$\mathbb{Z}_2$. All results are confirmed with our previous works in which the crystalline fSPT phases with wallpaper group symmetry were constructed and classified by real-space constructions \cite{wallpaper}. 

\begin{table}[t]
\renewcommand\arraystretch{1.2}
\begin{tabular}{|c|c|c|c|c|c|c|}
\hline
$~~G_b~~$&~~~spinless~~~&~~~spin-1/2~~~\\
\hline
$p1$&${\mathbb{Z}_2\times\mathbb{Z}_4}$&${\mathbb{Z}_2\times\mathbb{Z}_4}$\\
\hline
$p2$&${\mathbb{Z}_2^4}$&$\mathbb{Z}_4\times\mathbb{Z}_8^3$\\
\hline
$pm$&${\mathbb{Z}_2^6}$&$\mathbb{Z}_4\times\mathbb{Z}_8$\\
\hline
$pg$&${\mathbb{Z}_2\times\mathbb{Z}_4}$&${\mathbb{Z}_4\times\mathbb{Z}_2}$\\
\hline
$cm$&${\mathbb{Z}_2^4}$&${\mathbb{Z}_2}\times{\mathbb{Z}_4}$\\
\hline
$pmm$&${\mathbb{Z}_2^8}$&${\mathbb{Z}_2^8}$\\
\hline
$pmg$&${\mathbb{Z}_2^5}$&${\mathbb{Z}_4\times\mathbb{Z}_8^2}$\\
\hline
$pgg$&${\mathbb{Z}_2^3}$&$~~{\mathbb{Z}_2}\times{\mathbb{Z}_4}\times\mathbb{Z}_8~~$\\
\hline
$cmm$&${\mathbb{Z}_2^5}$&${\mathbb{Z}_8}\times{\mathbb{Z}_2^4}$\\
\hline
$p4$&$~~{\mathbb{Z}_2^3}\times{\mathbb{Z}_4}~~$&${\mathbb{Z}_2}\times{\mathbb{Z}_8^3}$\\
\hline
$p4m$&${\mathbb{Z}_2^7}$&${\mathbb{Z}_2^6}$\\
\hline
$p4g$&${\mathbb{Z}_2^4}$&${\mathbb{Z}_8}\times{\mathbb{Z}_2^3}$\\
\hline
$p3$&$~{\mathbb{Z}_2}\times{\mathbb{Z}_3^3}~$&${\mathbb{Z}_2}\times{\mathbb{Z}_3^3}$\\
\hline
$~~~p3m1~~~$&${\mathbb{Z}_2^3}$&${\mathbb{Z}_4}$\\
\hline
$p31m$&${\mathbb{Z}_2^3}\times{\mathbb{Z}_3}$&${\mathbb{Z}_4\times\mathbb{Z}_3}$\\
\hline
$p6$&${\mathbb{Z}_2^2}\times{\mathbb{Z}_3^2}$&${\mathbb{Z}_{12}\times\mathbb{Z}_8}\times{\mathbb{Z}_3}$\\
\hline
$p6m$&${\mathbb{Z}_2^4}$&${\mathbb{Z}_2^4}$\\
\hline
\end{tabular}
\caption{Classifications for 2D crystalline fSPT phases protected by wallpaper group symmetry from tensor network representations, for both spinless and spin-1/2 fermions.}
\label{classification}
\end{table}

\begin{figure*}
\begin{tikzpicture}
\usetikzlibrary{shapes.geometric}
\tikzstyle{sergio}=[rectangle,draw=none]
\draw[thin] (0,4.25) -- (0,3.75);
\draw[thin] (-0.5,3.75) -- (0,3.75);
\draw[thin] (-0.5,4.25) -- (0,4.25);
\draw[thin] (-0.5,4.25) -- (-0.5,3.75);
\draw[thin] (-0.25,4.25) -- (-0.25,4.75);
\draw[thin] (-0.25,3.25) -- (-0.25,3.75);
\draw[thin] (-0.5,4) -- (-1,4);
\draw[thin] (0.5,4) -- (0,4);
\path (-0.25,4) node [style=sergio] {\scriptsize$\mu$};
\path (0.4,4.25) node [style=sergio] {\scriptsize$\tau_1$};
\path (-0.5,4.65) node [style=sergio] {\scriptsize$\tau_2$};
\path (-12.25,2.35) node [style=sergio] {$1.~p1$};
\draw[thin] (-5,1.1) -- (-5,0.6);
\draw[thin] (-5.5,0.6) -- (-5,0.6);
\draw[thin] (-5.5,1.1) -- (-5,1.1);
\draw[thin] (-5.5,1.1) -- (-5.5,0.6);
\draw[double] (-5.25,1.1) -- (-5.25,1.6);
\draw[double] (-5.25,0.1) -- (-5.25,0.6);
\draw[thin] (-5.5,0.85) -- (-6,0.85);
\draw[thin] (-4.5,0.85) -- (-5,0.85);
\draw[thin] (-4.5,1.1) -- (-4,1.1);
\draw[thin] (-4.5,1.1) -- (-4.5,0.6);
\draw[thin] (-4,0.6) -- (-4,1.1);
\draw[thin] (-4,0.6) -- (-4.5,0.6);
\draw[thin] (-5,0.1) -- (-5.5,0.1);
\draw[thin] (-5.5,-0.4) -- (-5.5,0.1);
\draw[thin] (-5.5,-0.4) -- (-5,-0.4);
\draw[thin] (-5,0.1) -- (-5,-0.4);
\draw[thin] (-5.5,-0.15) -- (-6,-0.15);
\draw[thin] (-3.5,0.85) -- (-4,0.85);
\draw[double] (-5.25,-0.4) -- (-5.25,-0.9);
\draw[thin] (-3.5,0.6) -- (-3.5,1.1);
\draw[thin] (-3,1.1) -- (-3.5,1.1);
\draw[thin] (-3,1.1) -- (-3,0.6);
\draw[thin] (-3.5,0.6) -- (-3,0.6);
\draw[thin] (-2.5,0.85) -- (-3,0.85);
\draw[double] (-3.25,0.6) -- (-3.25,0.1);
\draw[double] (-3.25,1.6) -- (-3.25,1.1);
\draw[thin] (-4.5,-0.15) -- (-5,-0.15);
\draw[thin] (-4.5,-0.4) -- (-4.5,0.1);
\draw[thin] (-4,0.1) -- (-4.5,0.1);
\draw[thin] (-4,0.1) -- (-4,-0.4);
\draw[thin] (-4.5,-0.4) -- (-4,-0.4);
\draw[thin] (-3.5,-0.15) -- (-4,-0.15);
\draw[thin] (-3.5,0.1) -- (-3,0.1);
\draw[thin] (-3.5,0.1) -- (-3.5,-0.4);
\draw[thin] (-3,-0.4) -- (-3.5,-0.4);
\draw[thin] (-3,-0.4) -- (-3,0.1);
\draw[double] (-3.25,-0.4) -- (-3.25,-0.9);
\draw[thin] (-2.5,-0.15) -- (-3,-0.15);
\path (-4.25,-1.4) node [style=sergio] {$7.~pmg$};
\path (-5.25,0.85) node [style=sergio] {\scriptsize$\mu_1$};
\path (-5.25,-0.15) node [style=sergio] {\scriptsize$\mu_2$};
\path (-4.25,0.85) node [style=sergio] {\scriptsize$\mu_3$};
\path (-4.25,-0.15) node [style=sergio] {\scriptsize$\mu_4$};
\path (-3.25,0.85) node [style=sergio] {\scriptsize$\mu_1$};
\path (-3.25,-0.15) node [style=sergio] {\scriptsize$\mu_2$};
\path (-5.5,1.5) node [style=sergio] {\scriptsize$\tau_1$};
\path (-5.5,0.35) node [style=sergio] {\scriptsize$\tau_2$};
\path (-5.5,-0.8) node [style=sergio] {\scriptsize$\tau_1$};
\path (-5.9,1.1) node [style=sergio] {\scriptsize$\tau_4$};
\path (-4.75,1.1) node [style=sergio] {\scriptsize$\tau_4$};
\path (-3.75,1.1) node [style=sergio] {\scriptsize$\tau_4$};
\path (-2.6,1.1) node [style=sergio] {\scriptsize$\tau_4$};
\path (-5.9,0.1) node [style=sergio] {\scriptsize$\tau_3$};
\path (-4.75,0.1) node [style=sergio] {\scriptsize$\tau_3$};
\path (-3.75,0.1) node [style=sergio] {\scriptsize$\tau_3$};
\path (-2.6,0.1) node [style=sergio] {\scriptsize$\tau_3$};
\path (-3,1.5) node [style=sergio] {\scriptsize$\tau_2$};
\path (-3,0.35) node [style=sergio] {\scriptsize$\tau_1$};
\path (-3,-0.8) node [style=sergio] {\scriptsize$\tau_2$};
\draw[thin] (-4.5,4.25) -- (-4.5,3.75);
\draw[thin] (-5,3.75) -- (-4.5,3.75);
\draw[thin] (-5,4.25) -- (-4.5,4.25);
\draw[thin] (-5,4.25) -- (-5,3.75);
\draw[double] (-4.75,4.25) -- (-4.75,4.75);
\draw[double] (-4.75,3.25) -- (-4.75,3.75);
\draw[thin] (-5,4) -- (-5.5,4);
\draw[thin] (-4,4) -- (-4.5,4);
\draw[thin] (-4,3.75) -- (-4,4.25);
\draw[thin] (-3.5,4.25) -- (-4,4.25);
\draw[thin] (-3.5,4.25) -- (-3.5,3.75);
\draw[thin] (-4,3.75) -- (-3.5,3.75);
\draw[thin] (-3,4) -- (-3.5,4);
\draw[double] (-3.75,3.75) -- (-3.75,3.25);
\draw[double] (-3.75,4.75) -- (-3.75,4.25);
\path (-4.25,2.35) node [style=sergio] {$3.~pm$};
\path (-4.75,4) node [style=sergio] {\scriptsize$\mu_1$};
\path (-3.75,4) node [style=sergio] {\scriptsize$\mu_2$};
\path (-5.4,4.25) node [style=sergio] {\scriptsize$\tau_1$};
\path (-4.25,4.25) node [style=sergio] {\scriptsize$\tau_1$};
\path (-3.1,4.25) node [style=sergio] {\scriptsize$\tau_1$};
\path (-5,3.35) node [style=sergio] {\scriptsize$\tau_2$};
\path (-5,4.65) node [style=sergio] {\scriptsize$\tau_2$};
\path (-3.5,3.35) node [style=sergio] {\scriptsize$\tau_3$};
\path (-3.5,4.65) node [style=sergio] {\scriptsize$\tau_3$};
\draw[thin] (-12,4.25) -- (-12,3.75);
\draw[thin] (-12.5,3.75) -- (-12,3.75);
\draw[thin] (-12.5,4.25) -- (-12,4.25);
\draw[thin] (-12.5,4.25) -- (-12.5,3.75);
\draw[thin] (-12.25,4.25) -- (-12.25,4.75);
\draw[thin] (-12.25,3.25) -- (-12.25,3.75);
\draw[thin] (-12.5,4) -- (-13,4);
\draw[thin] (-11.5,4) -- (-12,4);
\path (-12.25,4) node [style=sergio] {\scriptsize$\mu$};
\path (-11.6,4.25) node [style=sergio] {\scriptsize$\tau_1$};
\path (-12.5,4.65) node [style=sergio] {\scriptsize$\tau_2$};
\path (-0.25,2.35) node [style=sergio] {$4.~pg$};
\draw[thin] (-12,0.6) -- (-12,0.1);
\draw[thin] (-12.5,0.1) -- (-12,0.1);
\draw[thin] (-12.5,0.6) -- (-12,0.6);
\draw[thin] (-12.5,0.6) -- (-12.5,0.1);
\draw[double] (-12.25,0.6) -- (-12.25,1.1);
\draw[thin] (-12.25,0.1) -- (-12.25,0.1);
\draw[thin] (-12.5,0.25) -- (-13,0.1);
\draw[thin] (-12.5,0.45) -- (-13,0.6);
\draw[thin] (-12,0.25) -- (-11.5,0.1);
\draw[thin] (-12,0.45) -- (-11.5,0.6);
\draw[double] (-12.25,-0.4) -- (-12.25,0.1);
\path (-12.25,-1.4) node [style=sergio] {$5.~cm$};
\path (-12.25,0.35) node [style=sergio] {\scriptsize$\mu$};
\path (-12.25,1.35) node [style=sergio] {\scriptsize$\tau_2$};
\path (-12.25,-0.65) node [style=sergio] {\scriptsize$\tau_2$};
\path (-13,0.85) node [style=sergio] {\scriptsize$\tau_1$};
\path (-13,-0.15) node [style=sergio] {\scriptsize$\tau_1$};
\path (-11.5,0.85) node [style=sergio] {\scriptsize$\tau_1$};
\path (-11.5,-0.15) node [style=sergio] {\scriptsize$\tau_1$};
\draw[thin] (-8.5,1.1) -- (-8.5,0.6);
\draw[thin] (-9,0.6) -- (-8.5,0.6);
\draw[thin] (-9,1.1) -- (-8.5,1.1);
\draw[thin] (-9,1.1) -- (-9,0.6);
\draw[double] (-8.75,1.1) -- (-8.75,1.6);
\draw[double] (-8.75,0.1) -- (-8.75,0.6);
\draw[double] (-9,0.85) -- (-9.5,0.85);
\draw[double] (-8,0.85) -- (-8.5,0.85);
\draw[thin] (-8.5,0.1) -- (-9,0.1);
\draw[thin] (-9,-0.4) -- (-9,0.1);
\draw[thin] (-9,-0.4) -- (-8.5,-0.4);
\draw[thin] (-8.5,0.1) -- (-8.5,-0.4);
\draw[double] (-9,-0.15) -- (-9.5,-0.15);
\draw[double] (-8.75,-0.4) -- (-8.75,-0.9);
\draw[thin] (-8,0.6) -- (-8,1.1);
\draw[thin] (-7.5,1.1) -- (-8,1.1);
\draw[thin] (-7.5,1.1) -- (-7.5,0.6);
\draw[thin] (-8,0.6) -- (-7.5,0.6);
\draw[double] (-7,0.85) -- (-7.5,0.85);
\draw[double] (-7.75,0.6) -- (-7.75,0.1);
\draw[double] (-7.75,1.6) -- (-7.75,1.1);
\draw[double] (-8,-0.15) -- (-8.5,-0.15);
\draw[thin] (-8,0.1) -- (-7.5,0.1);
\draw[thin] (-8,0.1) -- (-8,-0.4);
\draw[thin] (-7.5,-0.4) -- (-8,-0.4);
\draw[thin] (-7.5,-0.4) -- (-7.5,0.1);
\draw[double] (-7.75,-0.4) -- (-7.75,-0.9);
\draw[double] (-7,-0.15) -- (-7.5,-0.15);
\path (-8.25,-1.4) node [style=sergio] {$6.~pmm$};
\path (-8.75,0.85) node [style=sergio] {\scriptsize$\mu_1$};
\path (-7.75,0.85) node [style=sergio] {\scriptsize$\mu_3$};
\path (-8.75,-0.15) node [style=sergio] {\scriptsize$\mu_2$};
\path (-7.75,-0.15) node [style=sergio] {\scriptsize$\mu_4$};
\path (-9.4,1.1) node [style=sergio] {\scriptsize$\tau_1$};
\path (-8.25,1.1) node [style=sergio] {\scriptsize$\tau_1$};
\path (-7.1,1.1) node [style=sergio] {\scriptsize$\tau_1$};
\path (-9.4,-0.4) node [style=sergio] {\scriptsize$\tau_3$};
\path (-8.25,-0.4) node [style=sergio] {\scriptsize$\tau_3$};
\path (-7.1,-0.4) node [style=sergio] {\scriptsize$\tau_3$};
\path (-9,1.5) node [style=sergio] {\scriptsize$\tau_2$};
\path (-9,0.35) node [style=sergio] {\scriptsize$\tau_2$};
\path (-9,-0.8) node [style=sergio] {\scriptsize$\tau_2$};
\path (-7.5,1.5) node [style=sergio] {\scriptsize$\tau_4$};
\path (-7.5,0.35) node [style=sergio] {\scriptsize$\tau_4$};
\path (-7.5,-0.8) node [style=sergio] {\scriptsize$\tau_4$};
\draw[thin] (-9,4.75) -- (-9,4.25);
\draw[thin] (-9.5,4.25) -- (-9,4.25);
\draw[thin] (-9.5,4.75) -- (-9,4.75);
\draw[thin] (-9.5,4.75) -- (-9.5,4.25);
\draw[thin] (-9.25,4.75) -- (-9.25,5.25);
\draw[thin] (-9.25,3.75) -- (-9.25,4.25);
\draw[thin] (-9.5,4.5) -- (-10,4.5);
\draw[thin] (-8.5,4.5) -- (-9,4.5);
\draw[thin] (-8.5,4.75) -- (-8,4.75);
\draw[thin] (-8.5,4.75) -- (-8.5,4.25);
\draw[thin] (-8,4.25) -- (-8,4.75);
\draw[thin] (-8,4.25) -- (-8.5,4.25);
\draw[thin] (-9,3.75) -- (-9.5,3.75);
\draw[thin] (-9.5,3.25) -- (-9.5,3.75);
\draw[thin] (-9.5,3.25) -- (-9,3.25);
\draw[thin] (-9,3.75) -- (-9,3.25);
\draw[thin] (-9.5,3.5) -- (-10,3.5);
\draw[thin] (-7.5,4.5) -- (-8,4.5);
\draw[thin] (-9.25,3.25) -- (-9.25,2.75);
\draw[thin] (-7.5,4.25) -- (-7.5,4.75);
\draw[thin] (-7,4.75) -- (-7.5,4.75);
\draw[thin] (-7,4.75) -- (-7,4.25);
\draw[thin] (-7.5,4.25) -- (-7,4.25);
\draw[thin] (-6.5,4.5) -- (-7,4.5);
\draw[thin] (-7.25,4.25) -- (-7.25,3.75);
\draw[thin] (-7.25,5.25) -- (-7.25,4.75);
\draw[thin] (-8.5,3.5) -- (-9,3.5);
\draw[thin] (-8.5,3.25) -- (-8.5,3.75);
\draw[thin] (-8,3.75) -- (-8.5,3.75);
\draw[thin] (-8,3.75) -- (-8,3.25);
\draw[thin] (-8.5,3.25) -- (-8,3.25);
\draw[thin] (-7.5,3.5) -- (-8,3.5);
\draw[thin] (-7.5,3.75) -- (-7,3.75);
\draw[thin] (-7.5,3.75) -- (-7.5,3.25);
\draw[thin] (-7,3.25) -- (-7.5,3.25);
\draw[thin] (-7,3.25) -- (-7,3.75);
\draw[thin] (-7.25,3.25) -- (-7.25,2.75);
\draw[thin] (-6.5,3.5) -- (-7,3.5);
\path (-8.25,2.35) node [style=sergio] {$2.~p2$};
\path (-9.25,4.5) node [style=sergio] {\scriptsize$\mu_1$};
\path (-9.25,3.5) node [style=sergio] {\scriptsize$\mu_2$};
\path (-8.25,4.5) node [style=sergio] {\scriptsize$\mu_3$};
\path (-8.25,3.5) node [style=sergio] {\scriptsize$\mu_4$};
\path (-7.25,4.5) node [style=sergio] {\scriptsize$\mu_1$};
\path (-7.25,3.5) node [style=sergio] {\scriptsize$\mu_2$};
\path (-9.5,5.15) node [style=sergio] {\scriptsize$\tau_1$};
\path (-9.5,4) node [style=sergio] {\scriptsize$\tau_1$};
\path (-9.5,2.85) node [style=sergio] {\scriptsize$\tau_1$};
\path (-9.9,4.75) node [style=sergio] {\scriptsize$\tau_2$};
\path (-8.75,4.75) node [style=sergio] {\scriptsize$\tau_2$};
\path (-7.75,4.75) node [style=sergio] {\scriptsize$\tau_2$};
\path (-6.6,4.75) node [style=sergio] {\scriptsize$\tau_2$};
\path (-9.9,3.75) node [style=sergio] {\scriptsize$\tau_3$};
\path (-8.75,3.75) node [style=sergio] {\scriptsize$\tau_3$};
\path (-7.75,3.75) node [style=sergio] {\scriptsize$\tau_3$};
\path (-6.6,3.75) node [style=sergio] {\scriptsize$\tau_3$};
\path (-7,5.15) node [style=sergio] {\scriptsize$\tau_1$};
\path (-7,4) node [style=sergio] {\scriptsize$\tau_1$};
\path (-7,2.85) node [style=sergio] {\scriptsize$\tau_1$};
\draw[thin] (-0.5,0.6) -- (-1,0.6);
\draw[thin] (-1,0.6) -- (-1,0.1);
\draw[thin] (-1,0.1) -- (-0.5,0.1);
\draw[thin] (-0.5,0.6) -- (-0.5,0.1);
\draw[thin] (-0.85,0.6) -- (-1,1.1);
\draw[thin] (-0.65,0.6) -- (-0.5,1.1);
\draw[thin] (-0.85,0.1) -- (-1,-0.4);
\draw[thin] (-0.65,0.1) -- (-0.5,-0.4);
\draw[thin] (-0.85,0.6) -- (-1,1.1);
\draw[thin] (0,0.35) -- (-0.5,0.35);
\draw[thin] (-1,0.35) -- (-1.5,0.35);
\draw[thin] (0,0.1) -- (0,0.6);
\draw[thin] (0.5,0.6) -- (0,0.6);
\draw[thin] (0.5,0.6) -- (0.5,0.1);
\draw[thin] (0,0.1) -- (0.5,0.1);
\draw[thin] (1,0.35) -- (0.5,0.35);
\path (0.25,0.35) node [style=sergio] {\scriptsize$\mu_2$};
\path (-0.75,0.35) node [style=sergio] {\scriptsize$\mu_1$};
\path (-1.2,1) node [style=sergio] {\scriptsize$\tau_1$};
\path (-0.3,1) node [style=sergio] {\scriptsize$\tau_1$};
\path (-1.2,-0.3) node [style=sergio] {\scriptsize$\tau_1$};
\path (-0.3,-0.3) node [style=sergio] {\scriptsize$\tau_1$};
\path (-1.2,1) node [style=sergio] {\scriptsize$\tau_1$};
\path (-1.5,0.1) node [style=sergio] {\scriptsize$\tau_2$};
\path (1,0.1) node [style=sergio] {\scriptsize$\tau_2$};
\path (-0.25,0.2) node [style=sergio] {\scriptsize$\tau_2$};
\path (-0.25,-1.4) node [style=sergio] {$8.~pgg$};
\draw[thin] (-12.5,-3.5) -- (-12.5,-4);
\draw[thin] (-12.5,-3.5) -- (-12,-3.5);
\draw[thin] (-12,-4) -- (-12.5,-4);
\draw[thin] (-12,-4) -- (-12,-3.5);
\draw[thin] (-12,-3.5) -- (-11.5,-3);
\draw[thin] (-13,-4.5) -- (-12.5,-4);
\draw[thin] (-13,-4.5) -- (-13.5,-4.5);
\draw[thin] (-13.5,-5) -- (-13.5,-4.5);
\draw[thin] (-13.5,-5) -- (-13,-5);
\draw[thin] (-13,-4.5) -- (-13,-5);
\draw[thin] (-13.5,-5) -- (-13.75,-5.25);
\draw[thin] (-13.5,-4.5) -- (-13.75,-4.25);
\draw[thin] (-13,-5) -- (-12.75,-5.25);
\draw[double] (-13.25,-5) -- (-13.25,-5.5);
\draw[double] (-13.25,-4) -- (-13.25,-4.5);
\draw[double] (-13.5,-4.75) -- (-14,-4.75);
\draw[double] (-11.5,-4.75) -- (-13,-4.75);
\draw[thin] (-11.5,-2.5) -- (-11.5,-3);
\draw[thin] (-11.5,-2.5) -- (-11,-2.5);
\draw[thin] (-11,-3) -- (-11,-2.5);
\draw[thin] (-11,-3) -- (-11.5,-3);
\draw[thin] (-10.75,-2.25) -- (-11,-2.5);
\draw[thin] (-11.75,-2.25) -- (-11.5,-2.5);
\draw[thin] (-11,-3) -- (-10.75,-3.25);
\draw[double] (-11.25,-2.5) -- (-11.25,-2);
\draw[double] (-11.25,-4.5) -- (-11.25,-3);
\draw[double] (-12,-2.75) -- (-11.5,-2.75);
\draw[double] (-11,-2.75) -- (-10.5,-2.75);
\draw[thin] (-11.5,-5) -- (-11.5,-4.5);
\draw[thin] (-11,-4.5) -- (-11.5,-4.5);
\draw[thin] (-11,-4.5) -- (-11,-5);
\draw[thin] (-11.5,-5) -- (-11,-5);
\draw[double] (-10.5,-4.75) -- (-11,-4.75);
\draw[double] (-11.25,-5) -- (-11.25,-5.5);
\path (-12.25,-3.75) node [style=sergio] {\scriptsize$\mu_1$};
\path (-13.25,-4.75) node [style=sergio] {\scriptsize$\mu_2$};
\path (-11.25,-2.75) node [style=sergio] {\scriptsize$\mu_2$};
\path (-11.25,-4.75) node [style=sergio] {\scriptsize$\mu_3$};
\path (-12.25,-6) node [style=sergio] {$9.~cmm$};
\path (-12.55,-4.3) node [style=sergio] {\scriptsize$\tau_1$};
\path (-11.6,-3.3) node [style=sergio] {\scriptsize$\tau_1$};
\path (-10.5,-2) node [style=sergio] {\scriptsize$\tau_1$};
\path (-14,-5.5) node [style=sergio] {\scriptsize$\tau_1$};
\path (-12,-2) node [style=sergio] {\scriptsize$\tau_1$};
\path (-10.5,-3.5) node [style=sergio] {\scriptsize$\tau_1$};
\path (-12.5,-5.5) node [style=sergio] {\scriptsize$\tau_1$};
\path (-14,-4) node [style=sergio] {\scriptsize$\tau_1$};
\path (-14.25,-4.75) node [style=sergio] {\scriptsize$\tau_2$};
\path (-10.25,-4.75) node [style=sergio] {\scriptsize$\tau_2$};
\path (-12.25,-5) node [style=sergio] {\scriptsize$\tau_2$};
\path (-10.25,-2.75) node [style=sergio] {\scriptsize$\tau_2$};
\path (-12.25,-2.75) node [style=sergio] {\scriptsize$\tau_2$};
\path (-13.25,-3.75) node [style=sergio] {\scriptsize$\tau_3$};
\path (-13.25,-5.75) node [style=sergio] {\scriptsize$\tau_3$};
\path (-11.25,-5.75) node [style=sergio] {\scriptsize$\tau_3$};
\path (-11,-2) node [style=sergio] {\scriptsize$\tau_3$};
\path (-11,-3.75) node [style=sergio] {\scriptsize$\tau_3$};
\draw[thin] (-9,-3) -- (-8.5,-3);
\draw[thin] (-9,-3) -- (-9,-3.5);
\draw[thin] (-8.5,-3.5) -- (-8.5,-3);
\draw[thin] (-8.5,-3.5) -- (-9,-3.5);
\draw[thin] (-8,-3.5) -- (-8,-3);
\draw[thin] (-7.5,-3) -- (-8,-3);
\draw[thin] (-8,-3.5) -- (-7.5,-3.5);
\draw[thin] (-7.5,-3) -- (-7.5,-3.5);
\draw[thin] (-8.75,-3) -- (-8.75,-2.5);
\draw[thin] (-8.75,-4) -- (-8.75,-3.5);
\draw[thin] (-9,-4) -- (-8.5,-4);
\draw[thin] (-9,-4) -- (-9,-4.5);
\draw[thin] (-8.5,-4.5) -- (-8.5,-4);
\draw[thin] (-8.5,-4.5) -- (-9,-4.5);
\draw[thin] (-8.75,-5) -- (-8.75,-4.5);
\draw[thin] (-9,-3.25) -- (-9.5,-3.25);
\draw[thin] (-8,-3.25) -- (-8.5,-3.25);
\draw[thin] (-7,-3.25) -- (-7.5,-3.25);
\draw[thin] (-9,-4.25) -- (-9.5,-4.25);
\draw[thin] (-8,-4.25) -- (-8.5,-4.25);
\draw[thin] (-7,-4.25) -- (-7.5,-4.25);
\draw[thin] (-7.75,-3) -- (-7.75,-2.5);
\draw[thin] (-7.75,-4) -- (-7.75,-3.5);
\draw[thin] (-7.75,-5) -- (-7.75,-4.5);
\draw[thin] (-7.5,-4) -- (-8,-4);
\draw[thin] (-8,-4.5) -- (-8,-4);
\draw[thin] (-7.5,-4) -- (-7.5,-4.5);
\draw[thin] (-8,-4.5) -- (-7.5,-4.5);
\path (-9,-5) node [style=sergio] {\scriptsize$\tau_1$};
\path (-9,-3.75) node [style=sergio] {\scriptsize$\tau_1$};
\path (-9,-2.5) node [style=sergio] {\scriptsize$\tau_1$};
\path (-9.75,-3.25) node [style=sergio] {\scriptsize$\tau_1$};
\path (-8.232,-3.3861) node [style=sergio] {\scriptsize$\tau_1$};
\path (-6.75,-3.25) node [style=sergio] {\scriptsize$\tau_1$};
\path (-9.75,-4.25) node [style=sergio] {\scriptsize$\tau_2$};
\path (-8.2411,-4.3785) node [style=sergio] {\scriptsize$\tau_2$};
\path (-7.5,-3.75) node [style=sergio] {\scriptsize$\tau_2$};
\path (-7.5,-2.5) node [style=sergio] {\scriptsize$\tau_2$};
\path (-8.75,-4.25) node [style=sergio] {\scriptsize$\mu_2$};
\path (-8.75,-3.25) node [style=sergio] {\scriptsize$\mu_1$};
\path (-7.75,-3.25) node [style=sergio] {\scriptsize$\mu_2$};
\path (-7.75,-4.25) node [style=sergio] {\scriptsize$\mu_3$};
\path (-6.75,-4.25) node [style=sergio] {\scriptsize$\tau_2$};
\path (-8.25,-6) node [style=sergio] {$10.~p4$};
\draw[thin] (-1,-3) -- (-0.5,-3);
\draw[thin] (-1,-3) -- (-1,-3.5);
\draw[thin] (-0.5,-3.5) -- (-0.5,-3);
\draw[thin] (-0.5,-3.5) -- (-1,-3.5);
\draw[thin] (0,-3.5) -- (0,-3);
\draw[thin] (0.5,-3) -- (0,-3);
\draw[thin] (0,-3.5) -- (0.5,-3.5);
\draw[thin] (0.5,-3) -- (0.5,-3.5);
\draw[thin] (-0.75,-3) -- (-0.75,-2.5);
\draw[thin] (-0.75,-4) -- (-0.75,-3.5);
\draw[thin] (-1,-4) -- (-0.5,-4);
\draw[thin] (-1,-4) -- (-1,-4.5);
\draw[thin] (-0.5,-4.5) -- (-0.5,-4);
\draw[thin] (-0.5,-4.5) -- (-1,-4.5);
\draw[thin] (-0.75,-5) -- (-0.75,-4.5);
\draw[thin] (-1,-3.25) -- (-1.5,-3.25);
\draw[thin] (0,-3.25) -- (-0.5,-3.25);
\draw[thin] (1,-3.25) -- (0.5,-3.25);
\draw[thin] (-1,-4.25) -- (-1.5,-4.25);
\draw[thin] (0,-4.25) -- (-0.5,-4.25);
\draw[thin] (1,-4.25) -- (0.5,-4.25);
\draw[thin] (0.25,-3) -- (0.25,-2.5);
\draw[thin] (0.25,-4) -- (0.25,-3.5);
\draw[thin] (0.25,-5) -- (0.25,-4.5);
\draw[thin] (0,-4.5) -- (0,-4);
\draw[thin] (0.5,-4) -- (0,-4);
\draw[thin] (0,-4.5) -- (0.5,-4.5);
\draw[thin] (0.5,-4) -- (0.5,-4.5);
\draw[double] (0,-3.5) -- (-0.5,-4);
\draw[double] (0.75,-2.75) -- (0.5,-3);
\draw[double] (-0.25,-2.75) -- (0,-3);
\draw[double] (0.5,-3.5) -- (0.75,-3.75);
\draw[double] (-1.25,-3.75) -- (-1,-4);
\draw[double] (-0.5,-4.5) -- (-0.25,-4.75);
\draw[double] (-1,-4.5) -- (-1.25,-4.75);
\path (-1,-5) node [style=sergio] {\scriptsize$\tau_1$};
\path (-0.8887,-3.7603) node [style=sergio] {\scriptsize$\tau_1$};
\path (-1,-2.5) node [style=sergio] {\scriptsize$\tau_1$};
\path (-1.5,-3.5) node [style=sergio] {\scriptsize$\tau_1$};
\path (-0.25,-3.5) node [style=sergio] {\scriptsize$\tau_1$};
\path (1,-3.5) node [style=sergio] {\scriptsize$\tau_1$};
\path (-1.5,-4.5) node [style=sergio] {\scriptsize$\tau_1$};
\path (-0.25,-4.5) node [style=sergio] {\scriptsize$\tau_1$};
\path (1,-4.5) node [style=sergio] {\scriptsize$\tau_1$};
\path (0.4255,-3.7682) node [style=sergio] {\scriptsize$\tau_1$};
\path (0.5,-5) node [style=sergio] {\scriptsize$\tau_1$};
\path (0.5,-2.5) node [style=sergio] {\scriptsize$\tau_1$};
\path (0.8468,-2.6333) node [style=sergio] {\scriptsize$\tau_2$};
\path (0.8707,-3.8655) node [style=sergio] {\scriptsize$\tau_2$};
\path (-0.1071,-3.8258) node [style=sergio] {\scriptsize$\tau_2$};
\path (-0.2473,-2.6192) node [style=sergio] {\scriptsize$\tau_2$};
\path (-1.3921,-3.756) node [style=sergio] {\scriptsize$\tau_2$};
\path (-0.2473,-4.8928) node [style=sergio] {\scriptsize$\tau_2$};
\path (-1.3841,-4.861) node [style=sergio] {\scriptsize$\tau_2$};
\path (-0.75,-4.25) node [style=sergio] {\scriptsize$\mu_2$};
\path (-0.75,-3.25) node [style=sergio] {\scriptsize$\mu_1$};
\path (0.25,-3.25) node [style=sergio] {\scriptsize$\mu_2$};
\path (0.25,-4.25) node [style=sergio] {\scriptsize$\mu_1$};
\path (-0.25,-6) node [style=sergio] {$12.~p4g$};
\draw[thin] (-5,-3) -- (-4.5,-3);
\draw[thin] (-5,-3) -- (-5,-3.5);
\draw[thin] (-4.5,-3.5) -- (-4.5,-3);
\draw[thin] (-4.5,-3.5) -- (-5,-3.5);
\draw[double] (-4.5,-3.5) -- (-4,-4);
\draw[double] (-3.25,-3.75) -- (-3.5,-4);
\draw[thin] (-4,-3.5) -- (-4,-3);
\draw[thin] (-3.5,-3) -- (-4,-3);
\draw[thin] (-4,-3.5) -- (-3.5,-3.5);
\draw[thin] (-3.5,-3) -- (-3.5,-3.5);
\draw[double] (-4.75,-3) -- (-4.75,-2.5);
\draw[double] (-4.75,-4) -- (-4.75,-3.5);
\draw[thin] (-5,-4) -- (-4.5,-4);
\draw[thin] (-5,-4) -- (-5,-4.5);
\draw[thin] (-4.5,-4.5) -- (-4.5,-4);
\draw[thin] (-4.5,-4.5) -- (-5,-4.5);
\draw[double] (-4.75,-5) -- (-4.75,-4.5);
\draw[double] (-5,-3.25) -- (-5.5,-3.25);
\draw[double] (-4,-3.25) -- (-4.5,-3.25);
\draw[double] (-3,-3.25) -- (-3.5,-3.25);
\draw[double] (-5,-4.25) -- (-5.5,-4.25);
\draw[double] (-4,-4.25) -- (-4.5,-4.25);
\draw[double] (-3,-4.25) -- (-3.5,-4.25);
\draw[double] (-3.75,-3) -- (-3.75,-2.5);
\draw[double] (-3.75,-4) -- (-3.75,-3.5);
\draw[double] (-3.75,-5) -- (-3.75,-4.5);
\draw[thin] (-3.5,-4) -- (-4,-4);
\draw[thin] (-4,-4.5) -- (-4,-4);
\draw[thin] (-3.5,-4) -- (-3.5,-4.5);
\draw[thin] (-4,-4.5) -- (-3.5,-4.5);
\draw[double] (-3.25,-4.75) -- (-3.5,-4.5);
\draw[double] (-4.25,-4.75) -- (-4,-4.5);
\draw[double] (-5,-3) -- (-5.25,-2.75);
\draw[double] (-5,-3.5) -- (-5.25,-3.75);
\draw[double] (-4.25,-2.75) -- (-4.5,-3);
\path (-5,-5) node [style=sergio] {\scriptsize$\tau_1$};
\path (-5,-3.75) node [style=sergio] {\scriptsize$\tau_1$};
\path (-5,-2.5) node [style=sergio] {\scriptsize$\tau_1$};
\path (-5.75,-3.25) node [style=sergio] {\scriptsize$\tau_1$};
\path (-4.232,-3.3861) node [style=sergio] {\scriptsize$\tau_1$};
\path (-2.75,-3.25) node [style=sergio] {\scriptsize$\tau_1$};
\path (-5.75,-4.25) node [style=sergio] {\scriptsize$\tau_2$};
\path (-4.2411,-4.3785) node [style=sergio] {\scriptsize$\tau_2$};
\path (-3.5,-3.75) node [style=sergio] {\scriptsize$\tau_2$};
\path (-3.5,-2.5) node [style=sergio] {\scriptsize$\tau_2$};
\path (-4.75,-4.25) node [style=sergio] {\scriptsize$\mu_2$};
\path (-4.75,-3.25) node [style=sergio] {\scriptsize$\mu_1$};
\path (-3.75,-3.25) node [style=sergio] {\scriptsize$\mu_2$};
\path (-3.75,-4.25) node [style=sergio] {\scriptsize$\mu_3$};
\path (-2.75,-4.25) node [style=sergio] {\scriptsize$\tau_2$};
\path (-3,-3.75) node [style=sergio] {\scriptsize$\tau_3$};
\path (-3,-5) node [style=sergio] {\scriptsize$\tau_3$};
\path (-4.25,-5) node [style=sergio] {\scriptsize$\tau_3$};
\path (-4.4065,-3.8265) node [style=sergio] {\scriptsize$\tau_3$};
\path (-5.4454,-2.7631) node [style=sergio] {\scriptsize$\tau_3$};
\path (-5.4363,-3.7502) node [style=sergio] {\scriptsize$\tau_3$};
\path (-4.2592,-2.6106) node [style=sergio] {\scriptsize$\tau_3$};
\path (-4.25,-6) node [style=sergio] {$11.~p4m$};
\end{tikzpicture}
\caption{The corresponding PEPS tensors for \#1 to \#12 wallpaper groups. Here $\mu$'s label different physical indices, and $\tau$'s label different virtual indices.}
\label{summarize1}
\end{figure*}
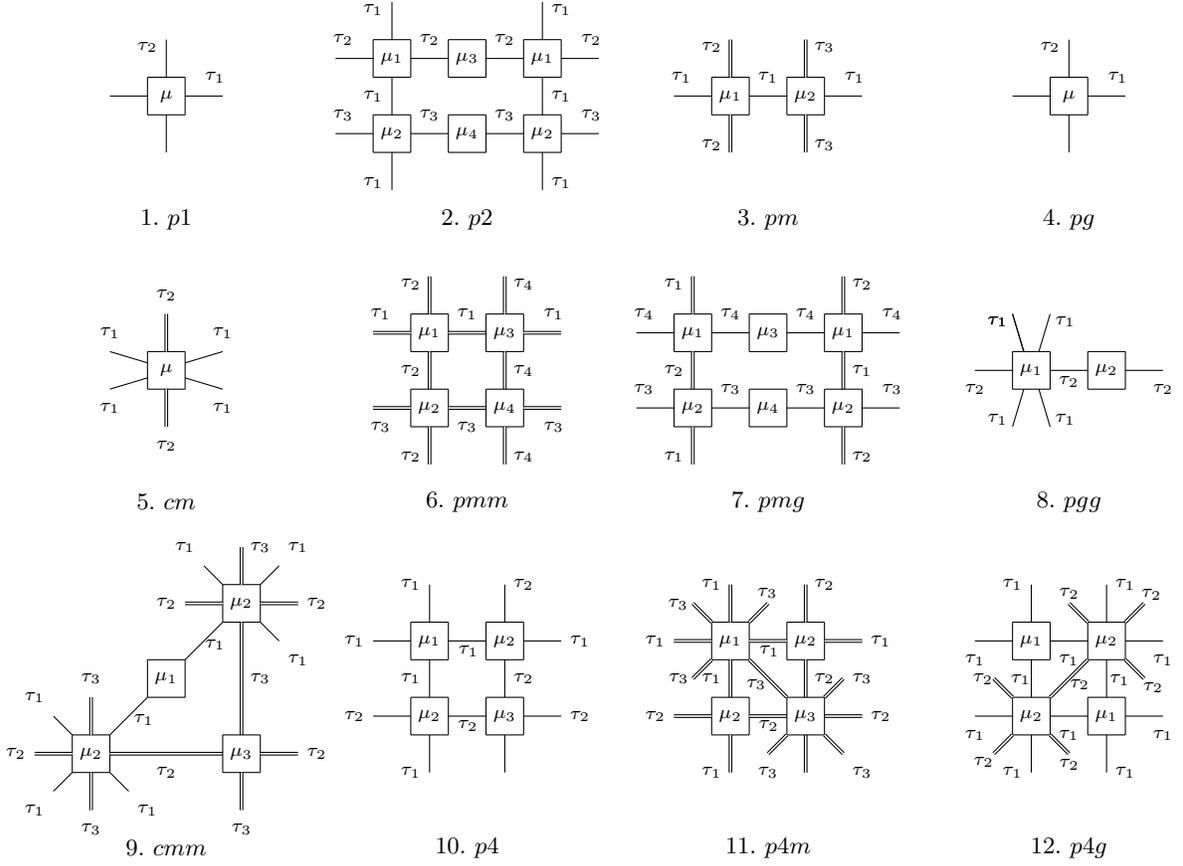

Tensor network representation of 2D crystalline fSPT phases provides a powerful tool for investigating various problems of crystalline topological phases. For instance, consider a 2D lattice with a specific wallpaper group $SG$, and jointly protected by an on-site symmetry group $G_0$, i.e., the total symmetry group of this lattice is $SG\times G_0$. The tensor network representation can also be applied to calculate the classification: the total on-site symmetry of physical/virtual indices are $G_0\times G_p/G_0\times G_v$, and the classification data is from the following species:
\begin{enumerate}[1.]
\item $F$-moves and super pentagon equations of fMPOs (the corresponding symmetry group is $G_0$);
\item Virtual indices, characterized by Eqs. (\ref{2-supercohomology}) and (\ref{2-twisted});
\item Physical indices, characterized by Eqs. (\ref{1-supercohomology}) and (\ref{1-twisted}).
\end{enumerate}
Then by investigating the obstruction, trivialization, and group structure as demonstrated in this paper, one can similarly obtain the classification of 2D fSPT phases jointly protected by crystalline and internal symmetries. 

Furthermore, Tensor network is one of the most powerful tools to investigate the topological phase transition, hence based on this work, by constructing the explicit PEPS ground state wavefunctions, the topological phase transition between different crystalline fSPT phases can be investigated under their tensor network representations.

\begin{acknowledgements}
We thank Zheng-Cheng Gu and Rui-Xing Zhang for enlightening discussions. JHZ is supported by Direct Grant No. 4053409 from The Chinese University of Hong Kong and funding from Hong Kong's Research Grants Council (GRF No.14306918, ANR/RGC Joint Research Scheme No. A-CUHK402/18). SY is supported by NSFC (Grant No. 11804181, No. 12174214) and the National Key R\&D Program of China (Grant No. 2018YFA0306504).
\end{acknowledgements}

\appendix

\section{Group extension, the spin of fermions and short exact sequence\label{SES}}
In the main text, we have discussed the tensor network states of 2D crystalline fSPT phases for both spinless and spin-1/2 fermions, and the group structure of the classification characterized by possible nontrivial extensions between virtual and physical indices. In this section, we discuss the mathematical foundation of these issues: the short exact sequence.

\subsection{Central extension of groups}
The following issues are characterized by different group extensions:
\begin{enumerate}[1.]
\item The spin of fermions (spinless or spin-1/2) are characterized by different central extensions of fermion parity $\mathbb{Z}_2^f$ by the physical symmetry group $G_b$:
\begin{align}
0\rightarrow\mathbb{Z}_2^f\rightarrow G_f\rightarrow G_b\rightarrow0
\label{spin of fermions}
\end{align}
different central extensions are characterized by different factor systems $\omega_2\in\mathcal{H}^2(G_b,\mathbb{Z}_2)$.
\item The group structure of the classification is central extension of degrees of freedom of virtual indices by physical indices.
\end{enumerate}
\paragraph{Lemma (Factor system)}For a group $(G,\cdot)$, an Abelian group $(A,+)$, and a short exact sequence:
\begin{align}
0\rightarrow A\rightarrow X\rightarrow G\rightarrow0
\label{AXG}
\end{align}
There is a factor system of the short exact sequence Eq. (\ref{AXG}) who consists the function $f$ and a homomorphism $\sigma$:
\begin{align}
\left.
\begin{aligned}
f:G\times G~&\lra~~~A\\
(g,h)~&\longmapsto f(g,h)
\end{aligned}
\right.~,~~\left.
\begin{aligned}
\sigma:G~~&\lra \text{End}(A)\\
g~~&\longmapsto~~~\sigma_g
\end{aligned}
\right.
\end{align}
where $\mathrm{End}(A)$ is the endomorphism of the Abelian group $A$. such that it makes the Cartesian product $G\times A$ a group $X$ with multiplication: \begin{align}
(g,a)*(h,b)=(g\cdot h,f(g,h)+a+\sigma_g(b))
\end{align}
And $f$ must be a group 2-cocycle which is classified by 2 group cohomology: $f\in\mathcal{H}^2(G,A)$. Where End$(A)$ is the endomorphism of group $A$.

\subsection{Physical understanding of spin of fermions}
From quantum mechanics, we know that for a fermion with well-defined spin, it should be an eigenstate of the spin angular momentum operator $\hat{S}_z$ with a well defined magnetic quantum number $\sigma_z$: $\sigma_z=0$ for spinless fermion and $\sigma_z=1/2$ for spin-1/2 fermion, with eigen-wavefunction proportional to the phase factor $e^{i\sigma_z\phi}$, $\phi\in[0,2\pi)$. For spinless fermions, if we rotate it by a lap (i.e., change $\phi$ from $0$ to $2\pi$), there is nothing changed because $\sigma_z=0$; for spin-1/2 fermions, if we rotate it by a lap, the phase factor $e^{i\sigma_z\phi}$ is changed by $-1$. They can be properly characterized by different factor systems $\omega_2$ of the short exact sequence (\ref{spin of fermions}). 

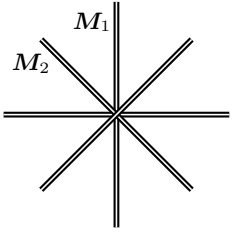
\begin{figure}
\begin{tikzpicture}
\tikzstyle{sergio}=[rectangle,draw=none]
\draw[double,thick] (-1,-0.5) -- (2,-0.5);
\draw[double,thick] (0.5,-2) -- (0.5,1);
\draw[double,thick] (1.5,-1.5) -- (-0.5,0.5);
\draw[double,thick] (-0.5,-1.5) -- (1.5,0.5);
\path (0.1792,0.7343) node [style=sergio] {$\boldsymbol{M}_1$};
\path (-0.6347,0.179) node [style=sergio] {$\boldsymbol{M}_2$};
\end{tikzpicture}
\caption{Two independent reflection axes as two generators of $D_4$ symmetry.}
\label{D4}
\end{figure}

For example, consider even-fold dihedral group $D_{2n}$ symmetry with two generators $\bs{R}$ (rotation) and $\bs{M}$ (reflection) satisfying $\bs{R}^{2n}=\bs{M}^2=I$ ($n=1,2,3$ and $I$ is identity). Different extensions of fermion parity are characterized by different 2-cocycles $\omega_2$:
\begin{align}
\omega_2\in\mathcal{H}^2(D_{2n},\mathbb{Z}_2)=\mathbb{Z}_2^3
\end{align}
In particular, the spinless fermions corresponding to the trivial 2-cocycle $\omega_2$ satisfying:
\begin{align}
\bs{R}^{2n}=1,~\bs{M}^2=1
\end{align}
while the spin-1/2 fermions corresponding to the 2-cocycle $\omega_2$ satisfying ($P_f$ is the fermion parity operator):
\begin{align}
\bs{R}^{2n}=\bs{M}^2=P_f,~\bs{M}\bs{R}\bs{M}^{-1}\bs{R}=1
\end{align}

\section{Physical indices in 2D systems\label{point group center}}
For a 2D lattice with arbitrary wallpaper group symmetry, a PEPS tensor should be aligned at the center of a specific 2D point group. In Sec. \ref{general} we have demonstrated that the physical index of a specific PEPS tensor is characterized by 1D irreducible representation of the \textit{total} symmetry group $G_f$, and classified by group 1-cohomology $\mathcal{H}^1[G_f,U(1)]$. We identify all possible physical indices of PEPS tensors aligned at the center of 2D point groups. 

We demonstrate the 4-order dihedral group $D_4$ as an example, and all other possibilities are summarized in Tables \ref{physical indices spinless} and \ref{physical indices spin-1/2}. 

For spinless fermions, different 1D irreducible representations of the total symmetry group $\mathbb{Z}_2^f\times D_4$ are classified by group 1-cohomology \cite{GAP,HAP}:
\begin{align}
\mathcal{H}^1\left[\mathbb{Z}_2^f\times D_4,U(1)\right]=\mathbb{Z}_2^3
\end{align}
Hence for spinless fermions, if we put a PEPS tensor at the center of $D_4$ symmetry, the dimension of corresponding physical index is $d=8$, with three generators: $c$ labels the complex fermions, $m_1$ and $m_2$ label the eigenvalues $-1$ of two independent reflection generators $\bs{M}_1$ and $\bs{M}_2$, of $D_4$ group (see Fig. \ref{D4}).

\begin{table}[t]
\renewcommand\arraystretch{1.2}
\begin{tabular}{|c|c|c|c|c|c|c|}
\hline
point group&~~dimension~~~&~~generator~~~\\
\hline
$C_1$&$d=2$&$\{c\}$\\
\hline
$C_2$&$d=4$&$\{c,r|r^2=1\}$\\
\hline
$C_3$&$d=6$&$\{c,r|r^3=1\}$\\
\hline
$C_4$&$d=8$&$\{c,r|r^4=1\}$\\
\hline
$C_6$&$d=12$&$\{c,r|r^6=1\}$\\
\hline
$D_1$&$d=4$&$\{c,m|m^2=1\}$\\
\hline
$D_2$&$d=8$&$\{c,m_1,m_2|m_1^2=m_2^2=1\}$\\
\hline
$D_3$&$d=4$&$\{c,m|m^2=1\}$\\
\hline
$D_4$&$d=8$&$\{c,m_1,m_2|m_1^2=m_2^2=1\}$\\
\hline
$D_6$&$d=8$&$\{c,m_1,m_2|m_1^2=m_2^2=1\}$\\
\hline
\end{tabular}
\caption{Physical indices of PEPS tensors located at centers of 2D point groups for spinless fermions, including their dimensions and generators. Here $c$ labels the complex fermion, $r$ labels the rotation eigenvalues and $m/m_1/m_2$ labels different reflection eigenvalues.}
\label{physical indices spinless}
\end{table}

For spin-1/2 fermions, different 1D irreducible representations of the total symmetry group $\mathbb{Z}_2^f\times_{\omega_2}D_4$ are classified by an alternative group 1-cohomology:
\begin{align}
\mathcal{H}^1\left[\mathbb{Z}_2^f\times_{\omega_2}D_4,U(1)\right]=\mathbb{Z}_2^2
\end{align}
Hence for spin-1/2 fermions, if we put a PEPS tensor at the center of $D_4$ symmetry, the dimension of corresponding physical index is $d=4$, with two generators: $m_1$ and $m_2$ label the eigenvalues $-1$ of two independent reflection generators $\bs{M}_1$ and $\bs{M}_2$ of $D_4$ group (see Fig. \ref{D4}).

\begin{table}[t]
\renewcommand\arraystretch{1.2}
\begin{tabular}{|c|c|c|c|c|c|c|}
\hline
point group&~~dimension~~~&~~generator~~~\\
\hline
$C_1$&$d=2$&$\{c\}$\\
\hline
$C_2$&$d=4$&$\{f^n|n=0,1,2,3\}$\\
\hline
$C_3$&$d=6$&$\{c,r|r^3=1\}$\\
\hline
$C_4$&$d=8$&$\{f^n|n=0,1,\cdot\cdot\cdot,7\}$\\
\hline
$C_6$&$d=12$&$\{f^n|n=0,1,\cdot\cdot\cdot,11\}$\\
\hline
$D_1$&$d=4$&$\{f^n|n=0,1,2,3\}$\\
\hline
$D_2$&$d=4$&$\{m_1,m_2|m_1^2=m_2^2=1\}$\\
\hline
$D_3$&$d=4$&$\{f^n|n=0,1,2,3\}$\\
\hline
$D_4$&$d=4$&$\{m_1,m_2|m_1^2=m_2^2=1\}$\\
\hline
$D_6$&$d=4$&$\{m_1,m_2|m_1^2=m_2^2=1\}$\\
\hline
\end{tabular}
\caption{Physical indices of PEPS tensors located at centers of 2D point groups for spin-1/2 fermions, including their dimensions and generators. Here $c$ labels the complex fermion, $r$ labels the rotation eigenvalues, $m_1/m_2$ labels different reflection eigenvalues and $f^n$ labels fermionic mode carries nontrivial eigenvalues of point group symmetry action.}
\label{physical indices spin-1/2}
\end{table}

\section{Superposition of MPOs with graded structure\label{AppMPO}}
In this section, we review the superposition of MPOs for both bosonic (bMPO) and fermionic (fMPO) systems, and deduce the 3-cohomology structure from the $F$-move and pentagon equation. In particular, because of the super (graded) structure of the fermionic PEPS and fMPOs, the mathematical structure of the fMPOs is 3-order group super-cohomology \cite{special,general1,general2}.

\subsection{Superposition of bMPOs}
Consider the following two bMPOs with symmetry $G$ ($g_1,g_2\in G$):
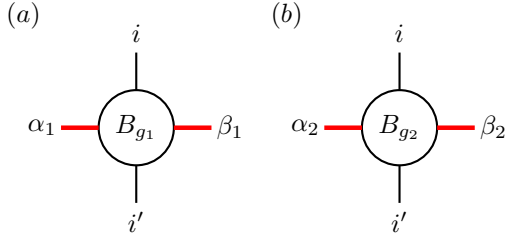
\begin{align}
\begin{tikzpicture}
\tikzstyle{sergio}=[rectangle,draw=none]
\filldraw[fill=none, draw=black, thick]  (-1.5,0) ellipse (0.5 and 0.5);
        \draw[color=red, line width=0.6mm] (-2,0) -- (-2.5,0);
        \draw[color=red, line width=0.6mm] (-1,0) -- (-0.5,0);
        \path (-2.75,0) node [style=sergio] {$\alpha_1$};
        \path (-0.25,0) node [style=sergio] {$\beta_1$};
        \draw[thick] (-1.5,0.5) -- (-1.5,1);
        \path (-1.5,1.25) node [style=sergio] {$i$};
		\draw[thick] (-1.5,-0.5) -- (-1.5,-1);
        \path (-1.5,-1.25) node [style=sergio] {$i'$};
                \path (-1.5,0) node [style=sergio] {$B_{g_1}$};
\path (-3,1.5) node [style=sergio] {$(a)$};
\path (0.5,1.5) node [style=sergio] {$(b)$};
\filldraw[fill=none, draw=black, thick]  (2,0) ellipse (0.5 and 0.5);
        \draw[color=red, line width=0.6mm] (1.5,0) -- (1,0);
        \draw[color=red, line width=0.6mm] (2.5,0) -- (3,0);
        \path (0.75,0) node [style=sergio] {$\alpha_2$};
        \path (3.25,0) node [style=sergio] {$\beta_2$};
        \draw[thick] (2,0.5) -- (2,1);
        \path (2,1.25) node [style=sergio] {$i$};
		\draw[thick] (2,-0.5) -- (2,-1);
        \path (2,-1.25) node [style=sergio] {$i'$};
                \path (2,0) node [style=sergio] {$B_{g_2}$};
\end{tikzpicture}
\nonumber
\end{align}

Multiplying the bMPOs $B_{g_1}$ and $B_{g_2}$ gives a new tensor $B_{g_1g_2}$:
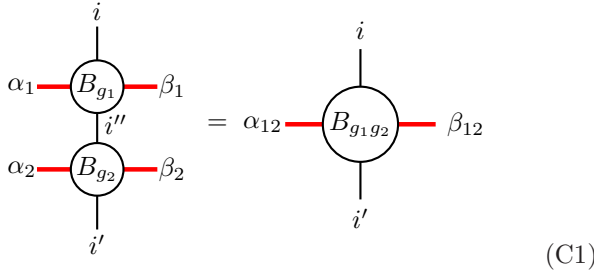
\begin{align}
\begin{tikzpicture}
\tikzstyle{sergio}=[rectangle,draw=none]
        \draw[color=red, line width=0.6mm] (-1.8,0.5) -- (-2.3,0.5);
        \draw[color=red, line width=0.6mm] (-1.2,0.5) -- (-0.7,0.5);
        \path (-2.5,0.5) node [style=sergio] {$\alpha_1$};
        \path (-0.5,0.5) node [style=sergio] {$\beta_1$};
        \draw[thick] (-1.5,0.8) -- (-1.5,1.3);
        \path (-1.5,1.5) node [style=sergio] {$i$};
        \path (-1.25,0) node [style=sergio] {$i''$};
        \draw[color=red, line width=0.6mm] (1.5,0) -- (1,0);
        \draw[color=red, line width=0.6mm] (2.5,0) -- (3,0);
        \path (0.7,0) node [style=sergio] {$\alpha_{12}$};
        \path (3.4,0) node [style=sergio] {$\beta_{12}$};
        \draw[thick] (2,0.5) -- (2,1);
		\draw[thick] (2,-1) -- (2,-0.5);
        \path (2,-1.25) node [style=sergio] {$i'$};
\filldraw[fill=white, draw=black, thick]  (2,0) ellipse (0.5 and 0.5);
                \path (2,0) node [style=sergio] {$B_{g_1g_2}$};
\path (0.1,0) node [style=sergio] {$=$};
        \draw[color=red, line width=0.6mm] (-1.8,-0.6) -- (-2.3,-0.6);
        \draw[color=red, line width=0.6mm] (-1.2,-0.6) -- (-0.7,-0.6);
        \path (-2.5,-0.6) node [style=sergio] {$\alpha_2$};
        \path (-0.5,-0.6) node [style=sergio] {$\beta_2$};
        \draw[thick] (-1.5,-0.3) -- (-1.5,0.2);
		\draw[thick] (-1.5,-0.9) -- (-1.5,-1.4);
        \path (-1.5,-1.6) node [style=sergio] {$i'$};
\filldraw[fill=white, draw=black, thick]  (-1.5,0.5) ellipse (0.35 and 0.35);
                \path (-1.5,0.5) node [style=sergio] {$B_{g_1}$};
\filldraw[fill=white, draw=black, thick]  (-1.5,-0.6) ellipse (0.35 and 0.35);
                \path (-1.5,-0.6) node [style=sergio] {$B_{g_2}$};
\path (2,1.25) node [style=sergio] {$i$};
\end{tikzpicture}
\label{superpose2}
\end{align}

Define a projection operator that can be illustrated graphically as:
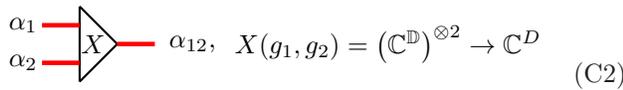
\begin{align}
\begin{tikzpicture}
\tikzstyle{sergio}=[rectangle,draw=none]
	 \draw[thick] (-2,0.5) -- (-2,-0.5);
\draw[thick] (-2,0.5) -- (-1.5,0);
\draw[thick] (-1.5,0) -- (-2,-0.5);
\path (-1.82,0) node [style=sergio] {$X$};
\draw[color=red, line width=0.6mm] (-2.5,0.25) -- (-2,0.25);
\draw[color=red, line width=0.6mm] (-2.5,-0.25) -- (-2,-0.25);
\draw[color=red, line width=0.6mm] (-1.5,0) -- (-1,0);
\path (-2.75,0.25) node [style=sergio] {$\alpha_{1}$};
\path (-2.75,-0.25) node [style=sergio] {$\alpha_{2}$};
\path (-0.5,0) node [style=sergio] {$\alpha_{12}$, };
\path (2.1,0) node [style=sergio] {$X(g_1,g_2)=\left(\mathbb{C^D}\right)^{\otimes2}\rightarrow\mathbb{C}^D$};
\end{tikzpicture}
\label{projection}
\end{align}
Then the superposition rule [cf. Eq. (\ref{superpose2})] can be rephrased in terms of the projection operators:
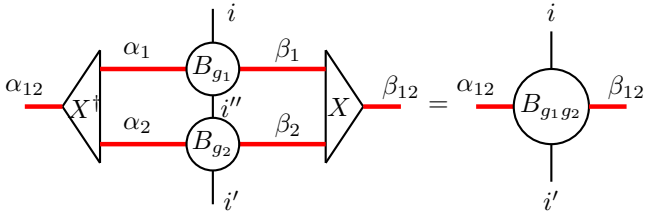
\[
\begin{tikzpicture}
\tikzstyle{sergio}=[rectangle,draw=none]
	 \draw[thick] (2,1) -- (2,-0.5);
\draw[thick] (2,1) -- (2.5,0.25);
\draw[thick] (2.5,0.25) -- (2,-0.5);
\path (2.18,0.25) node [style=sergio] {$X$};
\draw[color=red, line width=0.6mm] (0.8,0.75) -- (2,0.75);
\draw[color=red, line width=0.6mm] (0.8,-0.25) -- (2,-0.25);
\draw[color=red, line width=0.6mm] (2.5,0.25) -- (3,0.25);
\path (-0.5,1) node [style=sergio] {$\alpha_{1}$};
\path (-0.5,0) node [style=sergio] {$\alpha_{2}$};
	 \draw[thick] (-1,1) -- (-1,-0.5);
\draw[thick] (-1,1) -- (-1.5,0.25);
\draw[thick] (-1.5,0.25) -- (-1,-0.5);
\path (-1.2,0.25) node [style=sergio] {$X^\dag$};
\path (-2,0.5) node [style=sergio] {$\alpha_{12}$};
\draw[color=red, line width=0.6mm] (0.2,0.75) -- (-1,0.75);
\draw[color=red, line width=0.6mm] (-1,-0.25) -- (0.2,-0.25);
\draw[color=red, line width=0.6mm] (-1.5,0.25) -- (-2,0.25);
\path (3,0.5) node [style=sergio] {$\beta_{12}$};
\path (1.5,1) node [style=sergio] {$\beta_{1}$};
\path (1.5,0) node [style=sergio] {$\beta_{2}$};
\path (3.5,0.25) node [style=sergio] {$=$};
\draw[thick] (0.5,-0.55) -- (0.5,-1.05);
\draw[thick] (0.5,0.45) -- (0.5,0.05);
\draw[thick] (0.5,1.55) -- (0.5,1.05);
\path (0.75,1.5) node [style=sergio] {$i$};
\path (0.75,0.25) node [style=sergio] {$i''$};
\path (0.75,-1) node [style=sergio] {$i'$};
\draw[color=red, line width=0.6mm] (4.5,0.25) -- (4,0.25);
\filldraw[fill=white, draw=black, thick]  (0.5,0.75) ellipse (0.35 and 0.35);
\path (0.5,0.75) node [style=sergio] {$B_{g_1}$};
\filldraw[fill=white, draw=black, thick]  (0.5,-0.25) ellipse (0.35 and 0.35);
\path (0.5,-0.25) node [style=sergio] {$B_{g_2}$};
\filldraw[fill=white, draw=black, thick]  (5,0.25) ellipse (0.5 and 0.5);
\draw[color=red, line width=0.6mm] (5.5,0.25) -- (6,0.25);
\path (5,0.25) node [style=sergio] {$B_{g_1g_2}$};
\draw[thick] (5,0.75) -- (5,1.25);
\draw[thick] (5,-0.25) -- (5,-0.75);
\path (5,1.5) node [style=sergio] {$i$};
\path (5,-1) node [style=sergio] {$i'$};
\path (4,0.5) node [style=sergio] {$\alpha_{12}$};
\path (6,0.5) node [style=sergio] {$\beta_{12}$};
\end{tikzpicture}
\]

\subsubsection{Superpositions of three bMPOs: $F$-move}
After defining the projection operator $X$, we are ready to consider the superpositions of three bMPOs: $F$-move. For three virtual indices $\alpha_{j}$ ($j=1,2,3$), there are two different but equivalent ways of superposition: superpose $(\alpha_1,\alpha_2)/(\alpha_2,\alpha_3)$ first. Since the same bMPO will be obtained in two different ways, they can only differ by a $U(1)$ phase. Graphically:
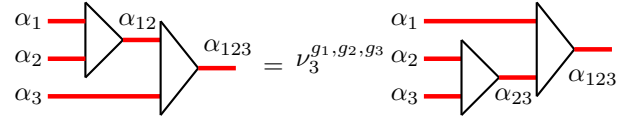
\[
\begin{tikzpicture}
\tikzstyle{sergio}=[rectangle,draw=none]
\draw[thick] (-2,0.5) -- (-2,-0.5);
\draw[thick] (-2,0.5) -- (-1.5,0);
\draw[thick] (-1.5,0) -- (-2,-0.5);
\draw[color=red, line width=0.6mm] (-2,0.25) -- (-2.5,0.25);
\draw[color=red, line width=0.6mm] (-2,-0.25) -- (-2.5,-0.25);
\draw[color=red, line width=0.6mm] (-1,0) -- (-1.5,0);
\draw[color=red, line width=0.6mm] (-1,-0.75) -- (-2.5,-0.75);
\draw[thick] (-1,-1) -- (-1,0.25);
\draw[thick] (-1,-1) -- (-0.5,-0.375);
\draw[thick] (-1,0.25) -- (-0.5,-0.375);
\draw[color=red, line width=0.6mm] (0,-0.375) -- (-0.5,-0.375);
\path (-2.75,0.25) node [style=sergio] {$\alpha_{1}$};
\path (-2.75,-0.25) node [style=sergio] {$\alpha_{2}$};
\path (-2.75,-0.75) node [style=sergio] {$\alpha_{3}$};
\path (-1.3,0.25) node [style=sergio] {$\alpha_{12}$};
\path (-0.1,-0.1) node [style=sergio] {$\alpha_{123}$};
\path (0.5,-0.375) node [style=sergio] {$=$};
\draw[color=red, line width=0.6mm] (2.5,0.25) -- (4,0.25);
\draw[color=red, line width=0.6mm] (2.5,-0.25) -- (3,-0.25);
\draw[color=red, line width=0.6mm] (2.5,-0.75) -- (3,-0.75);
\path (2.25,0.25) node [style=sergio] {$\alpha_{1}$};
\path (2.25,-0.25) node [style=sergio] {$\alpha_{2}$};
\path (2.25,-0.75) node [style=sergio] {$\alpha_{3}$};
\draw[thick] (3,0) -- (3,-1);
\draw[thick] (3,0) -- (3.5,-0.5);
\draw[thick] (3,-1) -- (3.5,-0.5);
\draw[color=red, line width=0.6mm] (3.5,-0.5) -- (4,-0.5);
\path (3.7,-0.75) node [style=sergio] {$\alpha_{23}$};
\draw[thick] (4,0.5) -- (4,-0.75);
\draw[thick] (4,0.5) -- (4.5,-0.125);
\draw[thick] (4,-0.75) -- (4.5,-0.125);
\draw[color=red, line width=0.6mm] (5,-0.125) -- (4.5,-0.125);
\path (4.75,-0.5) node [style=sergio] {$\alpha_{123}$};
\path (1.4,-0.25) node [style=sergio] {$\nu_3^{g_1,g_2,g_3}$};
\end{tikzpicture}
\]
where $\nu_3^{g_1,g_2,g_3}\in U(1)$ because in quantum mechanics, wavefunctions differ by a $U(1)$ phase characterize the same physical state. To further investigate the mathematical properties of this phase, we should consider the superpositions of more bMPOs.

\subsubsection{Superpositions of four bMPOs: Pentagon equation}
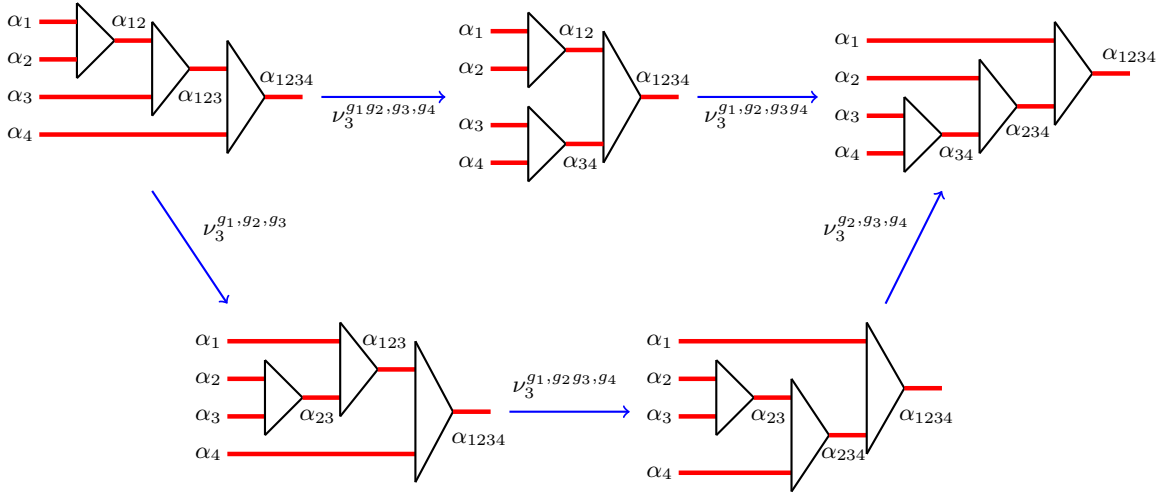
\begin{figure*}
\begin{tikzpicture}
\tikzstyle{sergio}=[rectangle,draw=none]
\draw[thick] (-2,0.5) -- (-2,-0.5);
\draw[thick] (-2,0.5) -- (-1.5,0);
\draw[thick] (-1.5,0) -- (-2,-0.5);
\draw[color=red, line width=0.6mm] (-2,0.25) -- (-2.5,0.25);
\draw[color=red, line width=0.6mm] (-2,-0.25) -- (-2.5,-0.25);
\draw[color=red, line width=0.6mm] (-1,0) -- (-1.5,0);
\draw[color=red, line width=0.6mm] (-1,-0.75) -- (-2.5,-0.75);
\draw[color=red, line width=0.6mm] (0,-1.25) -- (-2.5,-1.25);
\draw[thick] (-1,-1) -- (-1,0.25);
\draw[thick] (-1,-1) -- (-0.5,-0.375);
\draw[thick] (-1,0.25) -- (-0.5,-0.375);
\draw[color=red, line width=0.6mm] (0,-0.375) -- (-0.5,-0.375);
\path (-2.75,0.25) node [style=sergio] {$\alpha_{1}$};
\path (-2.75,-0.25) node [style=sergio] {$\alpha_{2}$};
\path (-2.75,-0.75) node [style=sergio] {$\alpha_{3}$};
\path (-2.75,-1.25) node [style=sergio] {$\alpha_{4}$};
\path (-1.3,0.25) node [style=sergio] {$\alpha_{12}$};
\path (-0.35,-0.75) node [style=sergio] {$\alpha_{123}$};
\draw[color=red, line width=0.6mm] (3.5,0.125) -- (4,0.125);
\draw[color=red, line width=0.6mm] (3.5,-0.375) -- (4,-0.375);
\draw[color=red, line width=0.6mm] (3.5,-1.125) -- (4,-1.125);
\draw[color=red, line width=0.6mm] (3.5,-1.625) -- (4,-1.625);
\path (3.25,0.125) node [style=sergio] {$\alpha_{1}$};
\path (3.25,-0.375) node [style=sergio] {$\alpha_{2}$};
\path (3.25,-1.125) node [style=sergio] {$\alpha_{3}$};
\path (3.25,-1.625) node [style=sergio] {$\alpha_{4}$};
\draw[thick] (4,0.375) -- (4,-0.625);
\draw[thick] (4,0.375) -- (4.5,-0.125);
\draw[thick] (4,-0.625) -- (4.5,-0.125);
\draw[color=red, line width=0.6mm] (4.5,-0.125) -- (5,-0.125);
\path (4.65,0.125) node [style=sergio] {$\alpha_{12}$};
\draw[thick] (5,0.125) -- (5,-1.625);
\draw[thick] (5,0.125) -- (5.5,-0.75);
\draw[thick] (5,-1.625) -- (5.5,-0.75);
\draw[color=red, line width=0.6mm] (5.5,-0.75) -- (6,-0.75);
\path (4.7,-1.625) node [style=sergio] {$\alpha_{34}$};
\path (2.1,-1) node [style=sergio] {$\nu_3^{g_1g_2,g_3,g_4}$};
\draw[thick] (0,-1.5) -- (0,0);
\draw[thick] (0.5,-0.75) -- (0,0);
\draw[thick] (0.5,-0.75) -- (0,-1.5);
\draw[color=red, line width=0.6mm] (0.5,-0.75) -- (1,-0.75);
\path (0.8,-0.5) node [style=sergio] {$\alpha_{1234}$};
\draw[->,thick,color=blue] (1.25,-0.75) -- (2.9,-0.75);
\draw[thick] (4,-0.875) -- (4,-1.875);
\draw[thick] (4,-0.875) -- (4.5,-1.375);
\draw[thick] (4,-1.875) -- (4.5,-1.375);
\draw[color=red, line width=0.6mm] (5,-1.375) -- (4.5,-1.375);
\path (5.8,-0.5) node [style=sergio] {$\alpha_{1234}$};
\draw[->,thick,color=blue] (6.25,-0.75) -- (7.85,-0.75);
\path (7.05,-1) node [style=sergio] {$\nu_3^{g_1,g_2,g_3g_4}$};
\draw[color=red, line width=0.6mm] (8.5,-1) -- (9,-1);
\draw[color=red, line width=0.6mm] (8.5,-1.5) -- (9,-1.5);
\draw[thick] (9,-0.75) -- (9,-1.75);
\draw[thick] (9,-0.75) -- (9.5,-1.25);
\draw[thick] (9,-1.75) -- (9.5,-1.25);
\draw[color=red, line width=0.6mm] (10,-1.25) -- (9.5,-1.25);
\path (8.25,-1.5) node [style=sergio] {$\alpha_{4}$};
\path (8.25,-1) node [style=sergio] {$\alpha_{3}$};
\path (8.25,-0.5) node [style=sergio] {$\alpha_{2}$};
\draw[color=red, line width=0.6mm] (8.5,-0.5) -- (10,-0.5);
\draw[thick] (10,-0.25) -- (10,-1.5);
\draw[thick] (10,-0.25) -- (10.5,-0.875);
\draw[thick] (10,-1.5) -- (10.5,-0.875);
\path (9.7,-1.5) node [style=sergio] {$\alpha_{34}$};
\draw[color=red, line width=0.6mm] (11,-0.875) -- (10.5,-0.875);
\path (8.25,0) node [style=sergio] {$\alpha_{1}$};
\draw[color=red, line width=0.6mm] (8.5,0) -- (11,0);
\draw[thick] (11,-1.125) -- (11,0.25);
\draw[thick] (11,-1.125) -- (11.5,-0.4375);
\draw[thick] (11,0.25) -- (11.5,-0.4375);
\draw[color=red, line width=0.6mm] (12,-0.4375) -- (11.5,-0.4375);
\path (10.65,-1.2) node [style=sergio] {$\alpha_{234}$};
\path (12,-0.2) node [style=sergio] {$\alpha_{1234}$};
\draw[->,thick,color=blue] (-1,-2) -- (0,-3.5);
\draw[color=red, line width=0.6mm] (1.5,-4) -- (0,-4);
\draw[color=red, line width=0.6mm] (0.5,-4.5) -- (0,-4.5);
\draw[color=red, line width=0.6mm] (0.5,-5) -- (0,-5);
\draw[color=red, line width=0.6mm] (2.5,-5.5) -- (0,-5.5);
\path (-0.25,-5.5) node [style=sergio] {$\alpha_{4}$};
\path (-0.25,-5) node [style=sergio] {$\alpha_{3}$};
\path (-0.25,-4.5) node [style=sergio] {$\alpha_{2}$};
\path (-0.25,-4) node [style=sergio] {$\alpha_{1}$};
\draw[thick] (0.5,-5.25) -- (0.5,-4.25);
\draw[thick] (0.5,-5.25) -- (1,-4.75);
\draw[thick] (0.5,-4.25) -- (1,-4.75);
\draw[color=red, line width=0.6mm] (1.5,-4.75) -- (1,-4.75);
\draw[thick] (1.5,-3.75) -- (1.5,-5);
\draw[thick] (1.5,-3.75) -- (2,-4.375);
\draw[thick] (1.5,-5) -- (2,-4.375);
\draw[color=red, line width=0.6mm] (2.5,-4.375) -- (2,-4.375);
\draw[thick] (2.5,-4) -- (2.5,-5.875);
\draw[thick] (3,-4.9375) -- (2.5,-5.875);
\draw[thick] (3,-4.9375) -- (2.5,-4);
\draw[color=red, line width=0.6mm] (3,-4.9375) -- (3.5,-4.9375);
\path (1.2,-5) node [style=sergio] {$\alpha_{23}$};
\path (2.1,-4) node [style=sergio] {$\alpha_{123}$};
\path (3.35,-5.3) node [style=sergio] {$\alpha_{1234}$};
\path (0.25,-2.5) node [style=sergio] {$\nu_3^{g_1,g_2,g_3}$};
\draw[->,thick,color=blue] (3.75,-4.9375) -- (5.35,-4.9375);
\path (4.5,-4.6) node [style=sergio] {$\nu_3^{g_1,g_2g_3,g_4}$};
\draw[color=red, line width=0.6mm] (6,-5.75) -- (7.5,-5.75);
\draw[color=red, line width=0.6mm] (6,-5) -- (6.5,-5);
\draw[color=red, line width=0.6mm] (6,-4.5) -- (6.5,-4.5);
\draw[color=red, line width=0.6mm] (6,-4) -- (8.5,-4);
\path (5.75,-4) node [style=sergio] {$\alpha_{1}$};
\path (5.75,-4.5) node [style=sergio] {$\alpha_{2}$};
\path (5.75,-5) node [style=sergio] {$\alpha_{3}$};
\path (5.75,-5.75) node [style=sergio] {$\alpha_{4}$};
\draw[thick] (6.5,-4.25) -- (6.5,-5.25);
\draw[thick] (6.5,-4.25) -- (7,-4.75);
\draw[thick] (6.5,-5.25) -- (7,-4.75);
\draw[color=red, line width=0.6mm] (7.5,-4.75) -- (7,-4.75);
\path (7.2,-5) node [style=sergio] {$\alpha_{23}$};
\draw[thick] (7.5,-4.5) -- (7.5,-6);
\draw[thick] (8,-5.25) -- (7.5,-6);
\draw[thick] (8,-5.25) -- (7.5,-4.5);
\draw[color=red, line width=0.6mm] (8,-5.25) -- (8.5,-5.25);
\path (8.2,-5.5) node [style=sergio] {$\alpha_{234}$};
\draw[thick] (8.5,-3.75) -- (8.5,-5.5);
\draw[thick] (9,-4.625) -- (8.5,-5.5);
\draw[thick] (9,-4.625) -- (8.5,-3.75);
\draw[color=red, line width=0.6mm] (9,-4.625) -- (9.5,-4.625);
\path (9.3,-5) node [style=sergio] {$\alpha_{1234}$};
\draw[->,thick,color=blue] (8.75,-3.5) -- (9.5,-2);
\path (8.5,-2.5) node [style=sergio] {$\nu_3^{g_2,g_3,g_4}$};
\end{tikzpicture}
\caption{The pentagon equation of superposing four bMPOs. Each step of $F$-move gives rise to a $U(1)$ phase.}
\label{pentagon}
\end{figure*}

To investigate the aforementioned $U(1)$ phase $\nu_3^{g_1,g_2,g_3}$, we should consider the pentagon equation of the superpositions of four bMPOs, because there are several different but equivalent ways to superpose 4 bMPOs up to a $U(1)$ phase. Equivalently, we can investigate the mathematical structure of the $U(1)$ phase $\nu_3^{g_1,g_2,g_3}$. The pentagon equation can be represented graphically in Fig. \ref{pentagon}.

From the pentagon equation we find the mathematical condition of the $U(1)$ phase $\nu_3$ ($g_j\in G$, $j=1,2,3,4$):
\begin{align}
\mathrm{d}\nu_3(g_1,g_2,g_3,g_4)=\frac{\nu_3^{g_1,g_2,g_3}\nu_3^{g_1,g_2g_3,g_4}\nu_3^{g_2,g_3,g_4}}{\nu_3^{g_1g_2,g_3,g_4}\nu_3^{g_1,g_2,g_3g_4}}=1
\end{align}
i.e., 3-cocycle condition. Equivalently, $\nu_3$ is classified by 3-cohomology with $U(1)$ coefficient:
\begin{align}
\nu_3\in\mathcal{H}^3\left[G,U(1)\right]
\end{align}
Thus the superposition of bMPOs gives the 3-cohomology mathematical structure, hence the 2D bosonic symmetry-protected topological (bSPT) phases are classified by 3-cohomology with $U(1)$ coefficient \cite{XieChenScience,CZX,cohomology}. 

\subsection{Superposition of fMPOs}
In the main text we have demonstrated that for arbitrary physical/virtual indices of a fermionic PEPS, there is an additional graded structure that describing the fermion parity. In addition, we should further identify the spin of fermions (spinless and spin-1/2) that is characterized by the extension of the physical symmetry group $G_b$ by fermion parity $\mathbb{Z}_2^f$ which is described by the following short-exact sequence:
\begin{align}
0\rightarrow\mathbb{Z}_2^f\rightarrow G_f\rightarrow G_b\rightarrow0
\end{align}
Spinless fermions is characterized by trivial 2-cocycle $\omega_2=0\in\mathcal{H}^2(G_b,\mathbb{Z}_2)$, and the total symmetry group is labeled by $G_f=G_b\times\mathbb{Z}_2^f$; spin-1/2 fermions is characterized by nontrivial 2-cocycle $\omega_2\in\mathcal{H}^2(G_b,\mathbb{Z}_2)$, and the total symmetry group is labeled by $G_f=G_b\times_{\omega_2}\mathbb{Z}_2^f$. Because all wallpaper group symmetry actions are unitary, we focus on the unitary symmetry groups exclusively. 

Inherit from the graded structure of the PEPS, the graded structure of the indices of fMPOs is characterized by a map $n_2: G_b\times G_b\rightarrow\mathbb{Z}_2$ and classified by 2-cohomology $\mathcal{H}^2(G_b,\mathbb{Z}_2)$, satisfying 2-cocycle condition (modulo 2, where $g_1,g_2,g_3\in G_b$):
\begin{align}
n_2(g_1,g_2)+n_2(g_1g_2,g_3)=n_2(g_1,g_2g_3)+n_2(g_2,g_3)
\label{spinless2}
\end{align}
Similar to the PEPS tensor, the projection operator defined in Eq. (\ref{projection}) should also be fermion parity even. Nevertheless, the fMPOs can be either even or odd of fermion parity.

We consider the PEPS tensors with even fMPOs first. The super $F$-move still gives rise to a $U(1)$ phase $\nu_3^{g_1,g_2,g_3}$. Nevertheless, because of the graded structure of the fMPOs, the super pentagon equation (see Fig. \ref{pentagon}) gives the twisted 3-cocycle conditions. For spinless fermions, the super pentagon equation gives the relation as follows:
\begin{align}
\frac{\nu_3^{g_1,g_2,g_3}\nu_3^{g_1,g_2g_3,g_4}\nu_3^{g_2,g_3,g_4}}{\nu_3^{g_1g_2,g_3,g_4}\nu_3^{g_1,g_2,g_3g_4}}=(-1)^{n_2(g_1,g_2)n_2(g_3,g_4)}
\label{spinless}
\end{align}
where $n_2(g_1,g_2)n_2(g_3,g_4)$ can be labeled by cup product:
\[
n_2(g_1,g_2)n_2(g_3,g_4)=n_2\smile n_2(g_1,g_2,g_3,g_4)
\]
For spin-1/2 fermions, the super pentagon equation gives the relation as following ($j=1,2,3,4$):
\begin{align}
\frac{\nu_3^{g_1,g_2,g_3}\nu_3^{g_1,g_2g_3,g_4}\nu_3^{g_2,g_3,g_4}}{\nu_3^{g_1g_2,g_3,g_4}\nu_3^{g_1,g_2,g_3g_4}}=(-1)^{(\omega_2+n_2)\smile n_2(\{g_j\})}
\label{spin-1/2}
\end{align}
where $\{g_j\}=g_1,g_2,g_3,g_4$. With this conditions, we notice that the 2D fermionic SPT (fSPT) phases with even fMPOs are classified by two indices (in the system without translation symmetry):
\begin{align}
\left\{
\begin{aligned}
&n_2\in\mathcal{H}^2(G_b,\mathbb{Z}_2)\\
&\nu_3 \in\mathcal{H}^3[G_b,U(1)]
\end{aligned}
\right.
\end{align}
with the twisted 3-cocycle conditions [Eq. (\ref{spinless}) for spinless fermions, Eq. (\ref{spin-1/2}) for spin-1/2 fermions].

The aforementioned fMPOs are fermion parity even, although the fermion parities of their physical or virtual indices might be odd. For fMPOs with odd fermion parity, there is an additional possibility: Majorana zero modes on $\partial R$ that was missed in the previous discussions \cite{Frank2017}. Equivalently, for this case, each virtual bond of the fermionic PEPS tensors represents the Majorana entanglement pair. The parity of the fMPOs that gives an additional index of the fSPT phases can be characterized by a map $n_1: G_b\rightarrow\mathbb{Z}_2$ and classified by 1-cohomology with $\mathbb{Z}_2$ coefficient: $\mathcal{H}^1(G,\mathbb{Z}_2)$, satisfying ($g_1,g_2\in G_b$):
\begin{align}
n_1(g_1)+n_1(g_2)-n_1(g_1g_2)=0
\end{align}
Furthermore, the parity of fMPOs gives rise to an additional twist to the graded structure of the physical and virtual indices for spin-1/2 fermions, which is reflected in the 2-cocycle condition of $n_2$ ($g_1,g_2,g_3\in G_b$):
\begin{align}
&n_2(g_1,g_2)+n_2(g_1g_2,g_3)-n_2(g_1,g_2g_3)-n_2(g_2,g_3)\nonumber\\
&=\omega_2\smile n_1(g_1,g_2,g_3)
\label{spin-1/22}
\end{align}
Finally, we conclude that the 2D fSPT phases are classified by the following three indices (in the system without translation symmetry):
\begin{align}
\left\{
\begin{aligned}
&n_1\in\mathcal{H}^1(G_b,\mathbb{Z}_2)\\
&n_2\in\mathcal{H}^2(G_b,\mathbb{Z}_2)\\
&\nu_3 \in\mathcal{H}^3[G_b,U(1)]
\end{aligned}
\right.
\end{align}
with the twisted cocycle conditions: Eqs. (\ref{spinless2}) and (\ref{spinless}) for spinless fermions, Eqs. (\ref{spin-1/2}) and (\ref{spin-1/22}) for spin-1/2 fermions.

\providecommand{\noopsort}[1]{}\providecommand{\singleletter}[1]{#1}%
%


\begin{thebibliography}{94}%
\makeatletter
\providecommand \@ifxundefined [1]{%
 \@ifx{#1\undefined}
}%
\providecommand \@ifnum [1]{%
 \ifnum #1\expandafter \@firstoftwo
 \else \expandafter \@secondoftwo
 \fi
}%
\providecommand \@ifx [1]{%
 \ifx #1\expandafter \@firstoftwo
 \else \expandafter \@secondoftwo
 \fi
}%
\providecommand \natexlab [1]{#1}%
\providecommand \enquote  [1]{``#1''}%
\providecommand \bibnamefont  [1]{#1}%
\providecommand \bibfnamefont [1]{#1}%
\providecommand \citenamefont [1]{#1}%
\providecommand \href@noop [0]{\@secondoftwo}%
\providecommand \href [0]{\begingroup \@sanitize@url \@href}%
\providecommand \@href[1]{\@@startlink{#1}\@@href}%
\providecommand \@@href[1]{\endgroup#1\@@endlink}%
\providecommand \@sanitize@url [0]{\catcode `\\12\catcode `\$12\catcode
  `\&12\catcode `\#12\catcode `\^12\catcode `\_12\catcode `\%12\relax}%
\providecommand \@@startlink[1]{}%
\providecommand \@@endlink[0]{}%
\providecommand \url  [0]{\begingroup\@sanitize@url \@url }%
\providecommand \@url [1]{\endgroup\@href {#1}{\urlprefix }}%
\providecommand \urlprefix  [0]{URL }%
\providecommand \Eprint [0]{\href }%
\providecommand \doibase [0]{http://dx.doi.org/}%
\providecommand \selectlanguage [0]{\@gobble}%
\providecommand \bibinfo  [0]{\@secondoftwo}%
\providecommand \bibfield  [0]{\@secondoftwo}%
\providecommand \translation [1]{[#1]}%
\providecommand \BibitemOpen [0]{}%
\providecommand \bibitemStop [0]{}%
\providecommand \bibitemNoStop [0]{.\EOS\space}%
\providecommand \EOS [0]{\spacefactor3000\relax}%
\providecommand \BibitemShut  [1]{\csname bibitem#1\endcsname}%
\let\auto@bib@innerbib\@empty
\bibitem [{\citenamefont {Haldane}(1983)}]{Haldane}%
  \BibitemOpen
  \bibfield  {author} {\bibinfo {author} {\bibfnamefont {F.~D.~M.}\
  \bibnamefont {Haldane}},\ }\bibfield  {title} {\enquote {\bibinfo {title}
  {Nonlinear field theory of large-spin heisenberg antiferromagnets:
  Semiclassically quantized solitons of the one-dimensional easy-axis néel
  state},}\ }\href {\doibase 10.1103/PhysRevLett.50.1153} {\bibfield  {journal}
  {\bibinfo  {journal} {Phys. Rev. Lett}\ }\textbf {\bibinfo {volume} {50}},\
  \bibinfo {pages} {1153} (\bibinfo {year} {1983})}\BibitemShut {NoStop}%
\bibitem [{\citenamefont {Affleck}\ \emph {et~al.}(1987)\citenamefont
  {Affleck}, \citenamefont {Kennedy}, \citenamefont {Lieb},\ and\ \citenamefont
  {Tasaki}}]{AKLT}%
  \BibitemOpen
  \bibfield  {author} {\bibinfo {author} {\bibfnamefont {Ian}\ \bibnamefont
  {Affleck}}, \bibinfo {author} {\bibfnamefont {Tom}\ \bibnamefont {Kennedy}},
  \bibinfo {author} {\bibfnamefont {Elliott~H.}\ \bibnamefont {Lieb}}, \ and\
  \bibinfo {author} {\bibfnamefont {Hal}\ \bibnamefont {Tasaki}},\ }\bibfield
  {title} {\enquote {\bibinfo {title} {Rigorous results on valence-bond ground
  states in antiferromagnets},}\ }\href@noop {} {\bibfield  {journal} {\bibinfo
   {journal} {Phys. Rev. Lett.}\ }\textbf {\bibinfo {volume} {59}},\ \bibinfo
  {pages} {799} (\bibinfo {year} {1987})}\BibitemShut {NoStop}%
\bibitem [{\citenamefont {Hasan}\ and\ \citenamefont {Kane}(2010)}]{KaneRMP}%
  \BibitemOpen
  \bibfield  {author} {\bibinfo {author} {\bibfnamefont {M.~Z.}\ \bibnamefont
  {Hasan}}\ and\ \bibinfo {author} {\bibfnamefont {C.~L.}\ \bibnamefont
  {Kane}},\ }\bibfield  {title} {\enquote {\bibinfo {title} {Colloquium:
  Topological insulators},}\ }\href
  {https://journals.aps.org/rmp/abstract/10.1103/RevModPhys.82.3045} {\bibfield
   {journal} {\bibinfo  {journal} {Rev. Mod. Phys.}\ }\textbf {\bibinfo
  {volume} {82}},\ \bibinfo {pages} {3045--3067} (\bibinfo {year}
  {2010})}\BibitemShut {NoStop}%
\bibitem [{\citenamefont {Qi}\ and\ \citenamefont {Zhang}(2011)}]{ZhangRMP}%
  \BibitemOpen
  \bibfield  {author} {\bibinfo {author} {\bibfnamefont {X.-L.}\ \bibnamefont
  {Qi}}\ and\ \bibinfo {author} {\bibfnamefont {S.-C.}\ \bibnamefont {Zhang}},\
  }\bibfield  {title} {\enquote {\bibinfo {title} {Topological insulators and
  superconductors},}\ }\href
  {https://journals.aps.org/rmp/abstract/10.1103/RevModPhys.83.1057} {\bibfield
   {journal} {\bibinfo  {journal} {Rev. Mod. Phys.}\ }\textbf {\bibinfo
  {volume} {83}},\ \bibinfo {pages} {1057--1110} (\bibinfo {year}
  {2011})}\BibitemShut {NoStop}%
\bibitem [{\citenamefont {Pollmann}\ \emph {et~al.}(2010)\citenamefont
  {Pollmann}, \citenamefont {Turner}, \citenamefont {Berg},\ and\ \citenamefont
  {Oshikawa}}]{pollmann10}%
  \BibitemOpen
  \bibfield  {author} {\bibinfo {author} {\bibfnamefont {F.}~\bibnamefont
  {Pollmann}}, \bibinfo {author} {\bibfnamefont {A.~M.}\ \bibnamefont
  {Turner}}, \bibinfo {author} {\bibfnamefont {E.}~\bibnamefont {Berg}}, \ and\
  \bibinfo {author} {\bibfnamefont {M.}~\bibnamefont {Oshikawa}},\ }\bibfield
  {title} {\enquote {\bibinfo {title} {Entanglement spectrum of a topological
  phase in one dimension},}\ }\href {\doibase 10.1103/PhysRevB.81.064439}
  {\bibfield  {journal} {\bibinfo  {journal} {Phys. Rev. B}\ }\textbf {\bibinfo
  {volume} {81}},\ \bibinfo {pages} {064439} (\bibinfo {year}
  {2010})}\BibitemShut {NoStop}%
\bibitem [{\citenamefont {Chen}\ \emph
  {et~al.}(2011{\natexlab{a}})\citenamefont {Chen}, \citenamefont {Gu},\ and\
  \citenamefont {Wen}}]{chen11a}%
  \BibitemOpen
  \bibfield  {author} {\bibinfo {author} {\bibfnamefont {X.}~\bibnamefont
  {Chen}}, \bibinfo {author} {\bibfnamefont {Z.-C.}\ \bibnamefont {Gu}}, \ and\
  \bibinfo {author} {\bibfnamefont {X.-G.}\ \bibnamefont {Wen}},\ }\bibfield
  {title} {\enquote {\bibinfo {title} {Classification of gapped symmetric
  phases in one-dimensional spin systems},}\ }\href {\doibase
  10.1103/PhysRevB.83.035107} {\bibfield  {journal} {\bibinfo  {journal} {Phys.
  Rev. B}\ }\textbf {\bibinfo {volume} {83}},\ \bibinfo {pages} {035107}
  (\bibinfo {year} {2011}{\natexlab{a}})}\BibitemShut {NoStop}%
\bibitem [{\citenamefont {Chen}\ \emph
  {et~al.}(2011{\natexlab{b}})\citenamefont {Chen}, \citenamefont {Gu},\ and\
  \citenamefont {Wen}}]{chen11b}%
  \BibitemOpen
  \bibfield  {author} {\bibinfo {author} {\bibfnamefont {X.}~\bibnamefont
  {Chen}}, \bibinfo {author} {\bibfnamefont {Z.-C.}\ \bibnamefont {Gu}}, \ and\
  \bibinfo {author} {\bibfnamefont {X.-G.}\ \bibnamefont {Wen}},\ }\bibfield
  {title} {\enquote {\bibinfo {title} {Complete classification of
  one-dimensional gapped quantum phases in interacting spin systems},}\ }\href
  {\doibase 10.1103/PhysRevB.84.235128} {\bibfield  {journal} {\bibinfo
  {journal} {Phys. Rev. B}\ }\textbf {\bibinfo {volume} {84}},\ \bibinfo
  {pages} {235128} (\bibinfo {year} {2011}{\natexlab{b}})}\BibitemShut
  {NoStop}%
\bibitem [{\citenamefont {Chen}\ \emph {et~al.}(2012)\citenamefont {Chen},
  \citenamefont {Gu}, \citenamefont {Liu},\ and\ \citenamefont
  {Wen}}]{XieChenScience}%
  \BibitemOpen
  \bibfield  {author} {\bibinfo {author} {\bibfnamefont {X.}~\bibnamefont
  {Chen}}, \bibinfo {author} {\bibfnamefont {Z.-C.}\ \bibnamefont {Gu}},
  \bibinfo {author} {\bibfnamefont {Z.-X.}\ \bibnamefont {Liu}}, \ and\
  \bibinfo {author} {\bibfnamefont {X.-G.}\ \bibnamefont {Wen}},\ }\bibfield
  {title} {\enquote {\bibinfo {title} {Symmetry-protected topological orders in
  interacting bosonic systems},}\ }\href
  {https://science.sciencemag.org/content/338/6114/1604} {\bibfield  {journal}
  {\bibinfo  {journal} {Science}\ }\textbf {\bibinfo {volume} {338}},\ \bibinfo
  {pages} {1604--1606} (\bibinfo {year} {2012})}\BibitemShut {NoStop}%
\bibitem [{\citenamefont {Chen}\ \emph {et~al.}(2013)\citenamefont {Chen},
  \citenamefont {Gu}, \citenamefont {Liu},\ and\ \citenamefont
  {Wen}}]{cohomology}%
  \BibitemOpen
  \bibfield  {author} {\bibinfo {author} {\bibfnamefont {X.}~\bibnamefont
  {Chen}}, \bibinfo {author} {\bibfnamefont {Z.-C.}\ \bibnamefont {Gu}},
  \bibinfo {author} {\bibfnamefont {Z.-X.}\ \bibnamefont {Liu}}, \ and\
  \bibinfo {author} {\bibfnamefont {X.-G.}\ \bibnamefont {Wen}},\ }\bibfield
  {title} {\enquote {\bibinfo {title} {Symmetry protected topological orders
  and the group cohomology of their symmetry group},}\ }\href
  {https://journals.aps.org/prb/abstract/10.1103/PhysRevB.87.155114} {\bibfield
   {journal} {\bibinfo  {journal} {Phys. Rev. B}\ }\textbf {\bibinfo {volume}
  {87}},\ \bibinfo {pages} {155114} (\bibinfo {year} {2013})}\BibitemShut
  {NoStop}%
\bibitem [{\citenamefont {Lu}\ and\ \citenamefont
  {Vishwanath}(2012)}]{invertible1}%
  \BibitemOpen
  \bibfield  {author} {\bibinfo {author} {\bibfnamefont {Y.-M.}\ \bibnamefont
  {Lu}}\ and\ \bibinfo {author} {\bibfnamefont {A.}~\bibnamefont
  {Vishwanath}},\ }\bibfield  {title} {\enquote {\bibinfo {title} {Theory and
  classification of interacting integer topological phases in two dimensions: A
  chern-simons approach},}\ }\href
  {https://journals.aps.org/prb/abstract/10.1103/PhysRevB.86.125119} {\bibfield
   {journal} {\bibinfo  {journal} {Phys. Rev. B}\ }\textbf {\bibinfo {volume}
  {86}},\ \bibinfo {pages} {125119} (\bibinfo {year} {2012})}\BibitemShut
  {NoStop}%
\bibitem [{\citenamefont {Freed}()}]{invertible2}%
  \BibitemOpen
  \bibfield  {author} {\bibinfo {author} {\bibfnamefont {D.~S.}\ \bibnamefont
  {Freed}},\ }\bibfield  {title} {\enquote {\bibinfo {title} {Short-range
  entanglement and invertible field theories},}\ }\href@noop {} {\ }\Eprint
  {http://arxiv.org/abs/1406.7278} {arXiv:1406.7278 [cond-mat.str-el]}
  \BibitemShut {NoStop}%
\bibitem [{\citenamefont {Freed}\ and\ \citenamefont
  {Hopkins}(2016)}]{invertible3}%
  \BibitemOpen
  \bibfield  {author} {\bibinfo {author} {\bibfnamefont {Daniel~S.}\
  \bibnamefont {Freed}}\ and\ \bibinfo {author} {\bibfnamefont {Michael~J.}\
  \bibnamefont {Hopkins}},\ }\bibfield  {title} {\enquote {\bibinfo {title}
  {{Reflection positivity and invertible topological phases}},}\ }\href@noop {}
  {\bibfield  {journal} {\bibinfo  {journal} {arXiv e-prints}\ } (\bibinfo
  {year} {2016})},\ \Eprint {http://arxiv.org/abs/1604.06527}
  {arXiv:1604.06527} \BibitemShut {NoStop}%
\bibitem [{\citenamefont {Kapustin}()}]{Kapustin2014}%
  \BibitemOpen
  \bibfield  {author} {\bibinfo {author} {\bibfnamefont {A.}~\bibnamefont
  {Kapustin}},\ }\bibfield  {title} {\enquote {\bibinfo {title} {Symmetry
  protected topological phases, anomalies, and cobordisms: Beyond group
  cohomology},}\ }\href@noop {} {\ }\Eprint {http://arxiv.org/abs/1403.1467}
  {arXiv:1403.1467 [cond-mat.str-el]} \BibitemShut {NoStop}%
\bibitem [{\citenamefont {Wen}(2015)}]{wen15}%
  \BibitemOpen
  \bibfield  {author} {\bibinfo {author} {\bibfnamefont {Xiao-Gang}\
  \bibnamefont {Wen}},\ }\bibfield  {title} {\enquote {\bibinfo {title}
  {Construction of bosonic symmetry-protected-trivial states and their
  topological invariants via
  $g\ifmmode\times\else\texttimes\fi{}so(\ensuremath{\infty})$ nonlinear
  $\ensuremath{\sigma}$ models},}\ }\href {\doibase 10.1103/PhysRevB.91.205101}
  {\bibfield  {journal} {\bibinfo  {journal} {Phys. Rev. B}\ }\textbf {\bibinfo
  {volume} {91}},\ \bibinfo {pages} {205101} (\bibinfo {year}
  {2015})}\BibitemShut {NoStop}%
\bibitem [{\citenamefont {Gu}\ and\ \citenamefont {Wen}(2014)}]{special}%
  \BibitemOpen
  \bibfield  {author} {\bibinfo {author} {\bibfnamefont {Z.-C.}\ \bibnamefont
  {Gu}}\ and\ \bibinfo {author} {\bibfnamefont {X.-G.}\ \bibnamefont {Wen}},\
  }\bibfield  {title} {\enquote {\bibinfo {title} {Symmetry-protected
  topological orders for interacting fermions: Fermionic topological nonlinear
  $\sigma$ models and a special group supercohomology theory},}\ }\href
  {\doibase 10.1103/PhysRevB.90.115141} {\bibfield  {journal} {\bibinfo
  {journal} {Phys. Rev. B}\ }\textbf {\bibinfo {volume} {90}},\ \bibinfo
  {pages} {115141} (\bibinfo {year} {2014})}\BibitemShut {NoStop}%
\bibitem [{\citenamefont {Wang}\ and\ \citenamefont {Gu}(2018)}]{general1}%
  \BibitemOpen
  \bibfield  {author} {\bibinfo {author} {\bibfnamefont {Q.-R.}\ \bibnamefont
  {Wang}}\ and\ \bibinfo {author} {\bibfnamefont {Z.-C.}\ \bibnamefont {Gu}},\
  }\bibfield  {title} {\enquote {\bibinfo {title} {Towards a complete
  classification of symmetry-protected topological phases for interacting
  fermions in three dimensions and a general group supercohomology theory},}\
  }\href {https://journals.aps.org/prx/abstract/10.1103/PhysRevX.8.011055}
  {\bibfield  {journal} {\bibinfo  {journal} {Phys. Rev. X}\ }\textbf {\bibinfo
  {volume} {8}},\ \bibinfo {pages} {011055} (\bibinfo {year}
  {2018})}\BibitemShut {NoStop}%
\bibitem [{\citenamefont {Wang}\ and\ \citenamefont {Gu}(2020)}]{general2}%
  \BibitemOpen
  \bibfield  {author} {\bibinfo {author} {\bibfnamefont {Q.-R.}\ \bibnamefont
  {Wang}}\ and\ \bibinfo {author} {\bibfnamefont {Z.-C.}\ \bibnamefont {Gu}},\
  }\bibfield  {title} {\enquote {\bibinfo {title} {Construction and
  classification of symmetry-protected topological phases in interacting
  fermion systems},}\ }\href {\doibase 10.1103/PhysRevX.10.031055} {\bibfield
  {journal} {\bibinfo  {journal} {Phys. Rev. X}\ }\textbf {\bibinfo {volume}
  {10}},\ \bibinfo {pages} {031055} (\bibinfo {year} {2020})},\ \Eprint
  {http://arxiv.org/abs/1811.00536} {arXiv:1811.00536 [cond-mat.str-el]}
  \BibitemShut {NoStop}%
\bibitem [{\citenamefont {Kapustin}\ \emph {et~al.}(2015)\citenamefont
  {Kapustin}, \citenamefont {Thorngren}, \citenamefont {Turzillo},\ and\
  \citenamefont {Wang}}]{Kapustin2015}%
  \BibitemOpen
  \bibfield  {author} {\bibinfo {author} {\bibfnamefont {Anton}\ \bibnamefont
  {Kapustin}}, \bibinfo {author} {\bibfnamefont {Ryan}\ \bibnamefont
  {Thorngren}}, \bibinfo {author} {\bibfnamefont {Alex}\ \bibnamefont
  {Turzillo}}, \ and\ \bibinfo {author} {\bibfnamefont {Zitao}\ \bibnamefont
  {Wang}},\ }\bibfield  {title} {\enquote {\bibinfo {title} {Fermionic symmetry
  protected topological phases and cobordisms},}\ }\href {\doibase
  https://link.springer.com/article/10.1007/JHEP12(2015)052} {\bibfield
  {journal} {\bibinfo  {journal} {JHEP}\ }\textbf {\bibinfo {volume} {1512}},\
  \bibinfo {pages} {052} (\bibinfo {year} {2015})}\BibitemShut {NoStop}%
\bibitem [{\citenamefont {Kapustin}\ and\ \citenamefont
  {Thorngren}(2017)}]{Kapustin2017}%
  \BibitemOpen
  \bibfield  {author} {\bibinfo {author} {\bibfnamefont {Anton}\ \bibnamefont
  {Kapustin}}\ and\ \bibinfo {author} {\bibfnamefont {Ryan}\ \bibnamefont
  {Thorngren}},\ }\bibfield  {title} {\enquote {\bibinfo {title} {Fermionic spt
  phases in higher dimensions and bosonization},}\ }\href {\doibase
  10.1007/JHEP10(2017)080} {\bibfield  {journal} {\bibinfo  {journal} {Journal
  of High Energy Physics}\ }\textbf {\bibinfo {volume} {2017}},\ \bibinfo
  {pages} {80} (\bibinfo {year} {2017})}\BibitemShut {NoStop}%
\bibitem [{\citenamefont {Fidkowski}\ and\ \citenamefont
  {Kitaev}(2010)}]{fidkowski10}%
  \BibitemOpen
  \bibfield  {author} {\bibinfo {author} {\bibfnamefont {Lukasz}\ \bibnamefont
  {Fidkowski}}\ and\ \bibinfo {author} {\bibfnamefont {Alexei}\ \bibnamefont
  {Kitaev}},\ }\bibfield  {title} {\enquote {\bibinfo {title} {Effects of
  interactions on the topological classification of free fermion systems},}\
  }\href {\doibase 10.1103/PhysRevB.81.134509} {\bibfield  {journal} {\bibinfo
  {journal} {Phys. Rev. B}\ }\textbf {\bibinfo {volume} {81}},\ \bibinfo
  {pages} {134509} (\bibinfo {year} {2010})}\BibitemShut {NoStop}%
\bibitem [{\citenamefont {Fidkowski}\ and\ \citenamefont
  {Kitaev}(2011{\natexlab{a}})}]{fidkowski11}%
  \BibitemOpen
  \bibfield  {author} {\bibinfo {author} {\bibfnamefont {L.}~\bibnamefont
  {Fidkowski}}\ and\ \bibinfo {author} {\bibfnamefont {A.}~\bibnamefont
  {Kitaev}},\ }\bibfield  {title} {\enquote {\bibinfo {title} {Topological
  phases of fermions in one dimension},}\ }\href {\doibase
  10.1103/PhysRevB.83.075103} {\bibfield  {journal} {\bibinfo  {journal} {Phys.
  Rev. B}\ }\textbf {\bibinfo {volume} {83}},\ \bibinfo {pages} {075103}
  (\bibinfo {year} {2011}{\natexlab{a}})}\BibitemShut {NoStop}%
\bibitem [{\citenamefont {{Wang}}\ \emph {et~al.}(2014)\citenamefont {{Wang}},
  \citenamefont {{Potter}},\ and\ \citenamefont {{Senthil}}}]{wangc-science}%
  \BibitemOpen
  \bibfield  {author} {\bibinfo {author} {\bibfnamefont {C.}~\bibnamefont
  {{Wang}}}, \bibinfo {author} {\bibfnamefont {A.~C.}\ \bibnamefont
  {{Potter}}}, \ and\ \bibinfo {author} {\bibfnamefont {T.}~\bibnamefont
  {{Senthil}}},\ }\bibfield  {title} {\enquote {\bibinfo {title}
  {{Classification of Interacting Electronic Topological Insulators in Three
  Dimensions}},}\ }\href {\doibase 10.1126/science.1243326} {\bibfield
  {journal} {\bibinfo  {journal} {Science}\ }\textbf {\bibinfo {volume}
  {343}},\ \bibinfo {pages} {629--631} (\bibinfo {year} {2014})},\ \Eprint
  {http://arxiv.org/abs/1306.3238} {arXiv:1306.3238} \BibitemShut {NoStop}%
\bibitem [{\citenamefont {Wang}\ and\ \citenamefont
  {Senthil}(2014)}]{ChongWang2014}%
  \BibitemOpen
  \bibfield  {author} {\bibinfo {author} {\bibfnamefont {C.}~\bibnamefont
  {Wang}}\ and\ \bibinfo {author} {\bibfnamefont {T.}~\bibnamefont {Senthil}},\
  }\bibfield  {title} {\enquote {\bibinfo {title} {Interacting fermionic
  topological insulators/superconductors in three dimensions},}\ }\href
  {\doibase 10.1103/PhysRevB.89.195124} {\bibfield  {journal} {\bibinfo
  {journal} {Phys. Rev. B}\ }\textbf {\bibinfo {volume} {89}},\ \bibinfo
  {pages} {195124} (\bibinfo {year} {2014})}\BibitemShut {NoStop}%
\bibitem [{\citenamefont {Witten}(2016)}]{Witten}%
  \BibitemOpen
  \bibfield  {author} {\bibinfo {author} {\bibfnamefont {E.}~\bibnamefont
  {Witten}},\ }\bibfield  {title} {\enquote {\bibinfo {title} {Fermion path
  integrals and topological phases},}\ }\href
  {https://link.aps.org/doi/10.1103/RevModPhys.88.035001} {\bibfield  {journal}
  {\bibinfo  {journal} {Rev. Mod. Phys.}\ }\textbf {\bibinfo {volume} {88}},\
  \bibinfo {pages} {035001} (\bibinfo {year} {2016})}\BibitemShut {NoStop}%
\bibitem [{\citenamefont {Levin}\ and\ \citenamefont {Gu}(2012)}]{LevinGu}%
  \BibitemOpen
  \bibfield  {author} {\bibinfo {author} {\bibfnamefont {M.}~\bibnamefont
  {Levin}}\ and\ \bibinfo {author} {\bibfnamefont {Z.-C.}\ \bibnamefont {Gu}},\
  }\bibfield  {title} {\enquote {\bibinfo {title} {Braiding statistics approach
  to symmetry-protected topological phases},}\ }\href
  {https://journals.aps.org/prb/abstract/10.1103/PhysRevB.86.115109} {\bibfield
   {journal} {\bibinfo  {journal} {Phys. Rev. B}\ }\textbf {\bibinfo {volume}
  {86}},\ \bibinfo {pages} {115109} (\bibinfo {year} {2012})}\BibitemShut
  {NoStop}%
\bibitem [{\citenamefont {Gu}\ and\ \citenamefont {Levin}(2014)}]{Gu-Levin}%
  \BibitemOpen
  \bibfield  {author} {\bibinfo {author} {\bibfnamefont {Z.-C.}\ \bibnamefont
  {Gu}}\ and\ \bibinfo {author} {\bibfnamefont {M.}~\bibnamefont {Levin}},\
  }\bibfield  {title} {\enquote {\bibinfo {title} {Effect of interactions on
  two-dimensional fermionic symmetry-protected topological phases with $z_2$
  symmetry},}\ }\href
  {https://journals.aps.org/prb/abstract/10.1103/PhysRevB.89.201113} {\bibfield
   {journal} {\bibinfo  {journal} {Phys. Rev. B}\ }\textbf {\bibinfo {volume}
  {89}},\ \bibinfo {pages} {201113(R)} (\bibinfo {year} {2014})}\BibitemShut
  {NoStop}%
\bibitem [{\citenamefont {Cheng}\ and\ \citenamefont {Gu}(2014)}]{gauging1}%
  \BibitemOpen
  \bibfield  {author} {\bibinfo {author} {\bibfnamefont {M.}~\bibnamefont
  {Cheng}}\ and\ \bibinfo {author} {\bibfnamefont {Z.-C.}\ \bibnamefont {Gu}},\
  }\bibfield  {title} {\enquote {\bibinfo {title} {Topological response theory
  of abelian symmetry-protected topological phases in two dimensions},}\ }\href
  {https://journals.aps.org/prl/abstract/10.1103/PhysRevLett.112.141602}
  {\bibfield  {journal} {\bibinfo  {journal} {Phys. Rev. Lett.}\ }\textbf
  {\bibinfo {volume} {112}},\ \bibinfo {pages} {141602} (\bibinfo {year}
  {2014})}\BibitemShut {NoStop}%
\bibitem [{\citenamefont {Wang}\ and\ \citenamefont {Levin}(2014)}]{threeloop}%
  \BibitemOpen
  \bibfield  {author} {\bibinfo {author} {\bibfnamefont {C.}~\bibnamefont
  {Wang}}\ and\ \bibinfo {author} {\bibfnamefont {M.}~\bibnamefont {Levin}},\
  }\bibfield  {title} {\enquote {\bibinfo {title} {Braiding statistics of loop
  excitations in three dimensions},}\ }\href {\doibase
  10.1103/PhysRevLett.113.080403} {\bibfield  {journal} {\bibinfo  {journal}
  {Phys. Rev. Lett.}\ }\textbf {\bibinfo {volume} {113}},\ \bibinfo {pages}
  {080403} (\bibinfo {year} {2014})}\BibitemShut {NoStop}%
\bibitem [{\citenamefont {Jiang}\ \emph {et~al.}(2014)\citenamefont {Jiang},
  \citenamefont {Mesaros},\ and\ \citenamefont {Ran}}]{ran14}%
  \BibitemOpen
  \bibfield  {author} {\bibinfo {author} {\bibfnamefont {S.}~\bibnamefont
  {Jiang}}, \bibinfo {author} {\bibfnamefont {A.}~\bibnamefont {Mesaros}}, \
  and\ \bibinfo {author} {\bibfnamefont {Y.}~\bibnamefont {Ran}},\ }\bibfield
  {title} {\enquote {\bibinfo {title} {Generalized modular transformations in
  $(3+1)\mathrm{D}$ topologically ordered phases and triple linking invariant
  of loop braiding},}\ }\href {\doibase 10.1103/PhysRevX.4.031048} {\bibfield
  {journal} {\bibinfo  {journal} {Phys. Rev. X}\ }\textbf {\bibinfo {volume}
  {4}},\ \bibinfo {pages} {031048} (\bibinfo {year} {2014})}\BibitemShut
  {NoStop}%
\bibitem [{\citenamefont {Wang}\ and\ \citenamefont {Wen}(2015)}]{wangj15}%
  \BibitemOpen
  \bibfield  {author} {\bibinfo {author} {\bibfnamefont {J.~C.}\ \bibnamefont
  {Wang}}\ and\ \bibinfo {author} {\bibfnamefont {X.-G.}\ \bibnamefont {Wen}},\
  }\bibfield  {title} {\enquote {\bibinfo {title} {Non-abelian string and
  particle braiding in topological order: Modular $\mathrm{SL}(3,\mathbb{Z})$
  representation and $(3+1)$-dimensional twisted gauge theory},}\ }\href
  {\doibase 10.1103/PhysRevB.91.035134} {\bibfield  {journal} {\bibinfo
  {journal} {Phys. Rev. B}\ }\textbf {\bibinfo {volume} {91}},\ \bibinfo
  {pages} {035134} (\bibinfo {year} {2015})}\BibitemShut {NoStop}%
\bibitem [{\citenamefont {Wang}\ and\ \citenamefont {Levin}(2015)}]{wangcj15}%
  \BibitemOpen
  \bibfield  {author} {\bibinfo {author} {\bibfnamefont {C.}~\bibnamefont
  {Wang}}\ and\ \bibinfo {author} {\bibfnamefont {M.}~\bibnamefont {Levin}},\
  }\bibfield  {title} {\enquote {\bibinfo {title} {Topological invariants for
  gauge theories and symmetry-protected topological phases},}\ }\href {\doibase
  10.1103/PhysRevB.91.165119} {\bibfield  {journal} {\bibinfo  {journal} {Phys.
  Rev. B}\ }\textbf {\bibinfo {volume} {91}},\ \bibinfo {pages} {165119}
  (\bibinfo {year} {2015})}\BibitemShut {NoStop}%
\bibitem [{\citenamefont {Lin}\ and\ \citenamefont {Levin}(2015)}]{lin15}%
  \BibitemOpen
  \bibfield  {author} {\bibinfo {author} {\bibfnamefont {C.-H.}\ \bibnamefont
  {Lin}}\ and\ \bibinfo {author} {\bibfnamefont {M.}~\bibnamefont {Levin}},\
  }\bibfield  {title} {\enquote {\bibinfo {title} {Loop braiding statistics in
  exactly soluble three-dimensional lattice models},}\ }\href {\doibase
  10.1103/PhysRevB.92.035115} {\bibfield  {journal} {\bibinfo  {journal} {Phys.
  Rev. B}\ }\textbf {\bibinfo {volume} {92}},\ \bibinfo {pages} {035115}
  (\bibinfo {year} {2015})}\BibitemShut {NoStop}%
\bibitem [{\citenamefont {Barkeshli}\ \emph {et~al.}(2019)\citenamefont
  {Barkeshli}, \citenamefont {Bonderson}, \citenamefont {Cheng},\ and\
  \citenamefont {Wang}}]{gauging3}%
  \BibitemOpen
  \bibfield  {author} {\bibinfo {author} {\bibfnamefont {M.}~\bibnamefont
  {Barkeshli}}, \bibinfo {author} {\bibfnamefont {P.}~\bibnamefont
  {Bonderson}}, \bibinfo {author} {\bibfnamefont {M.}~\bibnamefont {Cheng}}, \
  and\ \bibinfo {author} {\bibfnamefont {Z.}~\bibnamefont {Wang}},\ }\bibfield
  {title} {\enquote {\bibinfo {title} {Symmetry fractionalization, defects, and
  gauging of topological phases},}\ }\href {\doibase
  10.1103/PhysRevB.100.115147} {\bibfield  {journal} {\bibinfo  {journal}
  {Phys. Rev. B}\ }\textbf {\bibinfo {volume} {100}},\ \bibinfo {pages}
  {115147} (\bibinfo {year} {2019})},\ \Eprint {http://arxiv.org/abs/1410.4540}
  {arXiv:1410.4540 [cond-mat.str-el]} \BibitemShut {NoStop}%
\bibitem [{\citenamefont {Tantivasadakarn}(2017)}]{dimensionalreduction}%
  \BibitemOpen
  \bibfield  {author} {\bibinfo {author} {\bibfnamefont {N.}~\bibnamefont
  {Tantivasadakarn}},\ }\bibfield  {title} {\enquote {\bibinfo {title}
  {Dimensional reduction and topological invariants of symmetry-protected
  topological phases},}\ }\href
  {https://journals.aps.org/prb/abstract/10.1103/PhysRevB.96.195101} {\bibfield
   {journal} {\bibinfo  {journal} {Phys. Rev. B}\ }\textbf {\bibinfo {volume}
  {96}},\ \bibinfo {pages} {195101} (\bibinfo {year} {2017})}\BibitemShut
  {NoStop}%
\bibitem [{\citenamefont {Wang}\ \emph {et~al.}(2017)\citenamefont {Wang},
  \citenamefont {Lin},\ and\ \citenamefont {Gu}}]{gauging2}%
  \BibitemOpen
  \bibfield  {author} {\bibinfo {author} {\bibfnamefont {C.}~\bibnamefont
  {Wang}}, \bibinfo {author} {\bibfnamefont {C.-H.}\ \bibnamefont {Lin}}, \
  and\ \bibinfo {author} {\bibfnamefont {Z.-C.}\ \bibnamefont {Gu}},\
  }\bibfield  {title} {\enquote {\bibinfo {title} {Interacting fermionic
  symmetry-protected topological phases in two dimensions},}\ }\href
  {https://journals.aps.org/prb/abstract/10.1103/PhysRevB.95.195147} {\bibfield
   {journal} {\bibinfo  {journal} {Phys. Rev. B}\ }\textbf {\bibinfo {volume}
  {95}},\ \bibinfo {pages} {195147} (\bibinfo {year} {2017})}\BibitemShut
  {NoStop}%
\bibitem [{\citenamefont {Cheng}\ \emph
  {et~al.}(2018{\natexlab{a}})\citenamefont {Cheng}, \citenamefont {Bi},
  \citenamefont {You},\ and\ \citenamefont {Gu}}]{2DFSPT}%
  \BibitemOpen
  \bibfield  {author} {\bibinfo {author} {\bibfnamefont {M.}~\bibnamefont
  {Cheng}}, \bibinfo {author} {\bibfnamefont {Z.}~\bibnamefont {Bi}}, \bibinfo
  {author} {\bibfnamefont {Y.-Z.}\ \bibnamefont {You}}, \ and\ \bibinfo
  {author} {\bibfnamefont {Z.-C.}\ \bibnamefont {Gu}},\ }\bibfield  {title}
  {\enquote {\bibinfo {title} {Classification of symmetry-protected phases for
  interacting fermions in two dimensions},}\ }\href
  {https://journals.aps.org/prb/abstract/10.1103/PhysRevB.97.205109} {\bibfield
   {journal} {\bibinfo  {journal} {Phys. Rev. B}\ }\textbf {\bibinfo {volume}
  {97}},\ \bibinfo {pages} {205109} (\bibinfo {year}
  {2018}{\natexlab{a}})}\BibitemShut {NoStop}%
\bibitem [{\citenamefont {Cheng}\ \emph
  {et~al.}(2018{\natexlab{b}})\citenamefont {Cheng}, \citenamefont
  {Tantivasadakarn},\ and\ \citenamefont {Wang}}]{braiding}%
  \BibitemOpen
  \bibfield  {author} {\bibinfo {author} {\bibfnamefont {M.}~\bibnamefont
  {Cheng}}, \bibinfo {author} {\bibfnamefont {N.}~\bibnamefont
  {Tantivasadakarn}}, \ and\ \bibinfo {author} {\bibfnamefont {C.}~\bibnamefont
  {Wang}},\ }\bibfield  {title} {\enquote {\bibinfo {title} {Loop braiding
  statistics and interacting fermionic symmetry-protected topological phases in
  three dimensions},}\ }\href
  {https://journals.aps.org/prx/abstract/10.1103/PhysRevX.8.011054} {\bibfield
  {journal} {\bibinfo  {journal} {Phys. Rev. X}\ }\textbf {\bibinfo {volume}
  {8}},\ \bibinfo {pages} {011054} (\bibinfo {year}
  {2018}{\natexlab{b}})}\BibitemShut {NoStop}%
\bibitem [{\citenamefont {Vishwanath}\ and\ \citenamefont
  {Senthil}(2013)}]{Ashvin2013}%
  \BibitemOpen
  \bibfield  {author} {\bibinfo {author} {\bibfnamefont {A.}~\bibnamefont
  {Vishwanath}}\ and\ \bibinfo {author} {\bibfnamefont {T.}~\bibnamefont
  {Senthil}},\ }\bibfield  {title} {\enquote {\bibinfo {title} {Physics of
  three-dimensional bosonic topological insulators: Surface-deconfined
  criticality and quantized magnetoelectric effect},}\ }\href
  {https://journals.aps.org/prx/abstract/10.1103/PhysRevX.3.011016} {\bibfield
  {journal} {\bibinfo  {journal} {Phys. Rev. X}\ }\textbf {\bibinfo {volume}
  {3}},\ \bibinfo {pages} {011016} (\bibinfo {year} {2013})}\BibitemShut
  {NoStop}%
\bibitem [{\citenamefont {Wang}\ and\ \citenamefont
  {Senthil}(2013)}]{ChongWang2013}%
  \BibitemOpen
  \bibfield  {author} {\bibinfo {author} {\bibfnamefont {C.}~\bibnamefont
  {Wang}}\ and\ \bibinfo {author} {\bibfnamefont {T.}~\bibnamefont {Senthil}},\
  }\bibfield  {title} {\enquote {\bibinfo {title} {Boson topological
  insulators: A window into highly entangled quantum phases},}\ }\href
  {https://journals.aps.org/prb/abstract/10.1103/PhysRevB.87.235122} {\bibfield
   {journal} {\bibinfo  {journal} {Phys. Rev. B}\ }\textbf {\bibinfo {volume}
  {87}},\ \bibinfo {pages} {235122} (\bibinfo {year} {2013})}\BibitemShut
  {NoStop}%
\bibitem [{\citenamefont {Chen}\ \emph {et~al.}(2015)\citenamefont {Chen},
  \citenamefont {Burnell}, \citenamefont {Vishwanath},\ and\ \citenamefont
  {Fidkowski}}]{XieChen2015}%
  \BibitemOpen
  \bibfield  {author} {\bibinfo {author} {\bibfnamefont {X.}~\bibnamefont
  {Chen}}, \bibinfo {author} {\bibfnamefont {F.~J.}\ \bibnamefont {Burnell}},
  \bibinfo {author} {\bibfnamefont {A.}~\bibnamefont {Vishwanath}}, \ and\
  \bibinfo {author} {\bibfnamefont {L.}~\bibnamefont {Fidkowski}},\ }\bibfield
  {title} {\enquote {\bibinfo {title} {Anomalous symmetry fractionalization and
  surface topological order},}\ }\href
  {https://link.aps.org/doi/10.1103/PhysRevX.5.041013} {\bibfield  {journal}
  {\bibinfo  {journal} {Phys. Rev. X}\ }\textbf {\bibinfo {volume} {5}},\
  \bibinfo {pages} {041013} (\bibinfo {year} {2015})}\BibitemShut {NoStop}%
\bibitem [{\citenamefont {Wang}\ \emph {et~al.}(2016)\citenamefont {Wang},
  \citenamefont {Lin},\ and\ \citenamefont {Levin}}]{ChenjieWang2016}%
  \BibitemOpen
  \bibfield  {author} {\bibinfo {author} {\bibfnamefont {C.}~\bibnamefont
  {Wang}}, \bibinfo {author} {\bibfnamefont {C.-H.}\ \bibnamefont {Lin}}, \
  and\ \bibinfo {author} {\bibfnamefont {M.}~\bibnamefont {Levin}},\ }\bibfield
   {title} {\enquote {\bibinfo {title} {Bulk-boundary correspondence for
  three-dimensional symmetry-protected topological phases},}\ }\href
  {https://journals.aps.org/prx/abstract/10.1103/PhysRevX.6.021015} {\bibfield
  {journal} {\bibinfo  {journal} {Phys. Rev. X}\ }\textbf {\bibinfo {volume}
  {6}},\ \bibinfo {pages} {021015} (\bibinfo {year} {2016})}\BibitemShut
  {NoStop}%
\bibitem [{\citenamefont {Bonderson}\ \emph {et~al.}(2013)\citenamefont
  {Bonderson}, \citenamefont {Nayak},\ and\ \citenamefont {Qi}}]{XLQi2013}%
  \BibitemOpen
  \bibfield  {author} {\bibinfo {author} {\bibfnamefont {P.}~\bibnamefont
  {Bonderson}}, \bibinfo {author} {\bibfnamefont {C.}~\bibnamefont {Nayak}}, \
  and\ \bibinfo {author} {\bibfnamefont {X.-L.}\ \bibnamefont {Qi}},\
  }\bibfield  {title} {\enquote {\bibinfo {title} {A time-reversal invariant
  topological phase at the surface of a 3d topological insulator},}\ }\href
  {https://iopscience.iop.org/article/10.1088/1742-5468/2013/09/P09016/meta}
  {\bibfield  {journal} {\bibinfo  {journal} {Journal of Statistical Mechanics:
  Theory and Experiment}\ }\textbf {\bibinfo {volume} {2013}},\ \bibinfo
  {pages} {P09016} (\bibinfo {year} {2013})}\BibitemShut {NoStop}%
\bibitem [{\citenamefont {Wang}\ \emph {et~al.}(2013)\citenamefont {Wang},
  \citenamefont {Potter},\ and\ \citenamefont {Senthil}}]{Senthil2013}%
  \BibitemOpen
  \bibfield  {author} {\bibinfo {author} {\bibfnamefont {C.}~\bibnamefont
  {Wang}}, \bibinfo {author} {\bibfnamefont {A.~C.}\ \bibnamefont {Potter}}, \
  and\ \bibinfo {author} {\bibfnamefont {T.}~\bibnamefont {Senthil}},\
  }\bibfield  {title} {\enquote {\bibinfo {title} {Gapped symmetry preserving
  surface state for the electron topological insulator},}\ }\href
  {https://journals.aps.org/prb/abstract/10.1103/PhysRevB.88.115137} {\bibfield
   {journal} {\bibinfo  {journal} {Phys. Rev. B}\ }\textbf {\bibinfo {volume}
  {88}},\ \bibinfo {pages} {115137} (\bibinfo {year} {2013})}\BibitemShut
  {NoStop}%
\bibitem [{\citenamefont {Fidkowski}\ \emph {et~al.}(2013)\citenamefont
  {Fidkowski}, \citenamefont {Chen},\ and\ \citenamefont
  {Vishwanath}}]{Lukasz2013}%
  \BibitemOpen
  \bibfield  {author} {\bibinfo {author} {\bibfnamefont {L.}~\bibnamefont
  {Fidkowski}}, \bibinfo {author} {\bibfnamefont {X.}~\bibnamefont {Chen}}, \
  and\ \bibinfo {author} {\bibfnamefont {A.}~\bibnamefont {Vishwanath}},\
  }\bibfield  {title} {\enquote {\bibinfo {title} {Non-abelian topological
  order on the surface of a 3d topological superconductor from an exactly
  solved model},}\ }\href
  {https://journals.aps.org/prx/abstract/10.1103/PhysRevX.3.041016} {\bibfield
  {journal} {\bibinfo  {journal} {Phys. Rev. X}\ }\textbf {\bibinfo {volume}
  {3}},\ \bibinfo {pages} {041016} (\bibinfo {year} {2013})}\BibitemShut
  {NoStop}%
\bibitem [{\citenamefont {Chen}\ \emph {et~al.}(2014)\citenamefont {Chen},
  \citenamefont {Fidkowski},\ and\ \citenamefont {Vishwanath}}]{XieChen2014}%
  \BibitemOpen
  \bibfield  {author} {\bibinfo {author} {\bibfnamefont {X.}~\bibnamefont
  {Chen}}, \bibinfo {author} {\bibfnamefont {L.}~\bibnamefont {Fidkowski}}, \
  and\ \bibinfo {author} {\bibfnamefont {A.}~\bibnamefont {Vishwanath}},\
  }\bibfield  {title} {\enquote {\bibinfo {title} {Symmetry enforced
  non-abelian topological order at the surface of a topological insulator},}\
  }\href {https://journals.aps.org/prb/abstract/10.1103/PhysRevB.89.165132}
  {\bibfield  {journal} {\bibinfo  {journal} {Phys. Rev. B}\ }\textbf {\bibinfo
  {volume} {89}},\ \bibinfo {pages} {165132} (\bibinfo {year}
  {2014})}\BibitemShut {NoStop}%
\bibitem [{\citenamefont {Pollman}\ \emph {et~al.}(2012)\citenamefont
  {Pollman}, \citenamefont {Berg}, \citenamefont {Turner},\ and\ \citenamefont
  {Oshikawa}}]{Pollman2012}%
  \BibitemOpen
  \bibfield  {author} {\bibinfo {author} {\bibfnamefont {F.}~\bibnamefont
  {Pollman}}, \bibinfo {author} {\bibfnamefont {E.}~\bibnamefont {Berg}},
  \bibinfo {author} {\bibfnamefont {A.~M.}\ \bibnamefont {Turner}}, \ and\
  \bibinfo {author} {\bibfnamefont {M.}~\bibnamefont {Oshikawa}},\ }\bibfield
  {title} {\enquote {\bibinfo {title} {Symmetry protectionof topological phases
  in one-dimensional quantum spin systems},}\ }\href {\doibase
  10.1103/PhysRevB.85.075125} {\bibfield  {journal} {\bibinfo  {journal} {Phys.
  Rev. B}\ }\textbf {\bibinfo {volume} {85}},\ \bibinfo {pages} {075125}
  (\bibinfo {year} {2012})}\BibitemShut {NoStop}%
\bibitem [{\citenamefont {Chen}\ \emph
  {et~al.}(2011{\natexlab{c}})\citenamefont {Chen}, \citenamefont {Liu},\ and\
  \citenamefont {Wen}}]{CZX}%
  \BibitemOpen
  \bibfield  {author} {\bibinfo {author} {\bibfnamefont {X.}~\bibnamefont
  {Chen}}, \bibinfo {author} {\bibfnamefont {Z.-X.}\ \bibnamefont {Liu}}, \
  and\ \bibinfo {author} {\bibfnamefont {X.-G.}\ \bibnamefont {Wen}},\
  }\bibfield  {title} {\enquote {\bibinfo {title} {Two-dimensional
  symmetry-protected topological orders and their protected gapless edge
  excitations},}\ }\href
  {https://journals.aps.org/prb/abstract/10.1103/PhysRevB.84.235141} {\bibfield
   {journal} {\bibinfo  {journal} {Phys. Rev. B}\ }\textbf {\bibinfo {volume}
  {84}},\ \bibinfo {pages} {235141} (\bibinfo {year}
  {2011}{\natexlab{c}})}\BibitemShut {NoStop}%
\bibitem [{\citenamefont {Schuch}\ \emph {et~al.}(2011)\citenamefont {Schuch},
  \citenamefont {P\'erez-Garc\'{\i}a},\ and\ \citenamefont {Cirac}}]{schuch11}%
  \BibitemOpen
  \bibfield  {author} {\bibinfo {author} {\bibfnamefont {N.}~\bibnamefont
  {Schuch}}, \bibinfo {author} {\bibfnamefont {D.}~\bibnamefont
  {P\'erez-Garc\'{\i}a}}, \ and\ \bibinfo {author} {\bibfnamefont
  {I.}~\bibnamefont {Cirac}},\ }\bibfield  {title} {\enquote {\bibinfo {title}
  {Classifying quantum phases using matrix product states and projected
  entangled pair states},}\ }\href {\doibase 10.1103/PhysRevB.84.165139}
  {\bibfield  {journal} {\bibinfo  {journal} {Phys. Rev. B}\ }\textbf {\bibinfo
  {volume} {84}},\ \bibinfo {pages} {165139} (\bibinfo {year}
  {2011})}\BibitemShut {NoStop}%
\bibitem [{\citenamefont {Wahl}\ \emph {et~al.}(2013)\citenamefont {Wahl},
  \citenamefont {Tu}, \citenamefont {Schuch},\ and\ \citenamefont
  {Cirac}}]{Cirac2013}%
  \BibitemOpen
  \bibfield  {author} {\bibinfo {author} {\bibfnamefont {T.~B.}\ \bibnamefont
  {Wahl}}, \bibinfo {author} {\bibfnamefont {H.-H.}\ \bibnamefont {Tu}},
  \bibinfo {author} {\bibfnamefont {N.}~\bibnamefont {Schuch}}, \ and\ \bibinfo
  {author} {\bibfnamefont {J.~I.}\ \bibnamefont {Cirac}},\ }\bibfield  {title}
  {\enquote {\bibinfo {title} {Projected entangled-pair states can describe
  chiral topological states},}\ }\href {\doibase
  10.1103/PhysRevLett.111.236805} {\bibfield  {journal} {\bibinfo  {journal}
  {Phys. Rev. Lett.}\ }\textbf {\bibinfo {volume} {111}},\ \bibinfo {pages}
  {236805} (\bibinfo {year} {2013})}\BibitemShut {NoStop}%
\bibitem [{\citenamefont {Wahl}\ \emph {et~al.}(2014)\citenamefont {Wahl},
  \citenamefont {Haßler}, \citenamefont {Tu}, \citenamefont {Cirac},\ and\
  \citenamefont {Schuch}}]{Schuch2014}%
  \BibitemOpen
  \bibfield  {author} {\bibinfo {author} {\bibfnamefont {Thorsten~B.}\
  \bibnamefont {Wahl}}, \bibinfo {author} {\bibfnamefont {Stefan~T.}\
  \bibnamefont {Haßler}}, \bibinfo {author} {\bibfnamefont {Hong-Hao}\
  \bibnamefont {Tu}}, \bibinfo {author} {\bibfnamefont {J.~Ignacio}\
  \bibnamefont {Cirac}}, \ and\ \bibinfo {author} {\bibfnamefont {Norbert}\
  \bibnamefont {Schuch}},\ }\bibfield  {title} {\enquote {\bibinfo {title}
  {Symmetries and boundary theories for chiral projected entangled pair
  states},}\ }\href {\doibase 10.1103/PhysRevB.90.115133} {\bibfield  {journal}
  {\bibinfo  {journal} {Phys. Rev. B}\ }\textbf {\bibinfo {volume} {90}},\
  \bibinfo {pages} {115133} (\bibinfo {year} {2014})}\BibitemShut {NoStop}%
\bibitem [{\citenamefont {Dubail}\ and\ \citenamefont {Read}(2015)}]{Read2015}%
  \BibitemOpen
  \bibfield  {author} {\bibinfo {author} {\bibfnamefont {J.}~\bibnamefont
  {Dubail}}\ and\ \bibinfo {author} {\bibfnamefont {N.}~\bibnamefont {Read}},\
  }\bibfield  {title} {\enquote {\bibinfo {title} {Tensor network trial states
  for chiral topological phases in two dimensions and a no-go theorem in any
  dimension},}\ }\href {\doibase 10.1103/PhysRevB.92.205307} {\bibfield
  {journal} {\bibinfo  {journal} {Phys. Rev. B}\ }\textbf {\bibinfo {volume}
  {92}},\ \bibinfo {pages} {205307} (\bibinfo {year} {2015})}\BibitemShut
  {NoStop}%
\bibitem [{\citenamefont {Williamson}\ \emph {et~al.}(2016)\citenamefont
  {Williamson}, \citenamefont {Bultinck}, \citenamefont {Mariën},
  \citenamefont {Sahinoglu}, \citenamefont {Haegeman},\ and\ \citenamefont
  {Verstraete}}]{Frank2016}%
  \BibitemOpen
  \bibfield  {author} {\bibinfo {author} {\bibfnamefont {Dominic~J.}\
  \bibnamefont {Williamson}}, \bibinfo {author} {\bibfnamefont {Nick}\
  \bibnamefont {Bultinck}}, \bibinfo {author} {\bibfnamefont {Michael}\
  \bibnamefont {Mariën}}, \bibinfo {author} {\bibfnamefont {Mehmet~B.}\
  \bibnamefont {Sahinoglu}}, \bibinfo {author} {\bibfnamefont {Jutho}\
  \bibnamefont {Haegeman}}, \ and\ \bibinfo {author} {\bibfnamefont {Frank}\
  \bibnamefont {Verstraete}},\ }\bibfield  {title} {\enquote {\bibinfo {title}
  {Matrix product operators for symmetry-protected topological phases: Gauging
  and edge theories},}\ }\href {\doibase 10.1103/PhysRevB.94.205150} {\bibfield
   {journal} {\bibinfo  {journal} {Phys. Rev. B}\ }\textbf {\bibinfo {volume}
  {94}},\ \bibinfo {pages} {205150} (\bibinfo {year} {2016})}\BibitemShut
  {NoStop}%
\bibitem [{\citenamefont {Wille}\ \emph {et~al.}(2017)\citenamefont {Wille},
  \citenamefont {Buerschaper}, ,\ and\ \citenamefont {Eisert}}]{Eisert2017}%
  \BibitemOpen
  \bibfield  {author} {\bibinfo {author} {\bibfnamefont {C.}~\bibnamefont
  {Wille}}, \bibinfo {author} {\bibfnamefont {O.}~\bibnamefont {Buerschaper}},
  , \ and\ \bibinfo {author} {\bibfnamefont {J.}~\bibnamefont {Eisert}},\
  }\bibfield  {title} {\enquote {\bibinfo {title} {Fermionic topological
  quantum states as tensor networks},}\ }\href {\doibase
  10.1103/PhysRevB.95.245127} {\bibfield  {journal} {\bibinfo  {journal} {Phys.
  Rev. B}\ }\textbf {\bibinfo {volume} {95}},\ \bibinfo {pages} {245127}
  (\bibinfo {year} {2017})}\BibitemShut {NoStop}%
\bibitem [{\citenamefont {Bultinck}\ \emph {et~al.}(2017)\citenamefont
  {Bultinck}, \citenamefont {Williamson}, \citenamefont {Haegeman},\ and\
  \citenamefont {Verstraete}}]{Frank2017}%
  \BibitemOpen
  \bibfield  {author} {\bibinfo {author} {\bibfnamefont {Nick}\ \bibnamefont
  {Bultinck}}, \bibinfo {author} {\bibfnamefont {Dominic~J.}\ \bibnamefont
  {Williamson}}, \bibinfo {author} {\bibfnamefont {Jutho}\ \bibnamefont
  {Haegeman}}, \ and\ \bibinfo {author} {\bibfnamefont {Frank}\ \bibnamefont
  {Verstraete}},\ }\bibfield  {title} {\enquote {\bibinfo {title} {Fermionic
  projected entangled-pair states and topological phases},}\ }\href {\doibase
  10.1088/1751-8121/aa99cc} {\bibfield  {journal} {\bibinfo  {journal} {J.
  Phys. A: Math. Theor.}\ }\textbf {\bibinfo {volume} {51}},\ \bibinfo {pages}
  {025202} (\bibinfo {year} {2017})}\BibitemShut {NoStop}%
\bibitem [{\citenamefont {Molnar}\ \emph {et~al.}(2018)\citenamefont {Molnar},
  \citenamefont {Ge}, \citenamefont {Schuch},\ and\ \citenamefont
  {Cirac}}]{Cirac2018}%
  \BibitemOpen
  \bibfield  {author} {\bibinfo {author} {\bibfnamefont {Andras}\ \bibnamefont
  {Molnar}}, \bibinfo {author} {\bibfnamefont {Yimin}\ \bibnamefont {Ge}},
  \bibinfo {author} {\bibfnamefont {Norbert}\ \bibnamefont {Schuch}}, \ and\
  \bibinfo {author} {\bibfnamefont {J.~Ignacio}\ \bibnamefont {Cirac}},\
  }\bibfield  {title} {\enquote {\bibinfo {title} {A generalization of the
  injectivity condition for projected entangled pair states},}\ }\href@noop {}
  {\bibfield  {journal} {\bibinfo  {journal} {J. Math. Phys.}\ }\textbf
  {\bibinfo {volume} {59}},\ \bibinfo {pages} {021902} (\bibinfo {year}
  {2018})}\BibitemShut {NoStop}%
\bibitem [{\citenamefont {Kapustin}\ \emph {et~al.}(2018)\citenamefont
  {Kapustin}, \citenamefont {Turzillo},\ and\ \citenamefont
  {You}}]{Kapustin2018}%
  \BibitemOpen
  \bibfield  {author} {\bibinfo {author} {\bibfnamefont {Anton}\ \bibnamefont
  {Kapustin}}, \bibinfo {author} {\bibfnamefont {Alex}\ \bibnamefont
  {Turzillo}}, \ and\ \bibinfo {author} {\bibfnamefont {Minyoung}\ \bibnamefont
  {You}},\ }\bibfield  {title} {\enquote {\bibinfo {title} {Spin topological
  field theory and fermionic matrix product states},}\ }\href {\doibase
  10.1103/PhysRevB.98.125101} {\bibfield  {journal} {\bibinfo  {journal} {Phys.
  Rev. B}\ }\textbf {\bibinfo {volume} {98}},\ \bibinfo {pages} {125101}
  (\bibinfo {year} {2018})}\BibitemShut {NoStop}%
\bibitem [{\citenamefont {Şahinoğlu}\ \emph {et~al.}(2021)\citenamefont
  {Şahinoğlu}, \citenamefont {Williamson}, \citenamefont {Bultinck},
  \citenamefont {Mariën}, \citenamefont {Haegeman}, \citenamefont {Schuch},\
  and\ \citenamefont {Verstraete}}]{Frank2021}%
  \BibitemOpen
  \bibfield  {author} {\bibinfo {author} {\bibfnamefont {Mehmet~Burak}\
  \bibnamefont {Şahinoğlu}}, \bibinfo {author} {\bibfnamefont {Dominic}\
  \bibnamefont {Williamson}}, \bibinfo {author} {\bibfnamefont {Nick}\
  \bibnamefont {Bultinck}}, \bibinfo {author} {\bibfnamefont {Michael}\
  \bibnamefont {Mariën}}, \bibinfo {author} {\bibfnamefont {Jutho}\
  \bibnamefont {Haegeman}}, \bibinfo {author} {\bibfnamefont {Norbert}\
  \bibnamefont {Schuch}}, \ and\ \bibinfo {author} {\bibfnamefont {Frank}\
  \bibnamefont {Verstraete}},\ }\bibfield  {title} {\enquote {\bibinfo {title}
  {Characterizing topological order with matrix product operators},}\ }\href
  {\doibase 10.1007/s00023-020-00992-4} {\bibfield  {journal} {\bibinfo
  {journal} {Ann. Henri Poincare}\ }\textbf {\bibinfo {volume} {22}},\ \bibinfo
  {pages} {563--592} (\bibinfo {year} {2021})}\BibitemShut {NoStop}%
\bibitem [{\citenamefont {Fu}(2011)}]{TCI}%
  \BibitemOpen
  \bibfield  {author} {\bibinfo {author} {\bibfnamefont {L.}~\bibnamefont
  {Fu}},\ }\bibfield  {title} {\enquote {\bibinfo {title} {Topological
  crystalline insulators},}\ }\href
  {https://journals.aps.org/prl/abstract/10.1103/PhysRevLett.106.106802}
  {\bibfield  {journal} {\bibinfo  {journal} {Phys. Rev. Lett.}\ }\textbf
  {\bibinfo {volume} {106}},\ \bibinfo {pages} {106802} (\bibinfo {year}
  {2011})}\BibitemShut {NoStop}%
\bibitem [{\citenamefont {Hsieh}\ \emph {et~al.}(2012)\citenamefont {Hsieh},
  \citenamefont {Lin}, \citenamefont {Liu}, \citenamefont {Duan}, \citenamefont
  {Bansil},\ and\ \citenamefont {Fu}}]{Fu2012}%
  \BibitemOpen
  \bibfield  {author} {\bibinfo {author} {\bibfnamefont {T.~H.}\ \bibnamefont
  {Hsieh}}, \bibinfo {author} {\bibfnamefont {H.}~\bibnamefont {Lin}}, \bibinfo
  {author} {\bibfnamefont {J.}~\bibnamefont {Liu}}, \bibinfo {author}
  {\bibfnamefont {W.}~\bibnamefont {Duan}}, \bibinfo {author} {\bibfnamefont
  {A.}~\bibnamefont {Bansil}}, \ and\ \bibinfo {author} {\bibfnamefont
  {L.}~\bibnamefont {Fu}},\ }\bibfield  {title} {\enquote {\bibinfo {title}
  {Topological crystalline insulators in the snte material class},}\ }\href
  {\doibase 10.1038/ncomms1969} {\bibfield  {journal} {\bibinfo  {journal}
  {Nat. Commun.}\ }\textbf {\bibinfo {volume} {3}},\ \bibinfo {pages} {982}
  (\bibinfo {year} {2012})}\BibitemShut {NoStop}%
\bibitem [{\citenamefont {Isobe}\ and\ \citenamefont {Fu}(2015)}]{ITCI}%
  \BibitemOpen
  \bibfield  {author} {\bibinfo {author} {\bibfnamefont {H.}~\bibnamefont
  {Isobe}}\ and\ \bibinfo {author} {\bibfnamefont {L.}~\bibnamefont {Fu}},\
  }\bibfield  {title} {\enquote {\bibinfo {title} {Theory of interacting
  topological crystalline insulators},}\ }\href
  {https://journals.aps.org/prb/abstract/10.1103/PhysRevB.92.081304} {\bibfield
   {journal} {\bibinfo  {journal} {Phys. Rev. B}\ }\textbf {\bibinfo {volume}
  {92}},\ \bibinfo {pages} {081304(R)} (\bibinfo {year} {2015})}\BibitemShut
  {NoStop}%
\bibitem [{\citenamefont {Song}\ \emph {et~al.}(2017)\citenamefont {Song},
  \citenamefont {Huang}, \citenamefont {Fu},\ and\ \citenamefont
  {Hermele}}]{reduction}%
  \BibitemOpen
  \bibfield  {author} {\bibinfo {author} {\bibfnamefont {H.}~\bibnamefont
  {Song}}, \bibinfo {author} {\bibfnamefont {S.-J.}\ \bibnamefont {Huang}},
  \bibinfo {author} {\bibfnamefont {L.}~\bibnamefont {Fu}}, \ and\ \bibinfo
  {author} {\bibfnamefont {M.}~\bibnamefont {Hermele}},\ }\bibfield  {title}
  {\enquote {\bibinfo {title} {Topological phases protected by point group
  symmetry},}\ }\href
  {https://journals.aps.org/prx/abstract/10.1103/PhysRevX.7.011020} {\bibfield
  {journal} {\bibinfo  {journal} {Phys. Rev. X}\ }\textbf {\bibinfo {volume}
  {7}},\ \bibinfo {pages} {011020} (\bibinfo {year} {2017})}\BibitemShut
  {NoStop}%
\bibitem [{\citenamefont {Huang}\ \emph {et~al.}(2017)\citenamefont {Huang},
  \citenamefont {Song}, \citenamefont {Huang},\ and\ \citenamefont
  {Hermele}}]{building}%
  \BibitemOpen
  \bibfield  {author} {\bibinfo {author} {\bibfnamefont {S.-J.}\ \bibnamefont
  {Huang}}, \bibinfo {author} {\bibfnamefont {H.}~\bibnamefont {Song}},
  \bibinfo {author} {\bibfnamefont {Y.-P.}\ \bibnamefont {Huang}}, \ and\
  \bibinfo {author} {\bibfnamefont {M.}~\bibnamefont {Hermele}},\ }\bibfield
  {title} {\enquote {\bibinfo {title} {Building crystalline topological phases
  from lower-dimensional states},}\ }\href
  {https://journals.aps.org/prb/abstract/10.1103/PhysRevB.96.205106} {\bibfield
   {journal} {\bibinfo  {journal} {Phys. Rev. B}\ }\textbf {\bibinfo {volume}
  {96}},\ \bibinfo {pages} {205106} (\bibinfo {year} {2017})}\BibitemShut
  {NoStop}%
\bibitem [{\citenamefont {Thorngren}\ and\ \citenamefont
  {Else}(2018)}]{correspondence}%
  \BibitemOpen
  \bibfield  {author} {\bibinfo {author} {\bibfnamefont {Ryan}\ \bibnamefont
  {Thorngren}}\ and\ \bibinfo {author} {\bibfnamefont {Dominic~V.}\
  \bibnamefont {Else}},\ }\bibfield  {title} {\enquote {\bibinfo {title}
  {Gauging spatial symmetries and the classification of topological crystalline
  phases},}\ }\href
  {https://journals.aps.org/prx/abstract/10.1103/PhysRevX.8.011040} {\bibfield
  {journal} {\bibinfo  {journal} {Phys. Rev. X}\ }\textbf {\bibinfo {volume}
  {8}},\ \bibinfo {pages} {011040} (\bibinfo {year} {2018})}\BibitemShut
  {NoStop}%
\bibitem [{\citenamefont {Zou}(2018)}]{SET}%
  \BibitemOpen
  \bibfield  {author} {\bibinfo {author} {\bibfnamefont {L.}~\bibnamefont
  {Zou}},\ }\bibfield  {title} {\enquote {\bibinfo {title} {Bulk
  characterization of topological crystalline insulators: Stability under
  interactions and relations to symmetry enriched u (1) quantum spin
  liquids},}\ }\href
  {https://journals.aps.org/prb/abstract/10.1103/PhysRevB.97.045130} {\bibfield
   {journal} {\bibinfo  {journal} {Phys. Rev. B}\ }\textbf {\bibinfo {volume}
  {97}},\ \bibinfo {pages} {045130} (\bibinfo {year} {2018})}\BibitemShut
  {NoStop}%
\bibitem [{\citenamefont {Po}\ \emph {et~al.}(2017)\citenamefont {Po},
  \citenamefont {Vishwanath},\ and\ \citenamefont {Watanabe}}]{230}%
  \BibitemOpen
  \bibfield  {author} {\bibinfo {author} {\bibfnamefont {H.~C.}\ \bibnamefont
  {Po}}, \bibinfo {author} {\bibfnamefont {A.}~\bibnamefont {Vishwanath}}, \
  and\ \bibinfo {author} {\bibfnamefont {H.}~\bibnamefont {Watanabe}},\
  }\bibfield  {title} {\enquote {\bibinfo {title} {Symmetry-based indicators of
  band topology in the 230 space groups},}\ }\href
  {https://www.nature.com/articles/s41467-017-00133-2} {\bibfield  {journal}
  {\bibinfo  {journal} {Nature Communications}\ }\textbf {\bibinfo {volume}
  {8}},\ \bibinfo {pages} {50} (\bibinfo {year} {2017})}\BibitemShut {NoStop}%
\bibitem [{\citenamefont {Song}\ \emph
  {et~al.}(2020{\natexlab{a}})\citenamefont {Song}, \citenamefont {Xiong},\
  and\ \citenamefont {Huang}}]{BCSPT}%
  \BibitemOpen
  \bibfield  {author} {\bibinfo {author} {\bibfnamefont {H.}~\bibnamefont
  {Song}}, \bibinfo {author} {\bibfnamefont {C.~Z.}\ \bibnamefont {Xiong}}, \
  and\ \bibinfo {author} {\bibfnamefont {S.-J.}\ \bibnamefont {Huang}},\
  }\bibfield  {title} {\enquote {\bibinfo {title} {Bosonic crystalline symmetry
  protected topological phases beyond the group cohomology proposal},}\ }\href
  {\doibase 10.1103/PhysRevB.101.165129} {\bibfield  {journal} {\bibinfo
  {journal} {Phys. Rev. B}\ }\textbf {\bibinfo {volume} {101}},\ \bibinfo
  {pages} {165129} (\bibinfo {year} {2020}{\natexlab{a}})},\ \Eprint
  {http://arxiv.org/abs/1811.06558} {arXiv:1811.06558 [cond-mat.str-el]}
  \BibitemShut {NoStop}%
\bibitem [{\citenamefont {Jiang}\ and\ \citenamefont {Ran}(2017)}]{Jiang2017}%
  \BibitemOpen
  \bibfield  {author} {\bibinfo {author} {\bibfnamefont {S.}~\bibnamefont
  {Jiang}}\ and\ \bibinfo {author} {\bibfnamefont {Y.}~\bibnamefont {Ran}},\
  }\bibfield  {title} {\enquote {\bibinfo {title} {Anyon condensation and a
  generic tensor-network construction for symmetry-protected topological
  phases},}\ }\href {\doibase 10.1103/PhysRevB.95.125107} {\bibfield  {journal}
  {\bibinfo  {journal} {Phys. Rev. B}\ }\textbf {\bibinfo {volume} {95}},\
  \bibinfo {pages} {125107} (\bibinfo {year} {2017})}\BibitemShut {NoStop}%
\bibitem [{\citenamefont {Kruthoff}\ \emph {et~al.}(2017)\citenamefont
  {Kruthoff}, \citenamefont {de~Boer}, \citenamefont {van Wezel}, \citenamefont
  {Kane},\ and\ \citenamefont {Slager}}]{Kane2017}%
  \BibitemOpen
  \bibfield  {author} {\bibinfo {author} {\bibfnamefont {J.}~\bibnamefont
  {Kruthoff}}, \bibinfo {author} {\bibfnamefont {J.}~\bibnamefont {de~Boer}},
  \bibinfo {author} {\bibfnamefont {J.}~\bibnamefont {van Wezel}}, \bibinfo
  {author} {\bibfnamefont {C.~L.}\ \bibnamefont {Kane}}, \ and\ \bibinfo
  {author} {\bibfnamefont {R.-J.}\ \bibnamefont {Slager}},\ }\bibfield  {title}
  {\enquote {\bibinfo {title} {Topological classification of crystalline
  insulators through band structure combinatorics},}\ }\href {\doibase
  10.1103/PhysRevX.7.041069} {\bibfield  {journal} {\bibinfo  {journal} {Phys.
  Rev. X}\ }\textbf {\bibinfo {volume} {7}},\ \bibinfo {pages} {041069}
  (\bibinfo {year} {2017})}\BibitemShut {NoStop}%
\bibitem [{\citenamefont {Shiozak{i}}\ \emph {et~al.}()\citenamefont
  {Shiozak{i}}, \citenamefont {Sato},\ and\ \citenamefont
  {Gomi}}]{Shiozaki2018}%
  \BibitemOpen
  \bibfield  {author} {\bibinfo {author} {\bibfnamefont {Ken}\ \bibnamefont
  {Shiozak{i}}}, \bibinfo {author} {\bibfnamefont {Masatoshi}\ \bibnamefont
  {Sato}}, \ and\ \bibinfo {author} {\bibfnamefont {Kiyonori}\ \bibnamefont
  {Gomi}},\ }\bibfield  {title} {\enquote {\bibinfo {title} {Atiyah-hirzebruch
  spectral sequence in band topology: General formalism and topological
  invariants for 230 space groups},}\ }\href@noop {} {\ }\Eprint
  {http://arxiv.org/abs/1802.06694} {arXiv:1802.06694 [cond-mat.str-el]}
  \BibitemShut {NoStop}%
\bibitem [{\citenamefont {Son{g}}\ \emph {et~al.}(2019)\citenamefont {Son{g}},
  \citenamefont {Huang}, \citenamefont {Qi}, \citenamefont {Fang},\ and\
  \citenamefont {Hermele}}]{ZDSong2018}%
  \BibitemOpen
  \bibfield  {author} {\bibinfo {author} {\bibfnamefont {Zhida}\ \bibnamefont
  {Son{g}}}, \bibinfo {author} {\bibfnamefont {Sheng-Jie}\ \bibnamefont
  {Huang}}, \bibinfo {author} {\bibfnamefont {Yang}\ \bibnamefont {Qi}},
  \bibinfo {author} {\bibfnamefont {Chen}\ \bibnamefont {Fang}}, \ and\
  \bibinfo {author} {\bibfnamefont {Michael}\ \bibnamefont {Hermele}},\
  }\bibfield  {title} {\enquote {\bibinfo {title} {Topological states from
  topological crystals},}\ }\href {\doibase 10.1126/sciadv.aax2007} {\bibfield
  {journal} {\bibinfo  {journal} {Sci. Adv.}\ }\textbf {\bibinfo {volume}
  {5}},\ \bibinfo {pages} {eaax2007} (\bibinfo {year} {2019})},\ \Eprint
  {http://arxiv.org/abs/1810.02330} {arXiv:1810.02330 [cond-mat.mes-hall]}
  \BibitemShut {NoStop}%
\bibitem [{\citenamefont {Else}\ and\ \citenamefont
  {Thorngren}(2019)}]{defect}%
  \BibitemOpen
  \bibfield  {author} {\bibinfo {author} {\bibfnamefont {D.~V.}\ \bibnamefont
  {Else}}\ and\ \bibinfo {author} {\bibfnamefont {R.}~\bibnamefont
  {Thorngren}},\ }\bibfield  {title} {\enquote {\bibinfo {title} {Crystalline
  topological phases as defect networks},}\ }\href
  {https://journals.aps.org/prb/abstract/10.1103/PhysRevB.99.115116} {\bibfield
   {journal} {\bibinfo  {journal} {Phys. Rev. B}\ }\textbf {\bibinfo {volume}
  {99}},\ \bibinfo {pages} {115116} (\bibinfo {year} {2019})}\BibitemShut
  {NoStop}%
\bibitem [{\citenamefont {Song}\ \emph
  {et~al.}(2020{\natexlab{b}})\citenamefont {Song}, \citenamefont {Fang},\ and\
  \citenamefont {Qi}}]{realspace}%
  \BibitemOpen
  \bibfield  {author} {\bibinfo {author} {\bibfnamefont {Z.}~\bibnamefont
  {Song}}, \bibinfo {author} {\bibfnamefont {C.}~\bibnamefont {Fang}}, \ and\
  \bibinfo {author} {\bibfnamefont {Y.}~\bibnamefont {Qi}},\ }\bibfield
  {title} {\enquote {\bibinfo {title} {Real-space recipes for general
  topological crystalline states},}\ }\href {\doibase
  10.1038/s41467-020-17685-5} {\bibfield  {journal} {\bibinfo  {journal}
  {Nature Communications}\ }\textbf {\bibinfo {volume} {11}},\ \bibinfo {pages}
  {4197} (\bibinfo {year} {2020}{\natexlab{b}})},\ \Eprint
  {http://arxiv.org/abs/1810.11013} {arXiv:1810.11013 [cond-mat.str-el]}
  \BibitemShut {NoStop}%
\bibitem [{\citenamefont {Shiozaki}\ \emph {et~al.}()\citenamefont {Shiozaki},
  \citenamefont {Xiong},\ and\ \citenamefont {Gomi}}]{KenX}%
  \BibitemOpen
  \bibfield  {author} {\bibinfo {author} {\bibfnamefont {Ken}\ \bibnamefont
  {Shiozaki}}, \bibinfo {author} {\bibfnamefont {Charles~Zhaoxi}\ \bibnamefont
  {Xiong}}, \ and\ \bibinfo {author} {\bibfnamefont {Kiyonori}\ \bibnamefont
  {Gomi}},\ }\bibfield  {title} {\enquote {\bibinfo {title} {Generalized
  homology and atiyah-hirzebruch spectral sequence in crystalline symmetry
  protected topological phenomena},}\ }\href@noop {} {\ }\Eprint
  {http://arxiv.org/abs/1810.00801} {arXiv:1810.00801 [cond-mat.str-el]}
  \BibitemShut {NoStop}%
\bibitem [{\citenamefont {Cheng}\ and\ \citenamefont {Wang}()}]{rotation}%
  \BibitemOpen
  \bibfield  {author} {\bibinfo {author} {\bibfnamefont {M.}~\bibnamefont
  {Cheng}}\ and\ \bibinfo {author} {\bibfnamefont {C.}~\bibnamefont {Wang}},\
  }\bibfield  {title} {\enquote {\bibinfo {title} {Rotation symmetry-protected
  topological phases of fermions},}\ }\href@noop {} {\ }\Eprint
  {http://arxiv.org/abs/1810.12308} {arXiv:1810.12308 [cond-mat.str-el]}
  \BibitemShut {NoStop}%
\bibitem [{\citenamefont {Rasmussen}\ and\ \citenamefont {Lu}(2020)}]{LuX}%
  \BibitemOpen
  \bibfield  {author} {\bibinfo {author} {\bibfnamefont {Alex}\ \bibnamefont
  {Rasmussen}}\ and\ \bibinfo {author} {\bibfnamefont {Yuan-Ming}\ \bibnamefont
  {Lu}},\ }\bibfield  {title} {\enquote {\bibinfo {title} {Classification and
  construction of higher-order symmetry protected topological phases of
  interacting bosons},}\ }\href {\doibase 10.1103/PhysRevB.101.085137}
  {\bibfield  {journal} {\bibinfo  {journal} {Phys. Rev. B}\ }\textbf {\bibinfo
  {volume} {101}},\ \bibinfo {pages} {085137} (\bibinfo {year} {2020})},\
  \Eprint {http://arxiv.org/abs/1809.07325} {arXiv:1809.07325
  [cond-mat.str-el]} \BibitemShut {NoStop}%
\bibitem [{\citenamefont {Rasmussen}\ and\ \citenamefont {Lu}()}]{YMLu2018}%
  \BibitemOpen
  \bibfield  {author} {\bibinfo {author} {\bibfnamefont {A.}~\bibnamefont
  {Rasmussen}}\ and\ \bibinfo {author} {\bibfnamefont {Y.-M.}\ \bibnamefont
  {Lu}},\ }\bibfield  {title} {\enquote {\bibinfo {title} {Intrinsically
  interacting topological crystalline insulators and superconductors},}\
  }\href@noop {} {\ }\Eprint {http://arxiv.org/abs/1810.12317}
  {arXiv:1810.12317 [cond-mat.str-el]} \BibitemShut {NoStop}%
\bibitem [{\citenamefont {Cheng}(2019)}]{Cheng2018}%
  \BibitemOpen
  \bibfield  {author} {\bibinfo {author} {\bibfnamefont {M.}~\bibnamefont
  {Cheng}},\ }\bibfield  {title} {\enquote {\bibinfo {title} {Fermionic
  lieb-schultz-mattis theorems and weak symmetry-protected phases},}\ }\href
  {\doibase 10.1103/PhysRevB.99.075143} {\bibfield  {journal} {\bibinfo
  {journal} {Phys. Rev. B}\ }\textbf {\bibinfo {volume} {99}},\ \bibinfo
  {pages} {075143} (\bibinfo {year} {2019})}\BibitemShut {NoStop}%
\bibitem [{\citenamefont {Huang}\ and\ \citenamefont
  {Hermele}(2018)}]{Hermele2018}%
  \BibitemOpen
  \bibfield  {author} {\bibinfo {author} {\bibfnamefont {S.-J.}\ \bibnamefont
  {Huang}}\ and\ \bibinfo {author} {\bibfnamefont {M.}~\bibnamefont
  {Hermele}},\ }\bibfield  {title} {\enquote {\bibinfo {title} {Surface field
  theories of point group symmetry protected topological phases},}\ }\href
  {\doibase 10.1103/PhysRevB.97.075145} {\bibfield  {journal} {\bibinfo
  {journal} {Phys. Rev. B}\ }\textbf {\bibinfo {volume} {97}},\ \bibinfo
  {pages} {075145} (\bibinfo {year} {2018})}\BibitemShut {NoStop}%
\bibitem [{\citenamefont {Huang}(2020)}]{Huang2020PRR}%
  \BibitemOpen
  \bibfield  {author} {\bibinfo {author} {\bibfnamefont {S.-J.}\ \bibnamefont
  {Huang}},\ }\bibfield  {title} {\enquote {\bibinfo {title} {4d
  beyond-cohomology topologicalphase protected by $c_2$ symmetry and its
  boundary theories},}\ }\href {\doibase 10.1103/PhysRevResearch.2.033236}
  {\bibfield  {journal} {\bibinfo  {journal} {Phys. Rev. Research}\ }\textbf
  {\bibinfo {volume} {2}},\ \bibinfo {pages} {033236} (\bibinfo {year}
  {2020})}\BibitemShut {NoStop}%
\bibitem [{\citenamefont {Huang}\ and\ \citenamefont
  {Hsu}(2021)}]{Huang2021PRR}%
  \BibitemOpen
  \bibfield  {author} {\bibinfo {author} {\bibfnamefont {S.-J.}\ \bibnamefont
  {Huang}}\ and\ \bibinfo {author} {\bibfnamefont {Y.-T.}\ \bibnamefont
  {Hsu}},\ }\bibfield  {title} {\enquote {\bibinfo {title} {Faithful derivation
  of symmetry indicators: A case study for topological superconductors with
  time-reversal and inversion symmetries},}\ }\href {\doibase
  10.1103/PhysRevResearch.3.013243} {\bibfield  {journal} {\bibinfo  {journal}
  {Phys. Rev. Research}\ }\textbf {\bibinfo {volume} {3}},\ \bibinfo {pages}
  {013243} (\bibinfo {year} {2021})}\BibitemShut {NoStop}%
\bibitem [{\citenamefont {Shang-Qiang~Ning}\ and\ \citenamefont
  {Wang}(2021)}]{Ning2021}%
  \BibitemOpen
  \bibfield  {author} {\bibinfo {author} {\bibfnamefont {Zhengqiao~Li}\
  \bibnamefont {Shang-Qiang~Ning}, \bibfnamefont {Bin-Bin~Mao}}\ and\ \bibinfo
  {author} {\bibfnamefont {Chenjie}\ \bibnamefont {Wang}},\ }\bibfield  {title}
  {\enquote {\bibinfo {title} {Anomaly indicators and bulk-boundary
  correspondences for three-dimensional interacting topological crystalline
  phases with mirror and continuous symmetries},}\ }\href@noop {} {\bibfield
  {journal} {\bibinfo  {journal} {Phys. Rev. B}\ } (\bibinfo {year}
  {2021})}\BibitemShut {NoStop}%
\bibitem [{\citenamefont {Tanaka}\ \emph {et~al.}(2012)\citenamefont {Tanaka},
  \citenamefont {Ren}, \citenamefont {Sato}, \citenamefont {Nakayama},
  \citenamefont {Souma}, \citenamefont {Takahashi}, \citenamefont {Segawa},\
  and\ \citenamefont {Ando}}]{TCIrealization1}%
  \BibitemOpen
  \bibfield  {author} {\bibinfo {author} {\bibfnamefont {Y.}~\bibnamefont
  {Tanaka}}, \bibinfo {author} {\bibfnamefont {Z.}~\bibnamefont {Ren}},
  \bibinfo {author} {\bibfnamefont {T.}~\bibnamefont {Sato}}, \bibinfo {author}
  {\bibfnamefont {K.}~\bibnamefont {Nakayama}}, \bibinfo {author}
  {\bibfnamefont {S.}~\bibnamefont {Souma}}, \bibinfo {author} {\bibfnamefont
  {T.}~\bibnamefont {Takahashi}}, \bibinfo {author} {\bibfnamefont {Kouji}\
  \bibnamefont {Segawa}}, \ and\ \bibinfo {author} {\bibfnamefont {Yoichi}\
  \bibnamefont {Ando}},\ }\bibfield  {title} {\enquote {\bibinfo {title}
  {Experimental realization of a topological crystalline insulator in snte},}\
  }\href {\doibase 10.1038/nphys2442} {\bibfield  {journal} {\bibinfo
  {journal} {Nature Physics}\ }\textbf {\bibinfo {volume} {8}},\ \bibinfo
  {pages} {800} (\bibinfo {year} {2012})}\BibitemShut {NoStop}%
\bibitem [{\citenamefont {Dziawa}\ \emph {et~al.}(2012)\citenamefont {Dziawa},
  \citenamefont {Kowalski}, \citenamefont {Dybko}, \citenamefont {Buczko},
  \citenamefont {Szczerbakow}, \citenamefont {Szot}, \citenamefont
  {{\L}usakowska}, \citenamefont {Balasubramanian}, \citenamefont {Wojek},
  \citenamefont {Berntsen}, \citenamefont {Tjernberg},\ and\ \citenamefont
  {Story}}]{TCIrealization2}%
  \BibitemOpen
  \bibfield  {author} {\bibinfo {author} {\bibfnamefont {P.}~\bibnamefont
  {Dziawa}}, \bibinfo {author} {\bibfnamefont {B.~J.}\ \bibnamefont
  {Kowalski}}, \bibinfo {author} {\bibfnamefont {K.}~\bibnamefont {Dybko}},
  \bibinfo {author} {\bibfnamefont {R.}~\bibnamefont {Buczko}}, \bibinfo
  {author} {\bibfnamefont {A.}~\bibnamefont {Szczerbakow}}, \bibinfo {author}
  {\bibfnamefont {M.}~\bibnamefont {Szot}}, \bibinfo {author} {\bibfnamefont
  {E.}~\bibnamefont {{\L}usakowska}}, \bibinfo {author} {\bibfnamefont
  {T.}~\bibnamefont {Balasubramanian}}, \bibinfo {author} {\bibfnamefont
  {B.~M.}\ \bibnamefont {Wojek}}, \bibinfo {author} {\bibfnamefont {M.~H.}\
  \bibnamefont {Berntsen}}, \bibinfo {author} {\bibfnamefont {O.}~\bibnamefont
  {Tjernberg}}, \ and\ \bibinfo {author} {\bibfnamefont {T.}~\bibnamefont
  {Story}},\ }\bibfield  {title} {\enquote {\bibinfo {title} {Topological
  crystalline insulator states in pb$_{1-x}$sn$_x$se},}\ }\href {\doibase
  10.1038/nmat3449} {\bibfield  {journal} {\bibinfo  {journal} {Nature
  Materials}\ }\textbf {\bibinfo {volume} {11}},\ \bibinfo {pages} {1023}
  (\bibinfo {year} {2012})}\BibitemShut {NoStop}%
\bibitem [{\citenamefont {Okada}\ \emph {et~al.}(2013)\citenamefont {Okada},
  \citenamefont {Serbyn}, \citenamefont {Lin}, \citenamefont {Walkup},
  \citenamefont {Zhou}, \citenamefont {Dhital}, \citenamefont {Neupane},
  \citenamefont {Xu}, \citenamefont {Wang}, \citenamefont {Sankar},
  \citenamefont {Chou}, \citenamefont {Bansil}, \citenamefont {Hasan},
  \citenamefont {Wilson}, \citenamefont {Fu},\ and\ \citenamefont
  {Madhavan}}]{TCIrealization3}%
  \BibitemOpen
  \bibfield  {author} {\bibinfo {author} {\bibfnamefont {Y.}~\bibnamefont
  {Okada}}, \bibinfo {author} {\bibfnamefont {M.}~\bibnamefont {Serbyn}},
  \bibinfo {author} {\bibfnamefont {H.}~\bibnamefont {Lin}}, \bibinfo {author}
  {\bibfnamefont {D.}~\bibnamefont {Walkup}}, \bibinfo {author} {\bibfnamefont
  {W.}~\bibnamefont {Zhou}}, \bibinfo {author} {\bibfnamefont {C.}~\bibnamefont
  {Dhital}}, \bibinfo {author} {\bibfnamefont {M.}~\bibnamefont {Neupane}},
  \bibinfo {author} {\bibfnamefont {S.}~\bibnamefont {Xu}}, \bibinfo {author}
  {\bibfnamefont {Y.~J.}\ \bibnamefont {Wang}}, \bibinfo {author}
  {\bibfnamefont {R.}~\bibnamefont {Sankar}}, \bibinfo {author} {\bibfnamefont
  {F.}~\bibnamefont {Chou}}, \bibinfo {author} {\bibfnamefont {A.}~\bibnamefont
  {Bansil}}, \bibinfo {author} {\bibfnamefont {M.~Z.}\ \bibnamefont {Hasan}},
  \bibinfo {author} {\bibfnamefont {S.~D.}\ \bibnamefont {Wilson}}, \bibinfo
  {author} {\bibfnamefont {L.}~\bibnamefont {Fu}}, \ and\ \bibinfo {author}
  {\bibfnamefont {V.}~\bibnamefont {Madhavan}},\ }\bibfield  {title} {\enquote
  {\bibinfo {title} {Observation of dirac node formation and mass acquisition
  in a topological crystalline insulator},}\ }\href
  {https://science.sciencemag.org/content/341/6153/1496/tab-figures-data}
  {\bibfield  {journal} {\bibinfo  {journal} {Science}\ }\textbf {\bibinfo
  {volume} {341}},\ \bibinfo {pages} {1496} (\bibinfo {year}
  {2013})}\BibitemShut {NoStop}%
\bibitem [{\citenamefont {Ma}\ \emph {et~al.}(2017)\citenamefont {Ma},
  \citenamefont {Yi}, \citenamefont {Lv}, \citenamefont {Wang}, \citenamefont
  {Nie}, \citenamefont {Wang}, \citenamefont {Kong}, \citenamefont {Huang},
  \citenamefont {Richard}, \citenamefont {Zhang}, \citenamefont {Yaji},
  \citenamefont {Kurado}, \citenamefont {Shin}, \citenamefont {Weng},
  \citenamefont {Bernevig}, \citenamefont {Shi}, \citenamefont {Qian},\ and\
  \citenamefont {Ding}}]{TCIrealization4}%
  \BibitemOpen
  \bibfield  {author} {\bibinfo {author} {\bibfnamefont {J.}~\bibnamefont
  {Ma}}, \bibinfo {author} {\bibfnamefont {C.}~\bibnamefont {Yi}}, \bibinfo
  {author} {\bibfnamefont {B.}~\bibnamefont {Lv}}, \bibinfo {author}
  {\bibfnamefont {Z.}~\bibnamefont {Wang}}, \bibinfo {author} {\bibfnamefont
  {S.}~\bibnamefont {Nie}}, \bibinfo {author} {\bibfnamefont {L.}~\bibnamefont
  {Wang}}, \bibinfo {author} {\bibfnamefont {L.}~\bibnamefont {Kong}}, \bibinfo
  {author} {\bibfnamefont {Y.}~\bibnamefont {Huang}}, \bibinfo {author}
  {\bibfnamefont {P.}~\bibnamefont {Richard}}, \bibinfo {author} {\bibfnamefont
  {P.}~\bibnamefont {Zhang}}, \bibinfo {author} {\bibfnamefont
  {K.}~\bibnamefont {Yaji}}, \bibinfo {author} {\bibfnamefont {K.}~\bibnamefont
  {Kurado}}, \bibinfo {author} {\bibfnamefont {S.}~\bibnamefont {Shin}},
  \bibinfo {author} {\bibfnamefont {H.}~\bibnamefont {Weng}}, \bibinfo {author}
  {\bibfnamefont {B.~A.}\ \bibnamefont {Bernevig}}, \bibinfo {author}
  {\bibfnamefont {Y.}~\bibnamefont {Shi}}, \bibinfo {author} {\bibfnamefont
  {T.}~\bibnamefont {Qian}}, \ and\ \bibinfo {author} {\bibfnamefont
  {H.}~\bibnamefont {Ding}},\ }\bibfield  {title} {\enquote {\bibinfo {title}
  {Experimental evidence of hourglass fermion in the candidate nonsymmorphic
  topological insulator khgsb},}\ }\href
  {https://advances.sciencemag.org/content/3/5/e1602415} {\bibfield  {journal}
  {\bibinfo  {journal} {Sci. Adv.}\ }\textbf {\bibinfo {volume} {3}},\ \bibinfo
  {pages} {e1602415} (\bibinfo {year} {2017})}\BibitemShut {NoStop}%
\bibitem [{\citenamefont {Zhang}\ \emph {et~al.}(2020)\citenamefont {Zhang},
  \citenamefont {Wang}, \citenamefont {Yang}, \citenamefont {Qi},\ and\
  \citenamefont {Gu}}]{dihedral}%
  \BibitemOpen
  \bibfield  {author} {\bibinfo {author} {\bibfnamefont {J.-H.}\ \bibnamefont
  {Zhang}}, \bibinfo {author} {\bibfnamefont {Q.-R.}\ \bibnamefont {Wang}},
  \bibinfo {author} {\bibfnamefont {S.}~\bibnamefont {Yang}}, \bibinfo {author}
  {\bibfnamefont {Y.}~\bibnamefont {Qi}}, \ and\ \bibinfo {author}
  {\bibfnamefont {Z.-C.}\ \bibnamefont {Gu}},\ }\bibfield  {title} {\enquote
  {\bibinfo {title} {Construction and classification of point-group
  symmetry-protected topological phases in two-dimensional interacting
  fermionic systems},}\ }\href {\doibase 10.1103/PhysRevB.101.100501}
  {\bibfield  {journal} {\bibinfo  {journal} {Phys. Rev. B}\ }\textbf {\bibinfo
  {volume} {101}},\ \bibinfo {pages} {100501(R)} (\bibinfo {year}
  {2020})}\BibitemShut {NoStop}%
\bibitem [{\citenamefont {Zhang}\ \emph {et~al.}()\citenamefont {Zhang},
  \citenamefont {Yang}, \citenamefont {Qi},\ and\ \citenamefont
  {Gu}}]{wallpaper}%
  \BibitemOpen
  \bibfield  {author} {\bibinfo {author} {\bibfnamefont {J.-H.}\ \bibnamefont
  {Zhang}}, \bibinfo {author} {\bibfnamefont {S.}~\bibnamefont {Yang}},
  \bibinfo {author} {\bibfnamefont {Y.}~\bibnamefont {Qi}}, \ and\ \bibinfo
  {author} {\bibfnamefont {Z.-C.}\ \bibnamefont {Gu}},\ }\bibfield  {title}
  {\enquote {\bibinfo {title} {Real-space construction of crystalline
  topological superconductors and insulators in 2d interacting fermionic
  systems},}\ }\href@noop {} {\ }\Eprint {http://arxiv.org/abs/2012.15657}
  {arXiv:2012.15657 [cond-mat.str-el]} \BibitemShut {NoStop}%
\bibitem [{\citenamefont {Ouyang}\ \emph {et~al.}()\citenamefont {Ouyang},
  \citenamefont {Wang}, \citenamefont {Gu},\ and\ \citenamefont
  {Qi}}]{resolution}%
  \BibitemOpen
  \bibfield  {author} {\bibinfo {author} {\bibfnamefont {Y.}~\bibnamefont
  {Ouyang}}, \bibinfo {author} {\bibfnamefont {Q.-R.}\ \bibnamefont {Wang}},
  \bibinfo {author} {\bibfnamefont {Z.-C.}\ \bibnamefont {Gu}}, \ and\ \bibinfo
  {author} {\bibfnamefont {Y.}~\bibnamefont {Qi}},\ }\bibfield  {title}
  {\enquote {\bibinfo {title} {Computing classification of interacting
  fermionic symmetry-protected topological phases using topological
  invariants},}\ }\href@noop {} {\ }\Eprint {http://arxiv.org/abs/2005.06572}
  {arXiv:2005.06572 [cond-mat.str-el]} \BibitemShut {NoStop}%
\bibitem [{\citenamefont {Atiyah}\ and\ \citenamefont
  {Hirzebruch}(2003)}]{AHSS}%
  \BibitemOpen
  \bibfield  {author} {\bibinfo {author} {\bibfnamefont {M.~F.}\ \bibnamefont
  {Atiyah}}\ and\ \bibinfo {author} {\bibfnamefont {F.}~\bibnamefont
  {Hirzebruch}},\ }\href@noop {} {\emph {\bibinfo {title} {Topological Library:
  Part 3: Spectral Sequences in Topology}}}\ (\bibinfo  {publisher} {World
  Scientific},\ \bibinfo {year} {2003})\BibitemShut {NoStop}%
\bibitem [{\citenamefont {Kitaev}(2001)}]{Majorana}%
  \BibitemOpen
  \bibfield  {author} {\bibinfo {author} {\bibfnamefont {A.}~\bibnamefont
  {Kitaev}},\ }\href
  {https://iopscience.iop.org/article/10.1070/1063-7869/44/10S/S29/meta}
  {\bibfield  {journal} {\bibinfo  {journal} {Phys. Usp.}\ }\textbf {\bibinfo
  {volume} {44}},\ \bibinfo {pages} {131} (\bibinfo {year} {2001})}\BibitemShut
  {NoStop}%
\bibitem [{\citenamefont {Fidkowski}\ and\ \citenamefont
  {Kitaev}(2011{\natexlab{b}})}]{1Dfermion}%
  \BibitemOpen
  \bibfield  {author} {\bibinfo {author} {\bibfnamefont {L.}~\bibnamefont
  {Fidkowski}}\ and\ \bibinfo {author} {\bibfnamefont {A.}~\bibnamefont
  {Kitaev}},\ }\href
  {https://journals.aps.org/prb/abstract/10.1103/PhysRevB.83.075103} {\bibfield
   {journal} {\bibinfo  {journal} {Phys. Rev. B}\ }\textbf {\bibinfo {volume}
  {83}},\ \bibinfo {pages} {075103} (\bibinfo {year}
  {2011}{\natexlab{b}})}\BibitemShut {NoStop}%
\bibitem [{foo()}]{footnote}%
  \BibitemOpen
  \href@noop {} {\bibinfo  {journal} {For $G_v=0$, the only one possible
  nontrival dangling mode is Majorana zero mode; for $G_v=\mathbb{Z}_2$, there
  are two possible nontrival dangling modes: Majorana zero mode and complex
  fermion}\ }\BibitemShut {NoStop}%
\bibitem [{\citenamefont {GAP}(2021)}]{GAP}%
  \BibitemOpen
\bibfield  {journal} {  }\bibfield  {author} {\bibinfo {author} {\bibnamefont
  {GAP}},\ }\bibfield  {title} {\enquote {\bibinfo {title} {Gap - groups,
  algorithms, and programming, version 4.11.1},}\ }\href
  {https://www.gap-system.org/} {\bibfield  {journal} {\bibinfo  {journal}
  {https://www.gap-system.org/}\ } (\bibinfo {year} {2021})}\BibitemShut
  {NoStop}%
\bibitem [{\citenamefont {Ellis}(2020)}]{HAP}%
  \BibitemOpen
  \bibfield  {author} {\bibinfo {author} {\bibfnamefont {G.}~\bibnamefont
  {Ellis}},\ }\bibfield  {title} {\enquote {\bibinfo {title} {Hap, homological
  algebra programming, version 1.28},}\ }\href
  {http://hamilton.nuigalway.ie/Hap/www/} {\bibfield  {journal} {\bibinfo
  {journal} {http://hamilton.nuigalway.ie/Hap/www/}\ } (\bibinfo {year}
  {2020})}\BibitemShut {NoStop}%
\end{thebibliography}
\end{document}